\documentclass[twocolumn]{aastex63}

\usepackage{amsmath,amsthm,amssymb,amsfonts,ulem}

\received{\today}
\revised{\today}
\accepted{\today}
\submitjournal{AJ}

\shorttitle{SPHEREx redshift simulations}
\shortauthors{Feder et al.}
\newcommand{\lp}{\texttt{LePhare}}
\newcommand{\ha}{\textrm{H}$\alpha$}
\newcommand{\hb}{\textrm{H}$\beta$}
\newcommand{\pa}{P$\alpha$}
\newcommand{\oiii}{[\textrm{OIII}]}
\newcommand{\oii}{[O{\small II}]}
\newcommand{\nii}{[N{\small II}]}
\newcommand{\brown}{\texttt{B14}}
\newcommand{\logm}{$\log(M_\star/M_\odot)$}

\newcommand{\ebv}{E$(B-V)$}
\newcommand{\um}{$\mu$m}
\newcommand{\persqdeg}{deg$^{-2}$}
\newcommand{\sqdeg}{deg$^2$}

\newcommand{\ergcms}{erg cm$^{-2}$ s$^{-1}$}

\definecolor{Cyan}{rgb}{0.8, 0.13, 0.16}

\definecolor{Green}{rgb}{0.35,0.45,0.25}

\graphicspath{{./}{figures/}}

\begin{document}

\title{The Universe SPHEREx Will See: Empirically Based Galaxy Simulations and Redshift Predictions}

\correspondingauthor{Richard M. Feder}
\email{rfederst@caltech.edu}

\author[0000-0001-5382-6138]{Richard M. Feder}
\affiliation{California Institute of Technology, 1200 E California Blvd, 91125, USA}

\author[0000-0001-5382-6138]{Daniel C. Masters}
\affiliation{IPAC, California Institute of Technology, 1200 E California Blvd, 91125, USA}

\author[0000-0003-1954-5046]{Bomee Lee}
\affiliation{Korea Astronomy and Space Science Institue (KASI), Daejeon, Korea}

\author[0000-0002-5710-5212]{James J. Bock}
\affiliation{California Institute of Technology, 1200 E California Blvd, 91125, USA}

\author[0000-0003-1954-5046]{Yi-Kuan Chiang}
\affiliation{Institute of Astronomy \& Astrophysics, Academia Sinica, Taiwan}

\author[0000-0003-1954-5046]{Ami Choi}
\affiliation{NASA Goddard Space Flight Center, 8800 Greenbelt Rd, Greenbelt, MD 20771, USA}

\author[0000-0001-7432-2932]{Olivier Dor\'e}
\affiliation{Jet Propulsion Laboratory, California Institute of Technology, Pasadena, California 91109, USA}
\affiliation{California Institute of Technology, 1200 E California Blvd, 91125, USA}

\author[0000-0003-1954-5046]{Shoubaneh Hemmati}
\affiliation{IPAC, California Institute of Technology, 1200 E California Blvd, 91125, USA}

\author[0000-0002-7303-4397]{Olivier Ilbert}
\affiliation{Aix Marseille Univ, CNRS, CNES, LAM, Marseille, France}

\begin{abstract}

% Making predictions and testing methodologies for upcoming missions require excellent models of the universe. In the case of SPHEREx, the Spectro-Photometer for the History of the Universe, Epoch of Reionization, and Ices Explorer, this is particularly challenging as the properties galaxies will display at SPHEREx wavelength ($0.75-5\mu$m) and resolution ($R\sim35-130$) are currently poorly constrained. 
We simulate galaxy properties and redshift estimation for SPHEREx, the next NASA Medium Class Explorer. To make robust models of the galaxy population and test spectro-photometric redshift performance for SPHEREx, we develop a set of synthetic spectral energy distributions based on detailed fits to COSMOS2020 photometry spanning $0.1-8$ $\mu$m. Given that SPHEREx obtains low-resolution spectra, emission lines will be important for some fraction of galaxies. Here we expand on previous work, using better photometry and photometric redshifts from COSMOS2020, and tight empirical relations to predict robust emission line strengths and ratios. A second galaxy catalog derived from the GAMA survey is generated to ensure the bright ($m_{AB}<18$ in the $i$-band) sample is representative over larger areas. Using template fitting to estimate photometric continuum redshifts, we forecast recovery of 19 million galaxies over 30000 deg$^2$ with redshifts better than $\sigma_z<0.003(1+z)$, 445 million with $\sigma_z<0.1(1+z)$ and 810 million with $\sigma_z<0.2(1+z)$. We also find through idealized tests that emission line information from spectrally dithered flux measurements can yield redshifts with accuracy beyond that implied by the naive SPHEREx channel resolution, motivating the development of a hybrid continuum-line redshift estimation approach.
\end{abstract}

\keywords{Cosmology (343), Spectrophotometry (1556), Redshift surveys (1378), Emission line galaxies (459), Large-scale structure of the universe (902)}

\section{Introduction} \label{sec:intro}

The Spectro-Photometer for the History of the Universe, Epoch of Reionization, and Ices Explorer (SPHEREx) is the next NASA Medium Class Explorer mission which is planned for launch in early 2025. Using a wide-field 20 cm diameter telescope with an instantaneous field of view of $3.5^{\circ}\times11.3^{\circ}$ for each of two $1\times3$ detector mosaics, SPHEREx will conduct the first all-sky spectrophotometric survey in the near infrared (NIR) at wavelengths between 0.75 $\mu$m and 5 $\mu$m through four consecutive surveys over the nominal two year mission.

SPHEREx and other cosmology missions generally need to build and test methodologies using the best possible simulated data. SPHEREx aims to constrain the large-scale distribution of galaxies in order to put constraints on primordial non-Gaussianity \citep{dore14}. This measurement will need excellent redshifts for hundreds of millions of galaxies, with robust control and understanding of statistical and systematic errors. In the case of SPHEREx, this is particularly challenging as the properties galaxies will display at infrared wavelengths ($0.75-5$ $\mu$m) with low-resolution ($R\sim35-130$) spectroscopy are currently poorly constrained.

% \shooby{Mention first efforts on spherex simulation here maybe \citep{Stickley2016}}
\cite{Stickley2016} illustrated the use of synthetic SPHEREx spectrophotometry to measure redshifts of a large sample of bright galaxies to high accuracy ($\gtrsim 10^{7}$ galaxies over the full sky with redshift accuracy $\sigma_{z}<0.003(1+z)$, and many more with accuracy at the $\sim 1-10\%$ level). These simulations were performed on model spectra inferred from the Cosmological Evolution Survey field \citep[COSMOS;][]{COSMOS}, for which the complexity of the galaxy population is well constrained through deep, 30-band photometry spanning $0.1-8$ $\mu$m. 

Given the spectral resolution and infrared coverage of SPHEREx, emission line galaxies (ELGs) are an important population to consider. Nebular emission lines, typically rest-frame optical and UV (\ha, \oii, \oiii, Ly$\alpha$), are the targets of numerous existing and upcoming surveys such as the extended Baryon Acoustic Oscillations Spectroscopic Survey \citep[eBOSS;][]{eboss_elg}, the \emph{Euclid} Wide and Deep surveys \citep[using slitless spectroscopy,][]{euclid}, the Dark Energy Spectroscopic Instrument ELG survey \citep[DESI;][]{desi_elg}, the \emph{Roman} High Latitude Spectroscopic Survey \citep[HLSS;][]{wang21}, the Prime Focus Spectrograph Galaxy Evolution Survey \citep[PFS;][]{pfs}, the Physics of the Accelerating Universe Survey \citep[PAUS;][]{eriksen2019paus,alarcon2021paus}, and the Javalambre-Physics of the Accelerating Universe Astrophysical Survey \citep[J-PAS;][]{benitez09}, among others. These surveys promise to deliver emission line measurements for tens of millions of galaxies, increasing the size of existing samples by an order of magnitude. 

Properly identified emission lines serve as anchors for precise redshift measurements, making ELGs the target of modern large scale structure studies probing cosmic expansion and acceleration \citep{desi_elg, roman_elg}. Emission line strengths and ratios are also commonly used as observational proxies for intrinsic galaxy properties such as the star formation rate (SFR). Upcoming ELG surveys will chart galaxy evolution and  formation in unprecedented detail, enabling studies of galaxy properties both across a broad range of cosmic history and as a function of local environment \citep{pharo20, scoville13}. 

SPHEREx is unique in its potential to deliver precise redshifts using low-resolution spectroscopy. Both active and quiescent galaxies can yield high quality redshifts through accurate modeling of their continuua. SPHEREx will help in obtaining precise redshifts for a large number of sources by extending source flux measurements into the NIR. In the case where emission lines are used to improve redshifts for continuum-selected galaxies, there may be a large number of sources where proper modeling of low-significance emission lines is important. This places requirements on the accuracy of continuum modeling, however correlations between lines and continuua measured by SPHEREx can potentially break degeneracies of single-line identifications that plague existing spectroscopic surveys \citep{line_misidentify, vipers14, fmos}. Because SPHEREx is an imaging survey, no pre-selection is required to isolate ELG targets. This distinguishes SPHEREx from fiber-based spectroscopic measurements that are optimized with pre-selected targets. In this sense SPHEREx's survey strategy reduces the impact of target selection on ELG samples and opens up the prospect of blind searches for emission lines over large portions of the sky \citep[e.g.,][]{muse_blind}.

% This opens up the space of potential studies that can be done using SPHEREx spectra, which will be available across the full sky \richard{Mike: in what ways?}.

% \Oli{Needs a paragraph explaining a little more what we do and why we do it. We go straight from general intro to details.} \Oli{Needs to quote other efforts to produce synthetic spectral catalog, eg for DESI, eBOSS, COSMOS, Roman, etc. I guess what is specific to us is that we want the continuum AND multiple lines which is not the case for all these surveys. That's really one of the unique feature of this work, correct?}

% These requirements are common to narrowband photometric surveys. For example, J-PAS plans to measure galaxies over several thousand square degrees in 54 narrow bands (and two broad-bands) spanning 0.37-0.91 $\mu$m. Based on predictions from photometry taken across 1 deg$^2$ through the mini-JPAS survey, the full J-PAS is expected to reach depths of $m_{AB}\sim 22-23.5$ \citep{bonoli_JPAS, miniJPAS} across the narrow bands. SPHEREx extends this to longer wavelengths, where a paucity of measurements and large uncertainties in galaxy properties at higher redshifts complicate our ability to forecast what SPHEREx will observe.

The goal of this work is to create and use simulated galaxies to demonstrate the accurate measurement of redshifts using SPHEREx data. Our current best constraints on the distribution of galaxy spectra over the SPHEREx bandpass come from COSMOS2020 photometry. However, the resolution is poorly matched to SPHEREx because COSMOS photometry cannot precisely constrain the emission line properties of sources over a range of redshifts. To tackle this issue, we use empirical \citep[][hereafter \texttt{B14}]{Brown14} and model-based COSMOS templates to fit the COSMOS2020 photometry, which yields a realistic distribution of SED continua. To model emission lines, we turn to tight empirical relations for how line strengths and ratios scale with redshift, stellar mass, and spectral type. After developing the line prediction method, we test it through comparisons at the population level (line luminosity functions, line ratio distributions, line equivalent widths) and with individual source comparisons using spectroscopic catalogs covering the COSMOS field.
% \Oli{What is covered by etc. Say it explicitly.}

The catalog we develop is therefore a faithful representation of the galaxy population as constrained by COSMOS2020 \citep{cosmos2020} and numerous published studies on emission lines. Our catalog is similar in qualities to the synthetic Emission Line COSMOS catalog \citep[EL-COSMOS,][]{elcosmos_saito}, which is derived from COSMOS2015 photometry \citep{cosmos2015}. Emission lines in that catalog are modeled though a prescription for \hb\ based on estimated star formation rates and assumed metallicities $Z>0.2Z_{\odot}$ to obtain line ratios (which are effectively fixed aside from dust attenuation). The predictions in this work are similar to those in \cite{elcosmos_saito}, however our model differs by 1) using a combination of empirically measured and model-based templates, rather than a large grid of continuum models, and 2) combining empirical scaling relations for lines with locally calibrated ($z\sim 0$) emission line equivalent width (EW) measurements. 

Because SPHEREx is an all-sky mission, we complement the COSMOS catalog with wider survey data from the Galaxy and Mass Assembly survey \citep[GAMA,][]{GAMA} and associated multi-wavelength photometry to ensure that the distribution of bright galaxies in our simulated sample is representative of the full sky (our COSMOS catalog only covers $1.27 $ deg$^2 = 3.9\times 10^{-4}$ sr, while the GAMA footprint is 217 deg$^2$).

Our procedure for painting emission lines onto continuum estimates from template fits, implemented in the tool Conditional LIne Painting on Synthetic Spectra (\texttt{CLIPonSS}), is described in \S \ref{sec:meth}, and the synthetic line catalog is validated against several measurements in the literature in \S \ref{sec:validation}. The details of converting these SEDs into synthetic SPHEREx observations are presented in \S \ref{sec:sphx}, where we also consider the coverage of SPHEREx sources by external catalogs. In \S \ref{sec:lp} we forecast recovery of galaxy continuum redshifts by running the photometric redshift template fitting code from \cite{Stickley2016} on synthetic photometry, also showing initial demonstrations of redshift estimation that utilize low-resolution, spectrally dithered line flux measurements. 

All apparent magnitudes are specified in the AB magnitude system \citep{okegunn83}.

\section{Synthetic Spectral Library}
\label{sec:meth}
% In this section we describe the set of empirical and model-based galaxy templates which constitute our SED library (\S \ref{sec:sedlib}), our procedure fitting these templates to multi-wavelength photometry (\S \ref{sec:tempfit}) and our prescription for adding emission lines which couples these template fits with empirical scaling relations constrained by existing measurements (\S \ref{sec:elmodl}).

Our approach for predicting galaxy emission-line properties is to combine observations from the COSMOS survey with empirical models. The COSMOS photometry has sufficient depth and wavelength coverage to constrain galaxy spectral energy distributions (SEDs) over the SPHEREx bandpass. Our model then predicts emission lines expected for galaxies based on their stellar masses, redshifts, and best-fit spectral templates. Predicting emission line strengths is feasible because of the empirical observed correlations between the lines and other galaxy properties arising from the evolution of the mass-metallicity relation \citep[][]{tremonti2004}, the galaxy main sequence \citep{daddi07, whitaker12, speagle_ms} and the global star formation evolution \citep{Connolly97, MadauDickinson}.

\begin{figure*}
    \centering
    \includegraphics[width=0.9\linewidth]{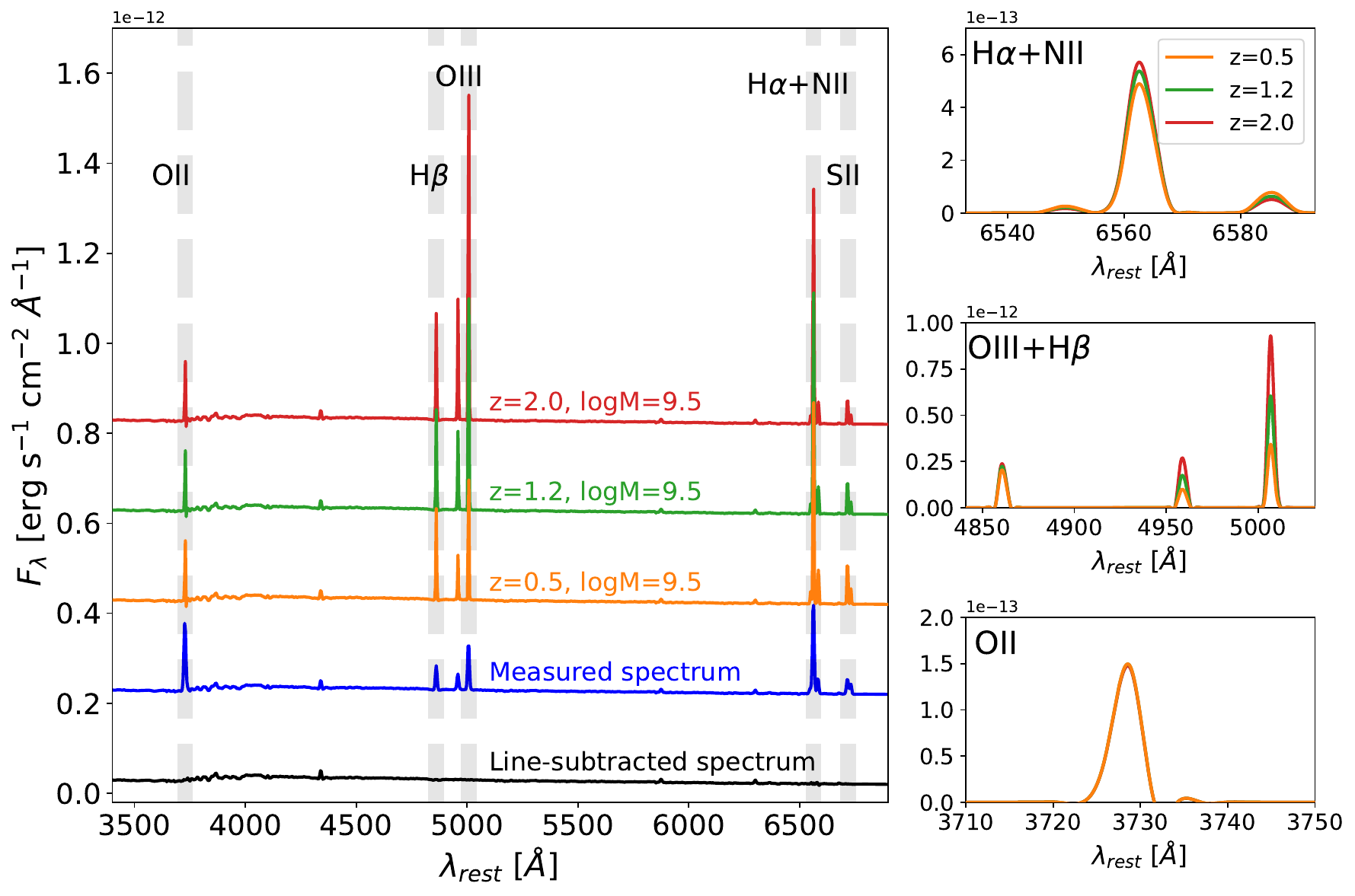}
    \caption{Track of synthetic galaxy SEDs as a function of redshift demonstrating our model procedure. For a given source, we start with the best-fit measured spectrum (blue) fit as a template to broad band photometry. We then subtract the existing lines (black), and insert new lines according to our empirical prescription. This is shown for three redshifts: $z=0.5$ (orange); 1.2 (green); and 2.0 (red), for a fixed stellar mass \logm = 9.5. Spectra are shifted vertically for visual purposes. The right hand panels show the three lines/line complexes at higher resolution for the three redshifts. \oii\ is modeled with a single Gaussian rather than as a doublet.}
    \label{fig:sed_tracks}
\end{figure*}

The procedure for assigning emission lines to galaxy continuua is illustrated in Fig. \ref{fig:sed_tracks}. Unlike the COSMOS templates, which lack emission lines, many \cite{Brown14} templates have lines which we rescale when painting new lines onto the synthetic SEDs. In this way, the line strengths in our empirical model are calibrated to observed galaxies at low redshift ($z<0.1$).

\subsection{Multi-wavelength photometry}

\subsubsection{COSMOS2020}

The galaxy simulations rely on multi-wavelength photometry from COSMOS2020, which comprises two catalogs with 30-band photometry spanning $0.1-8$ $\mu$m \citep{cosmos2020}. In particular we use the COSMOS2020 catalog derived from \texttt{The Farmer}, a profile-fitting tool for multi-wavelength photometry, along with associated photometric redshifts.

We select on classified galaxies with $i<25$ and flux measurements in at least one NIR band (UltraVISTA $J$ and $H$ bands). We further select on sources with consistent redshift measurements from both sets of COSMOS2020 photometry ($\hat{\sigma_{z}}< 0.2(1+z)$,  $|\hat{z}_{\textrm{Classic}} - \hat{z}_{\textrm{Farmer}}|/(1+\hat{z}_{\textrm{Farmer}}) < 0.15$), selecting galaxies as classified by \lp\ \citep[see \S 5.1 in][]{cosmos2020}. The majority of sources with this selection have accurate redshifts ($\sigma_{z}< 0.02(1+z)$) and sufficiently precise photometry to estimate stellar masses. In total our selection yields a sample of 166,014 galaxies covering an effective area of 1.27 deg$^2$. 

% The median reduced chi-squared of the fits is $\chi^2_{red}=2.2$, with less than 1\% of objects having $\chi^2_{red} > 10$. As seen  can be understood as coming from underestimated flux uncertainties \richard{From Jamie: how do you know this large $\chi^2$ isn't due to model shortcomings? Seems more plausible to me!}, perhaps due to unaccounted correlated noise in creating the mosaics. Our results are roughly consistent with \cite{cosmos2020}, in which the flux uncertainties were inflated by a factor of two to yield calibrated redshift uncertainties.

\subsubsection{Bright, low-redshift galaxies from the GAMA survey}
\label{sec:gama}
There are limitations in building a representative synthetic catalog from COSMOS2020 alone. Given the size of the COSMOS field, the diversity of spectra will be underestimated due to cosmic variance. Likewise, the limited volume probed by COSMOS suggests a limited low redshift sample, especially for massive galaxies. The COSMOS2020 galaxy catalog is limited to sources with $i>18$, with any sources brighter than this classified as stars.

For these reasons we supplement the COSMOS2020 catalog using a combination of spectroscopic measurements from the Galaxy and Mass Assembly (GAMA) survey \citep{GAMA} and corresponding twelve-band photometry ranging from the far UV to the infrared. We select sources with $i<18$ and a designated science class (\texttt{SC}) of 8. This is the selection for the primary science catalog used by GAMA -- the catalog has spectroscopic completeness of 98 per cent down to $r < 19.5$ \citep{GAMAphot}. In total we obtain 44,135 sources across four fields with effective areas of 55, 57, 57 and 48 square degrees. We fit the same template library described in \S\ref{sec:tempfit} to the available broad-band photometry with fixed redshifts.

\subsection{Galaxy Template SEDs}
\label{sec:sedlib}

In total we utilize 160 templates to fit observed galaxy photometry (\S \ref{sec:tempfit}) and then generate synthetic SEDs and SPHEREx spectrophotometry (\S \ref{sec:sphx}). The two sets of templates described in this section complement each other in terms of reproducing observed colors and galaxy types.

\subsubsection{\brown\ templates}
% \Oli{So Brown et al is based on MAGPHYS template fitted to data}
The collection of 129 measured galaxy SEDs from \cite{Brown14} comprise a broad range of galaxy archetypes from different stages of evolution and environment. The SEDs are constructed from a combination of optical \citep{moustakas1, moustakas2} and infrared spectroscopy from the \emph{Spitzer} Space Telescope Infrared Spectrograph (IRS) for $5.3-38$ $\mu$m \citep{spitzer, houck04} and \emph{Akari} \citep{akari} Infrared Camera (IRC) spectroscopy for $2.5-5$ $\mu$m (when available), spanning beyond the wavelength range of SPHEREx observations. Regions of the SEDs without coverage are interpolated using model spectra fit using the \texttt{MAGPHYS} model \citep{magphys}, which utilizes a stellar population synthesis (SPS) model (derived from the same set of \texttt{BC03} templates described in the next sub-section) and a self-consistent prescription for dust emission, absorption and polycyclic aromatic hydrocarbon (PAH) emission in the infrared. While the measured spectra have little to no coverage in the near infrared, the sections of interpolated spectra are calibrated against existing broadband photometry covering the range from 2MASS \citep{2MASS}, \emph{Spitzer} and the Wide-field Infrared Space Explorer \citep[WISE;][]{WISE}. The \brown\ galaxy SEDs constitute a diverse template basis for fitting galaxy photometry and reproducing observed galaxy colors. Many of the galaxies in the \brown\ sample have well measured optical emission lines, which we use to calibrate our line model locally (i.e., $z\sim 0$) before extrapolating to higher redshifts.

\subsubsection{COSMOS templates}
Thirty one of the templates are SEDs used in \cite{ilbert09}, which include templates from \cite{poletta07} and twelve model-based templates made from \cite{BC03} (\texttt{BC03} templates, hereafter), which are generated from SPS models along the starburst track and for passive elliptical galaxies. These templates were initially used in \cite{cosmos2020} to fit the photometric redshifts of COSMOS2020 galaxies. They complement the \brown\ templates, as the \brown\ templates come from low-redshift galaxies that may not be representative of higher redshift populations.

% The Brown template SEDs span a wide range of galaxy types, however they may not be representative of higher redshift galaxies. and in particular active/star-forming galaxies; \richard{moreover, they include intrinsic dust attenuation. From Jamie: if from the galaxy, this is a good thing not a shortcoming. Do you mean from the Milky Way?} 

% \ilbert{This section needs to be clarified. I am not sure of the models you are using. I have the impression that they are my models from Ilbert+09? if yes, that's a mix of Pollett+03 templates and selected ages of some BC03 templates. if you send me the list of templates, I can tell you.}
% \subsection{Continuum fits}

\subsection{Template fits to multi-band photometry}
\label{sec:tempfit}

The templates described in the previous section are fit to multi-wavelength photometry from the COSMOS and GAMA extragalactic surveys. We use the SED fitter \texttt{Fitcat}, used previously in \cite{Stickley2016}, to derive continuum models and galaxy properties. These properties include stellar mass \logm, dust extinction \ebv, dust law and index of the best-fit galaxy template. Note that the photometric redshifts from COSMOS2020 are taken as fixed within \texttt{Fitcat}.

% \Oli{$M$ or $M_\star$ for notation?}
% This evolution with $i$ is expected given that the \brown\ SEDs are calibrated on bright, nearby objects, which is what in part motivates the use of \texttt{BC03} templates altogether. 

% \Oli{Maybe say why the median $\chi^2$ is bad?}\Oli{And this evolution with $i$ is expected given that Brown et al. is calibrated on bright objects and you added BC03 exactly for that purpose? Maybe say so?}

% We select all sources with $i<25$ \richard{what is final selection?}~\dan{Selection we used was: HSC\_i\_MAG $<$ 25 \& (UVISTA\_H\_FLUX $>$ 0 OR UVISTA\_J\_FLUX $>$ 0) \& (lp\_zPDF\_u68-lp\_zPDF\_l68)/(1+lp\_zbest\_tractor) $<$ 0.2 \& lp\_type == 0 \& abs(lp\_zbest\_tractor - lp\_zbest\_classic) / (1+lp\_zbest\_tractor) $<$ 0.15. Basically brighter than i=25, with consistent redshift estimates from both phot catalogs, and flux measurement in at least one NIR band}, 
%  \Oli{True. Maybe put that later or say what we do about it now?}

% \Oli{You need a little more intro material and for each section and subsection to explain what is coming as it reads rather disjoint now. A little bit of perspective would help. We aim to do this... and to do so we will first do this, then that, ...}

\begin{figure*}
  \centering
    \includegraphics[width=0.95\linewidth]{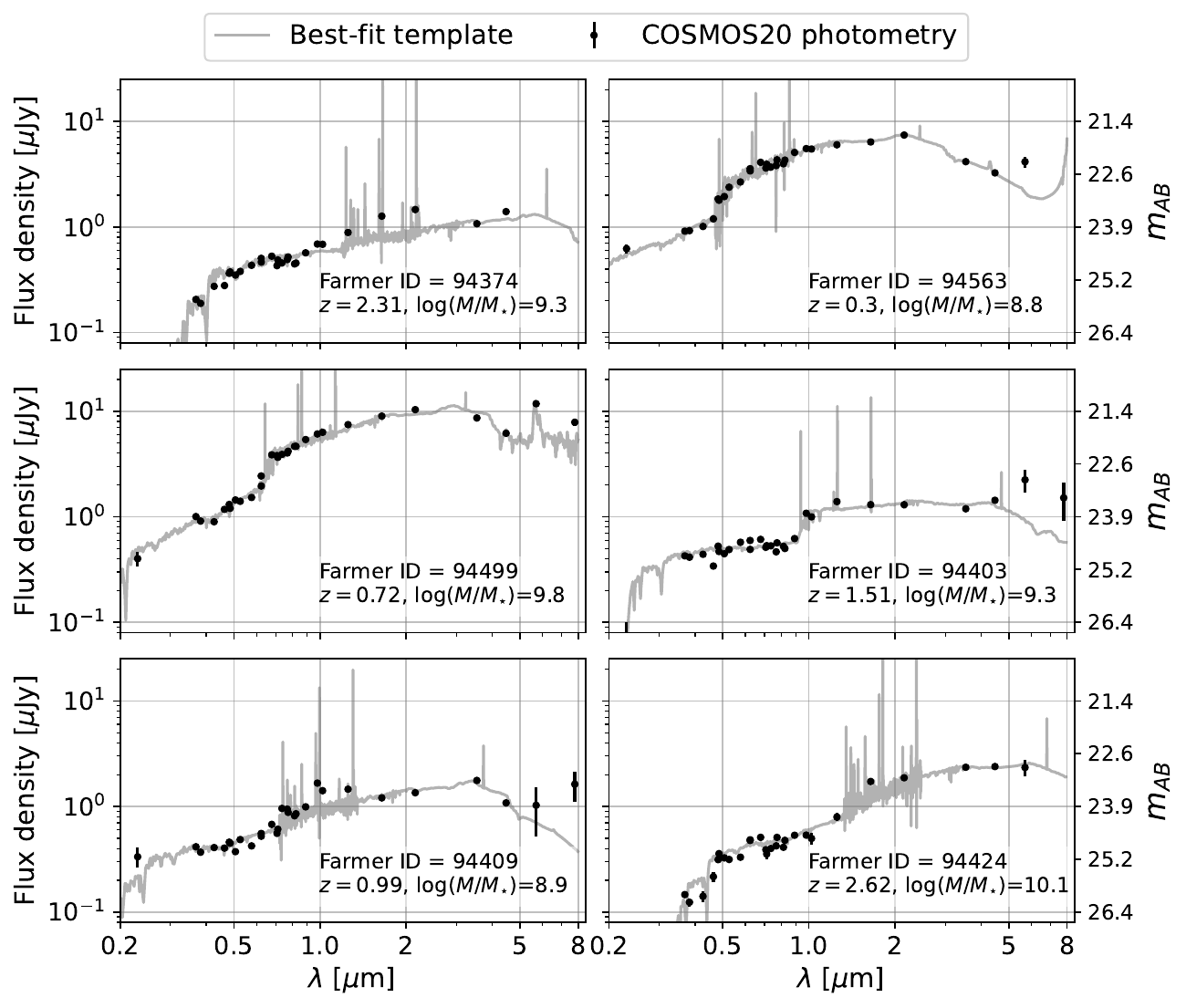}
    \caption{A selection of six spectral energy distribution (SED) fits to multi-band COSMOS2020 photometry for galaxies of various redshifts and stellar masses, indexed by Tractor IDs from the COSMOS2020 \texttt{Farmer} catalog. In some cases the COSMOS2020 photometry are sensitive to strong emission lines, for example in the top and bottom left panels.}
    \label{fig:cosmosfitcat}
\end{figure*}

Figure \ref{fig:cosmosfitcat} shows SED fits to COSMOS2020 photometry for six example galaxies. As can be seen, the SED templates are able to capture the properties of both star-forming and quiescent galaxies in our sample. The median reduced chi-squared of the fits is $\chi^2_{red}=2.2$, with less than 1\% of objects having $\chi^2_{red} > 10$. The higher than expected median $\chi^2$ is driven in part by the inability of the model templates to capture the diversity of galaxy properties, in particular at long wavelengths where emission from PAHs is difficult to model. In addition, the flux uncertainties for bright sources may tend to be underestimated \citep{cosmos2020} and certain correlated uncertainties across wavelengths are neglected. A large fraction of catalog sources are best fit by a small subset ($<10$) of the 160 templates, where the ``best-fit template" refers to the template corresponding to the smallest model $\chi^2$ compared with the data. The fraction of sources best fit by the COSMOS templates increases from thirty per cent at $i=18$ to forty percent at $i=25$. Likewise, for the GAMA sample, the fraction of sources best fit by COSMOS templates ranges from 35\% to 55\% across $13<i<18$. These trends are consistent with our expectations, as the \brown\ templates are calibrated to (local) bright objects. 

% Our results are roughly consistent with \cite{cosmos2020}, in which the flux uncertainties were inflated by a factor of two to yield calibrated redshift uncertainties.

\subsection{Emission line model}
\label{sec:elmodl}
The method used to assign emission lines to template continua that are fit to COSMOS2020 and GAMA sources consists of:
\begin{itemize}
    \item Inferred \ha\ and \oii\ equivalent widths for each galaxy based on its best-fit spectral template and/or UV continuum;
    \item Observed scaling relations of \nii/\ha\ and \oiii/\hb\ with redshift and stellar mass; and
    \item The (redshift-dependent) nebular extinction correction from \cite{elcosmos_saito}.
\end{itemize}
With these measured empirical relations we derive realistic emission lines expected for the COSMOS galaxies, which are then painted onto the best-fit template continua. 
% This is the starting point of the simulation.

% The collection of measured galaxy SEDs from \cite{Brown14} comprise a broad range of galaxy archetypes from different stages of evolution and environment. The SEDs are constructed from a combination of optical \citep{moustakas1, moustakas2} and infrared spectroscopy from \emph{Spitzer} \citep{spitzer, houck04} and \emph{Akari} \citep{akari}, spanning the full wavelength range of SPHEREx observations. Regions of the SEDs without coverage \Oli{So Brown et al is based on MAGPHYS template fitted to data} are interpolated using model spectra fit using the \texttt{MAGPHYS} model \citep{magphys}, which utilizes a stellar population synthesis (SPS) model and a self-consistent prescription for dust emission, absorption and PAH emission. The galaxy SEDs from \cite{Brown14} serve as a diverse template basis for fitting galaxy photometry and reproducing observed galaxy colors. Many of the galaxies in the Brown sample have well measured emission lines, which can be used to calibrate our line model locally (i.e., $z\sim 0$) before extrapolating to higher redshifts.

% \subsubsection{Measuring emission line properties}
The rest-frame optical lines in the SPHEREx wavelength range include: Balmer series lines \ha\ and \hb; singly- and doubly-ionized oxygen \oii\ [$\lambda$3728] and \oiii\ [$\lambda$5007]; nitrogen \nii\ [$\lambda$6584]; and the sulfur doublet SII [$\lambda\lambda$6716,6731]. Given SPHEREx NIR spectral coverage, we also include the less probed Paschen-$\alpha$ line at $\lambda_{rest}=1.87$ $\mu$m. Mid-infrared emission from PAHs will also be present, in particular the rest frame 3.3\um\ bump will be detectable for redshifts $z<0.5$. As we lack a realistic model of PAH emission in galaxies, we choose to leave any observed PAH emission from the \brown\ templates in our synthetic SEDs. 

We follow a procedure similar to \cite{speagle15} to extract emission line properties from the set of \brown\ templates. In regions of the observed SEDs where relevant SPHEREx emission lines are present, we fit a simple line + continuum model directly to the spectra. For each template, a small region centered on each line (widths varying from $\pm 50-100$ \AA\ for rest frame optical lines) is extracted and fit using a variable number of Gaussians depending on the number of spectral features, while the continuum is modeled locally with an offset and slope. From these models we estimate the flux and local continuum level associated with each line, from which we compute the equivalent width. We skip this step for any galaxies classified as passive using the Baldwin, Phillips \& Terlevich diagram \citep[BPT diagram;][]{bpt} in \brown. 

% Both the continuum templates (i.e. line-subtracted templates) and the \ha\ equivalent widths are an important ingredient in constructing our set of synthetic SEDs.

% \Oli{Where does this classification happen?}

% \ilbert{So, what do you do with the PAH? Do you include them at the end? Since they are in Brown, could be nice.}

\begin{table}[]
    \centering
    \begin{tabular}{c|c|c}
        Line & $\lambda_{\textrm{air}}$ [\AA] & SPHEREx coverage \\
        \hline
        H$\alpha$ & 6562.8 & $0.15<z<6.6$ \\
        $\left[\textrm{NII}\right]$ & 6583.5, 6548.0 & same as \ha\ \\
        H$\beta$ & 4861.4 & $z > 0.55$\\
        $\left[\textrm{OIII}\right]$ & 5006.8, 4958.9 & $z > 0.5$ \\
        \oii\ & 3728.8 & $z > 1.0$\\
        $\left[\textrm{SII}\right]$ & 6716.4, 6730.8 & $z > 0.1$\\
        Paschen-$\alpha$ & 18750.9 & $0<z<1.6$
    \end{tabular}
    \caption{Emission lines modeled in this work with rest-frame wavelengths in air. The third column indicates the redshift coverage for each observed line given the full SPHEREx bandpass.}
    \label{tab:my_label}
\end{table}

\subsubsection{Predicting emission line strengths}

Our procedure for generating sets of emission lines relies on tight empirical relations of lines and line ratios as a function of redshift and stellar mass. All of the empirical relations used in this section are obtained from measurements that have been corrected for dust extinction. Once the intrinsic line fluxes are computed, we then apply stellar and nebular extinction corrections to each source.

To predict \ha\ line fluxes, we use direct measurements of \ha\ from the \brown\ templates (described in \S 2.1) to determine each ``local" (low-redshift) equivalent width. For the set of active COSMOS templates, we compute the \ha\ equivalent width using scaling relations between the mean UV luminosity $L_{\textrm{UV}}$ and the star formation rate (SFR)
\begin{equation}
    \textrm{SFR}(M_{\odot}\textrm{yr}^{-1}) = 1.4\times 10^{-28} L_{\textrm{UV}}\quad (\textrm{erg s}^{-1}\textrm{Hz}^{-1}),
    \label{eq:sfr_uv}
\end{equation}
which can then be used to predict the \ha\ line luminosity $L_{\textrm{H}\alpha}$ \citep{kennicutt98};
\begin{equation}
    \textrm{SFR}(M_{\odot}\textrm{yr}^{-1})
    = 7.9\times 10^{-42} L_{\textrm{H}\alpha} \quad(\textrm{erg s}^{-1}).
    \label{eq:sfr_ha}
\end{equation}
For \brown\ templates with detectable \ha, the rest-frame equivalent width of \ha, denoted EW(\ha), is computed and scaled to higher redshifts using the relation of Fig. 6 from \cite{Brinchmann08} between $\Delta \textrm{EW(H}\alpha)$ and $\Delta \log\left([\textrm{OIII}]/\textrm{H}\beta\right)$, i.e., the deviations of equivalent width and line flux ratio from $z=0$. We model the redshift dependence of $\log\left(\oiii/\textrm{H}\beta \right)$ using Eq. 1 from \cite{Kewley13}

\begin{equation}
    \log\left(\frac{\oiii}{\textrm{H}\beta}\right) = \frac{0.61}{\log\left(\textrm{[NII]}/\textrm{H}\alpha\right) - \gamma} + \eta,
    \label{eq:kewley}
\end{equation}
where $\gamma =  0.02 + 0.1833z$ and $\eta = 1.2+0.03z$.
We calculate the flux ratio \nii/\ha\ as a function of \logm\ and redshift using an interpolation of Table 1 from \cite{faisst18}, which approximates the stellar mass vs. gas phase metallicity relation as a function of redshift. The line ratio NII$[\lambda 6585]$/NII$[\lambda 6549] = 3$ is applied to each \nii\ doublet.

Once the \ha-\nii\ complex is calculated, we apply the intrinsic ratios H$\alpha$/H$\beta$=2.86 and P$\alpha/H\alpha=0.123$ to obtain \hb\ and \pa\ line fluxes. We then use the extrapolated log[OIII]/H$\beta$ from (\ref{eq:kewley}) and OIII$[\lambda 5007]$/OIII$[\lambda 4959] = 3$ to obtain the two OIII line fluxes. The sulfur doublet [SII] is calculated using a best fit parabola to the local relation from SDSS DR12 of the O3S2 BPT diagram \citep[see Fig. 6 of ][]{Masters16}
\begin{equation}
     y = -0.44 - 0.3x - 0.66x^2,
\end{equation}
where $y=$log([SII]/\ha) and $x=$log([OIII]/\hb). 

We follow a similar procedure to predict \oii\ as for \ha. When available, we use \oii\ equivalent widths measured directly from the \brown\ templates. For COSMOS templates we use the mean SFR-L$_{[OII]}$ calibration from \cite{kewleyoii}, which assumes a fixed (de-reddened) \oii/\ha\ ratio, based on measurements from the Near Field Galaxy Survey (NFGS)
\begin{equation}
    \textrm{SFR} [\textrm{M}_{\odot}\textrm{yr}^{-1}] = (6.58 \pm 1.65) \times 10^{-42} \textrm{L}_{[OII]}.
\end{equation}

% To predict \oii\ we use the same relation between $M_{\textrm{UV}}$ and the star formation rate to predict \oii\ line luminosities. \cite{kewleyoii} builds on \cite{kennicutt98} by calibrating the SFR-$L_{OII}$ relation with corrections for reddening. Using the Near Field Galaxy Survey (NFGS) that work determined the following relation:

% \begin{equation}
%     \textrm{SFR} [\textrm{M}_{\odot}\textrm{yr}^{-1}] = (6.58 \pm 1.65) \times 10^{-42} \textrm{L([OII])}.
% \end{equation}
% \begin{figure*}
%     \centering
%    

\subsubsection{Dust extinction}
When fitting the set of galaxy templates to photometry of COSMOS2020 and GAMA sources, we apply dust attenuation using a grid of extinction curves, namely those derived in \cite{prevot84, calzetti2000, Seaton79, allen76, fitzpatrick84}. Although the template fits constrain each galaxy's stellar extinction well, it is known from measurements of the \ha/\hb\ Balmer decrement that the dust extinction in nebular regions differs from that of the stellar continuum and tends to be more pronounced at lower redshifts \citep{puglisi16, calzetti1994, jplus, reddy15, kashino19, faisst17}. To account for this differential nebular attenuation we scale the \texttt{Fitcat}-derived stellar extinction \ebv\ by the redshift-dependent differential extinction $f(z)$ derived in \cite{elcosmos_saito}, which is parameterized by a linear function capped at unity
\begin{equation}
    f(z) = \frac{E(B-V)_{star}}{E(B-V)_{neb}} = \textrm{min}(1, 0.44 + 0.2z).
\end{equation}
For all sources we apply nebular attenuation assuming the extinction curve from \cite{calzetti2000}.

\section{Line model Validation}
\label{sec:validation}
To test the fidelity of the empirical line model, we make a number of comparisons to existing line measurements. These include population-level comparisons of line equivalent widths (\ha+\nii), luminosity functions (\ha, \oii\ and \oiii) and line ratios as a function of stellar mass and redshift. Where external measurements in the COSMOS field are available, we also make direct, cross-matched comparisons of emission line fluxes.

\subsection{H-alpha+[NII] Line equivalent widths}
\begin{figure}
    \centering
    \includegraphics[width=\linewidth]{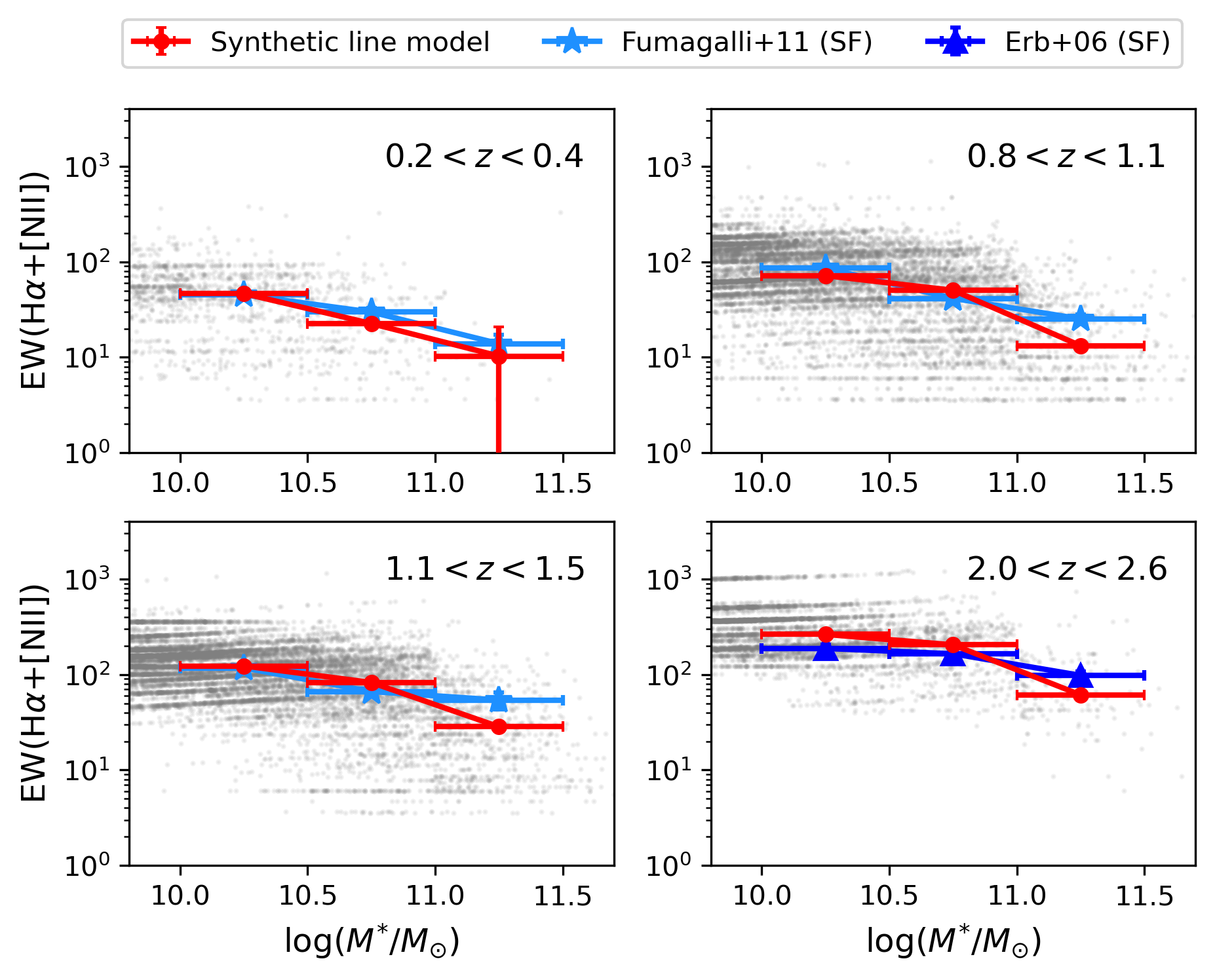}
    \caption{Predicted \ha\ + \nii\ equivalent widths of our synthetic line catalog as a function of stellar mass and redshift (grey points), with the binned average of the synthetic catalog plotted in red. These relations agree with previous spectroscopic measurements in the literature presented in \cite{3DHST} (light blue) from VVDS ($0.2<z<0.4$), 3D-HST ($0.8 < z < 1.5$) and \cite{Erb2006} (dark blue, $2.0 < z < 2.6$).} 
    \label{fig:ewhanii}
\end{figure}

We begin by comparing the predicted equivalent width of the \ha\ + \nii\ complex as a function of stellar mass with measurements from \cite{3DHST} using the VIMOS VLT Deep Survey at $0.2 < z < 0.4$ \citep[VVDS;][]{vvds}, 3D-HST ($0.8<z<1.5$) and \cite{Erb2006} ($2.0 < z < 2.6$). The evolution of EW(\ha) with redshift is often measured as an observational proxy for the specific star formation rate (sSFR)-redshift relation \citep{khostovan_ewha}. The comparison between our catalog and existing measurements is shown in Fig. \ref{fig:ewhanii}. To compare with the actively star-forming samples (SF), we select all sources with EW(\ha+\nii)$>3$ \AA, with the caveat that the samples we compare with come from surveys with varying selections and sensitivities. Nonetheless, the average equivalent widths within each stellar mass bin from our model are in close agreement with measurements. 

% The line flux of the individual sources (grey points) is obtained by assigning each source's ``local" equivalent width according to its best fit galaxy template.
% \richard{The model captures the average but not the dispersion in individual sources.}

It is understood that more massive galaxies undergo less vigorous star formation \citep{juneau05, zheng07}, leading to smaller \ha\ equivalent widths (at fixed redshift). This trend is captured by previous measurements and by our synthetic line catalog. Our catalog also captures the shift to higher equivalent widths at higher redshifts \citep{Brinchmann08}. For galaxies with $10 \leq$\logm$\leq 10.5$, the mean line equivalent width increases from $\langle$EW(\ha+\nii)$\rangle=$ 70 \AA\ for $z=0.2-0.4$ up to $\sim 200$ \AA\ for $z=2.0-2.6$. For the largest mass bin considered ($11 \leq$\logm$\leq 11.5$) the mean equivalent width goes from $\sim 15$ \AA\ to nearly 100 \AA\ over the same redshift range. 

% \Oli{Fig 2 is really nice! You should discuss it a little bit here.}

\subsection{Line luminosity functions}
From the catalog of line fluxes and redshifts we compute line luminosity functions (LFs). For each LF we select all catalog sources with estimated redshifts within 0.05 of $\overline{z}$, where $\overline{z}$ denotes the central redshift of the externally measured LFs. The width of the redshift bins is chosen to be small enough that redshift evolution is negligible within the bins but large enough to obtain good sample statistics. We use the total comoving volume within each redshift slice (i.e., within $[\overline{z}-0.05, \overline{z}+0.05]$ with $A_{eff}=1.27$ deg$^2$) to normalize the LFs. The synthetic LFs presented are not corrected for the Eddington biases sourced by flux uncertainties in the photometric catalogs \citep[c.f.][]{elcosmos_saito}, and the LFs include the effects of intrinsic dust attenuation.
\subsubsection{H-alpha}
\begin{figure*}
    \centering
    \includegraphics[width=0.85\linewidth]{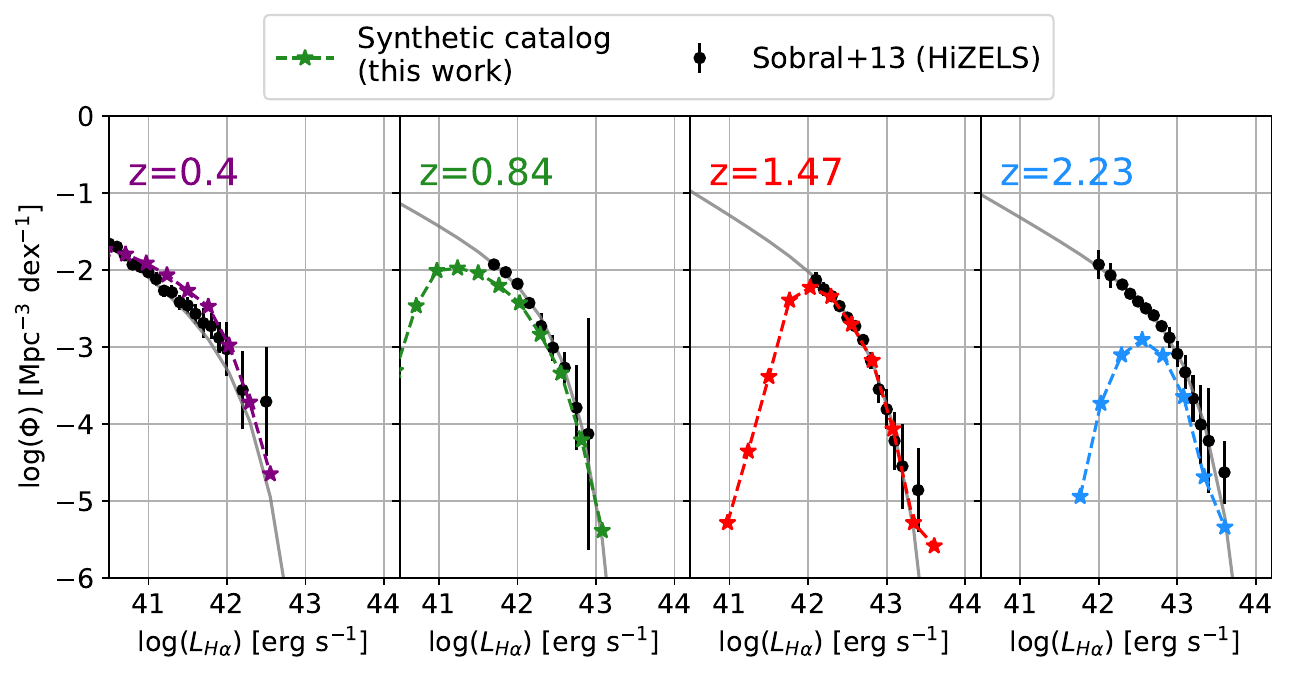}
    \caption{\ha\ line luminosity function evaluated at $z =$ 0.4 (purple), 0.84 (green), 1.47 (red) and 2.23 (blue), comparing best-fit Schechter functions from \cite{sobral2013} with empirically drawn fluxes from our model. The synthetic line LFs shown have not been corrected for catalog incompleteness.}
    \label{fig:ha_lf_compare}
\end{figure*}

We evaluate the synthetic \ha\ LF for redshift bins centered at $\overline{z}=\lbrace0.4, 0.84, 1.47, 2.23\rbrace$. Figure \ref{fig:ha_lf_compare} shows our derived \ha\ LFs compared with measurements and best-fit Schechter functions from \cite{sobral2013}, in which \ha\ measurements are corrected for \nii\ contamination following the relation from \cite{sobral2012}. On the bright end, our model is in close agreement with measurements in all redshift bins for $L \geq L_*$. Our LFs tend to fall steeply at fainter luminosities, underestimating $\phi$ for $L<L_*$. Unlike the LFs from \cite{sobral2013}, our derived LFs are uncorrected for luminosity incompleteness. The underprediction of $\phi$ is explained primarily by our selection on the COSMOS2020 catalog, rather than by the line flux modelling. Note that this behavior is common to all lines in our catalog. As the COSMOS catalog goes considerably deeper than SPHEREx photometry, this faint-source incompleteness starts to set in after the SPHEREx faint-source limit is reached, and as such is acceptable for modeling SPHEREx observations. Our line luminosities have a lower limit that is fainter than measurements in the \cite{sobral2013} sample (with the exception of the $z=2.23$ bin), which we understand as reflecting the fainter population probed by the COSMOS2020 catalog through broad-band continuum sensitivity. 
% \richard{From Matt: It is hard to tell from the way this last paragraph is worded (even though everyone on the SPHEREx team knows) whether the faint-source incompleteness in COSMOS starts to set in AFTER or BEFORE you reach the SPHEREx faint-source limit. It must be that the SPHEREx faint-source limit is reached well before that of COSMOS, stating that simply would be good.}\richard{Mike suggested making a table to illustrate this}

% \richard{In each redshift bin, we can relate the line luminosity limits of our COSMOS catalog along with the typical equivalent width to estimate the observed apparent magnitude of the source continuum. \ha\ is observed at [0.9, 1.2, 1.6, 2.1] $\mu$m for the four redshift bins above -- the COSMOS2020 3$\sigma$ depths within 3\arcsec\ apertures for YJH$K_s$ correspond to [26.1, 25.9, 25.5, 25.2] for the ultra deep survey and [24.8, 24.7, 24.4, 24.8] for the deep survey. Converting the mean \ha\ equivalent width from galaxies in the lower 10\% of line luminosity in our catalog (active galaxies only) along the mean line luminosity of that sample, the estimated continuum apparent magnitudes are 26.5, 25.6, 24.6 and 24.8, respectively. \cite{sobral2013} uses a combination of narrow band and broad-band NIR filters to select line emitters, corresponding to 3$\sigma$ AB magnitude depths of [25.1, 22.9, 22.3, 22.6] (COSMOS) and [27.2, 22.9, 22.3, 22.6] (UDS). These are used to get the narrow band excesses, so should not be directly compared with continuum depths, but the gist is that COSMOS goes fainter than Sobral.}

\subsubsection{[OII] doublet}
The \oii\ line will fall in SPHEREx's bandpass for galaxies with $z > 1$. While only the brightest \oii\ lines will be detectable at SPHEREx full-sky depth (see Section \ref{sec:elg_prevalence}), there are near-future telescopes that will measure the doublet through optical spectroscopy at lower redshift, for example \emph{Euclid} \citep{euclid} and \emph{Roman} \citep{wang21}. For this reason we validate \oii\ across a range of redshifts.

\begin{figure*}
    \centering
    \includegraphics[width=0.9\linewidth]{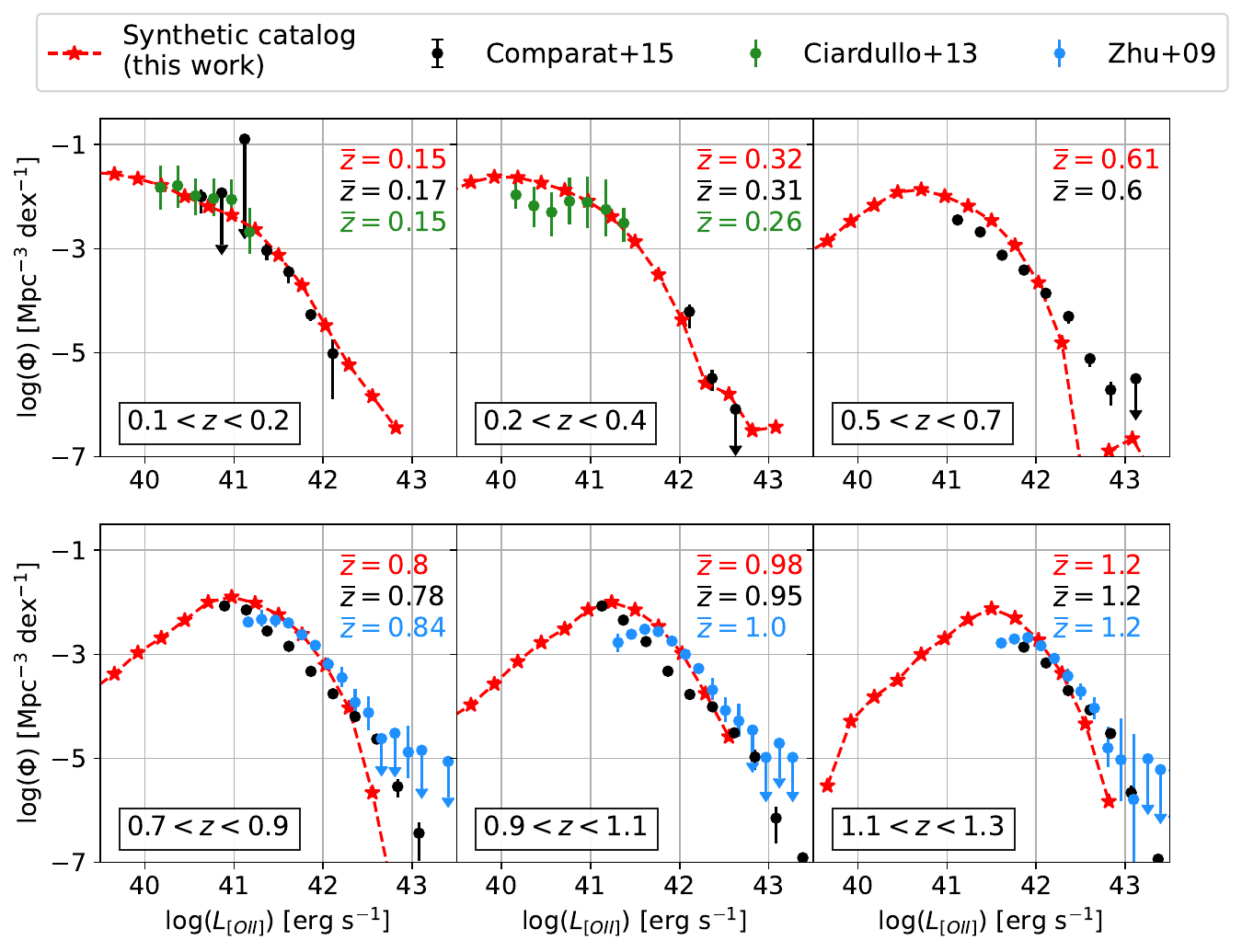}
    \caption{\oii\ luminosity function evaluated in six redshift bins between $z=0.1$ and $z=1.3$. Line fluxes from our model are compared with a combination of emission line measurements from \cite{comparat15} (black points), \cite{Zhu09} (blue) and \cite{ciardullo13} (green). The synthetic line LFs shown have not been corrected for catalog incompleteness.}
    \label{fig:lowz_oii}
\end{figure*}

To validate \oii\ for lower redshifts we compare our synthetic LFs with measurements from the FORS2 instrument at the Very Large Telescope (VLT) and the SDSS-III/BOSS spectrograph, along with measurements from GAMA, zCOSMOS and VVDS \citep{comparat15}. We plot LFs derived from our line catalog calculated in six redshift bins spanning $0<z< 1.3$ alongside these measurements in Fig. \ref{fig:lowz_oii}. We also include LF measurements from the HETDEX pilot survey for $0<z<0.4$ \citep{ciardullo13} and from the Deep Extragalactic Evolutionary Probe 2 (DEEP2) galaxy survey for $0.7<z<1.3$ \citep{Zhu09}. For redshift bins centered on $z=0.15$ and $z=0.3$, our LFs agree down to $L \sim 10^{40}$ erg s$^{-1}$. For $z>0.5$ the bright end LF is consistent with \cite{Zhu09}, though both are higher than measurements from \cite{comparat15}.

\begin{figure*}
    \centering
    \includegraphics[width=0.9\linewidth]{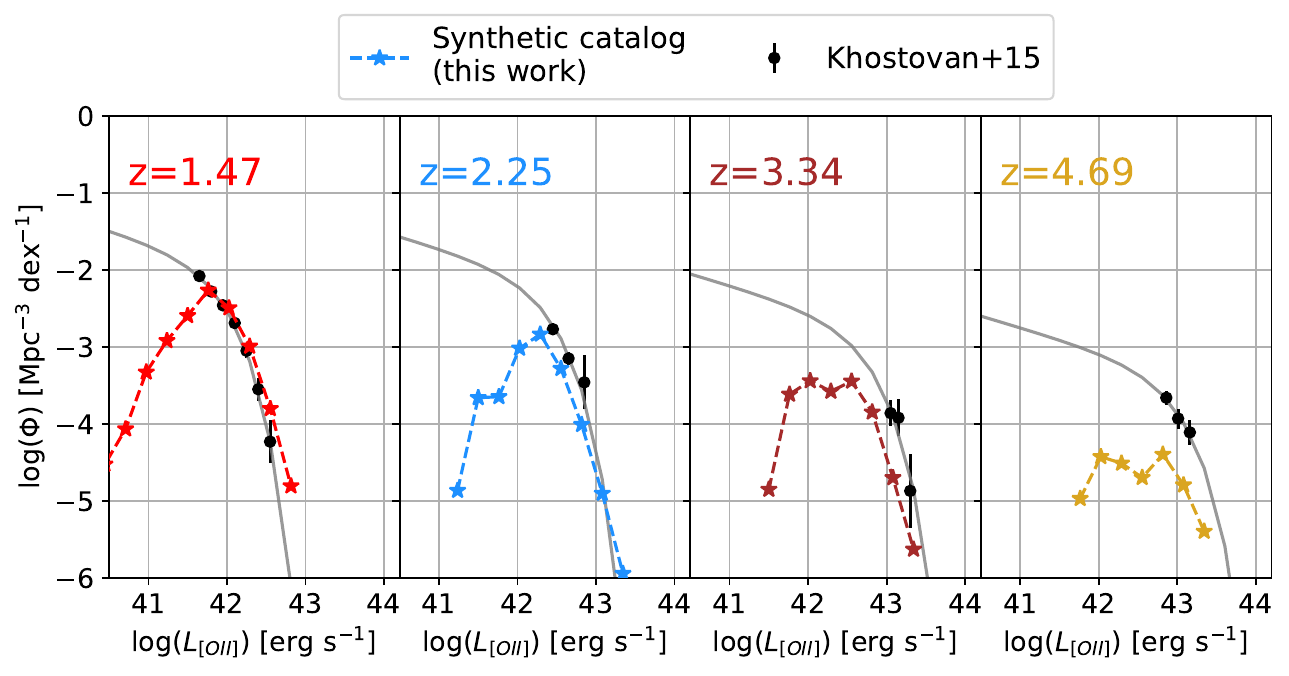}
    % {oii_lf_compare_Khostovan_vs_redshift_highz.pdf}
    \caption{\oii\ luminosity functions evaluated from our synthetic line catalog at $\overline{z} =$ 1.47 (red), 2.25 (blue), 3.34 (brown) and 4.69 (yellow), compared with measurements from HiZELS \citep[][black points]{khostovan} and their best-fit Schechter functions (grey lines). The incompleteness in the synthetic catalog becomes more pronounced at high redshifts, especially in the $z=4.69$ bin.}
    \label{fig:oii_lf_compare}
\end{figure*}

In Fig.~\ref{fig:oii_lf_compare} we compare \oii\ LFs with measured constraints from the HiZELS survey \citep{khostovan}. The bright-end LFs broadly agree with one another aside from in the highest redshift bins, where the synthetic COSMOS2020 catalog contains few sources. In general these results indicate that \oii\ is captured well at the population level. 

\subsubsection{[OIII]+H-beta}
\begin{figure*}
    \centering    \includegraphics[width=0.9\linewidth]{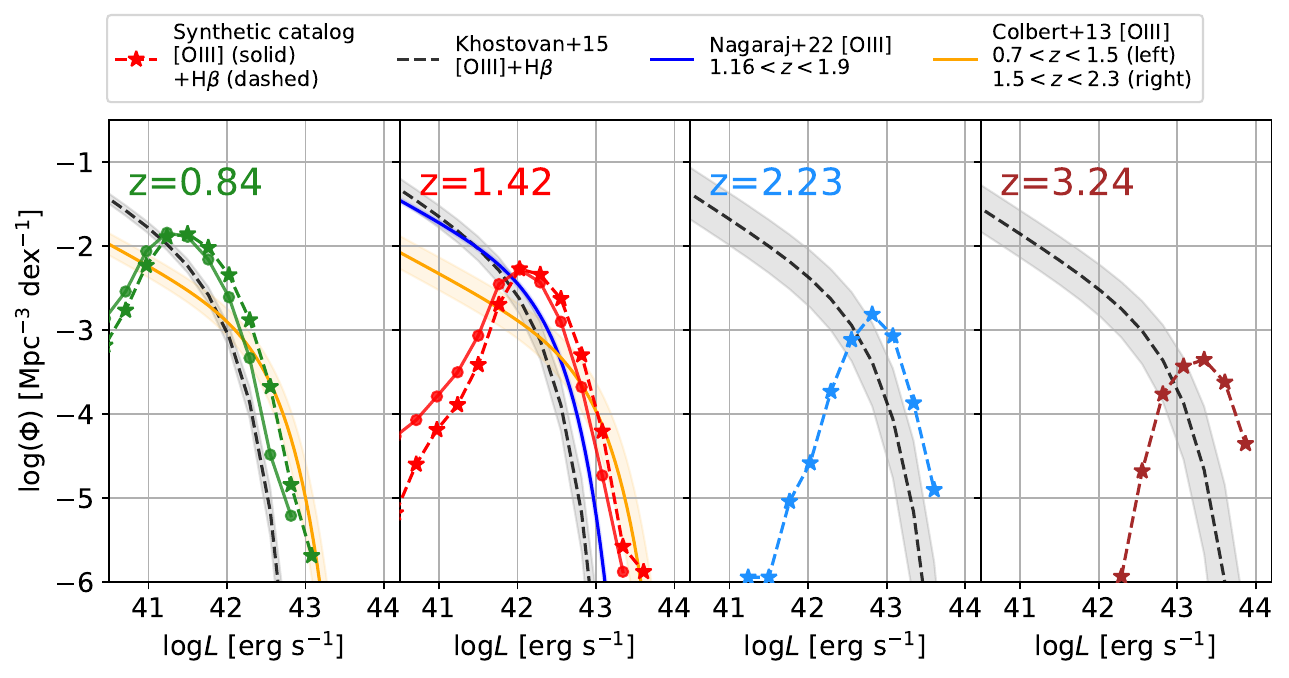}
    \caption{\oiii\ (solid) and \oiii+\hb\ (dashed) luminosity function evaluated from our synthetic line catalog at in four redshift bins spanning $z=0.84$ (left, green) and $z=3.24$ (right, brown). The black line and grey shaded regions show LF constraints from \cite{khostovan}. Similar constraints for \oiii\ are shown from \cite{colbert13} (yellow) and \cite{nagaraj22} (blue). The synthetic line LFs shown have not been corrected for catalog incompleteness.}
    \label{fig:lf_oiiihb}
\end{figure*}

The \oiii\ + \hb\ complex will be detectable by SPHEREx for galaxies with $z>0.55$ and can potentially aid redshift measurements if measured in addition to \ha\ and/or \oii. We show line LFs for \oiii\ and \oiii+\hb\ in Fig. \ref{fig:lf_oiiihb}. Our predicted LFs are higher on average than those derived from the HiZELS survey \citep{khostovan}. However, our predictions on the bright end are in close agreement with \cite{colbert13} and a recent study using $1.2<z<1.9$ emission line galaxies identified on the 3D-HST grism \citep{nagaraj22}. Given the disagreement in measurements between \cite{khostovan} and \cite{colbert13}, it is difficult to evaluate the significance of the model disagreement with \cite{khostovan}.

% Potential sources of discrepancy in the LF include over-prediction of flux by our empirical model, impure samples of narrow band emitters from photometric redshift errors, line mis-identifications, and cosmic variance, among other selection effects. Furthermore, the impact of dust extinction on emission lines at high redshift is not well understood, and line measurements can be further affected by the presence of AGNs. 

% \richard{checked Khostovan paper, there is a filter profile correction to account for emitters near the narrow band filter edges, so I think we have an apples to apples comparison. Another source of error they mention is line misidentifications, e.g., thinking \ha\ is \oiii. Median errors from photometric redshifts in comparison with spec-zs is $\Delta z / 1+z_{spec} = 0.047$.}

% Compared against measured LFs from the HiZELS survey \citep{khostovan}, our predicted LFs reside higher on the bright end as seen in Figure \ref{fig:lf_oiiihb}. One possibility is that our prescription for nebular attenuation is not appropriate for \oiii+\hb. 

\subsubsection{Paschen-alpha}
\begin{figure}
    \centering
\includegraphics[width=\linewidth]{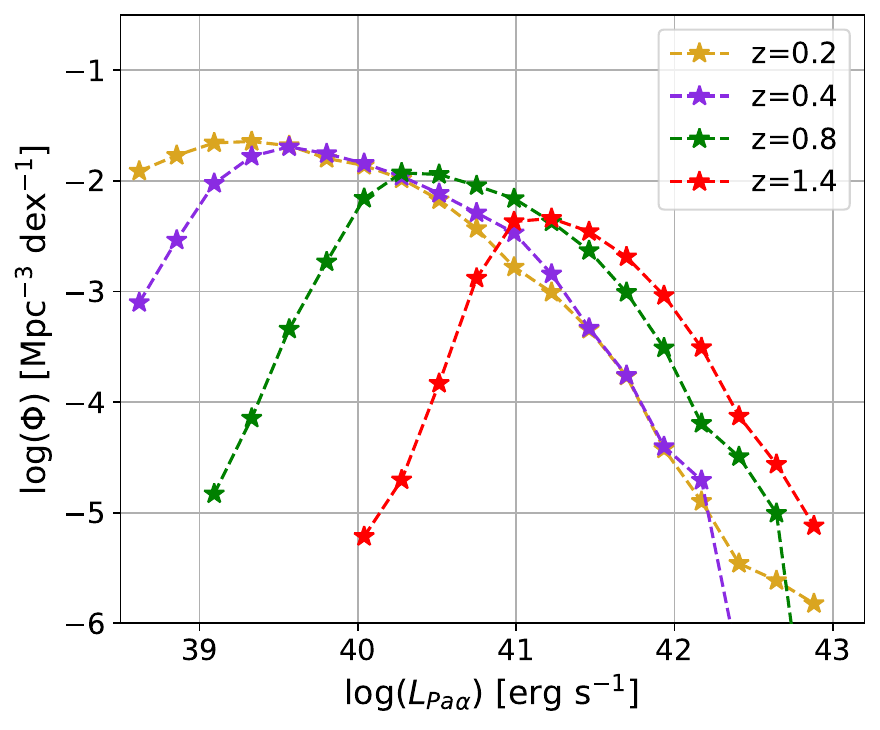}
    \caption{Paschen-$\alpha$ line luminosity function predicted from the synthetic catalog in four redshift bins spanning $0<z<1.6$, i.e., the redshift coverage for Paschen-$\alpha$ by SPHEREx. The synthetic line LFs shown have not been corrected for catalog incompleteness.}
    \label{fig:paschen_alpha}
\end{figure}

The broad spectral coverage and resolving power offered by SPHEREx in the near-infrared enable detection of the 1.87 $\mu$m Paschen-$\alpha$ line. Figure \ref{fig:paschen_alpha} shows LF predictions for Paschen-$\alpha$ in four redshift bins between $0<z<1.6$, taking the sum of COSMOS2020- and GAMA-derived LFs for $z=0.2$ and $z=0.4$. The bright end LF evolves mildly for redshifts $z>0.5$.

% \ilbert{But you will have no galaxies at $z>2$ in the wide, and probably really few in the deep from your fig.9. So, maybe you can explain that the prediction of OII is not really important. Expect if it is useful for the intensity mapping (no idea about that).}. 

\subsection{Direct line flux comparison in the COSMOS field}

We complement validation of our line model at the population level with direct comparisons to existing line flux measurements. One-to-one comparisons of cross-matched sources in the COSMOS field allow us to quantify any consistent biases as a function of line flux while controlling for the properties of the galaxies (assuming they are well constrained by one or both surveys).

\begin{figure}[ht]
    \centering
    \includegraphics[width=\linewidth]{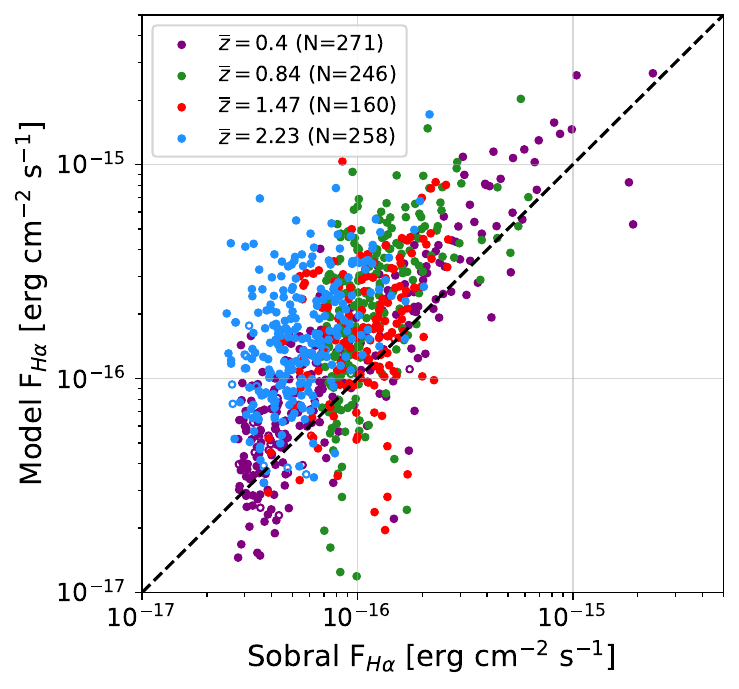}
    \caption{Direct \ha\ line fluxes compared with measurements from \cite{sobral2013} for bins with $\overline{z}=$ 0.4 (purple), 0.84 (green), 1.47 (red) and 2.23 (blue). Open circles indicate galaxies where the COSMOS2020 photometric redshift differs from that of the nominal \cite{sobral2013} redshift bin by more than 0.2.}
    \label{fig:ha_flux_compare}
\end{figure}

% While the line model employed in this work assumes mean trends with respect to stellar mass and redshift, one-to-one comparisons of cross-matched sources in the COSMOS field allow us to quantify any consistent biases as a function of line flux while controlling for the properties of the galaxies (assuming they are well constrained by one or both surveys).
The \ha\ sample from \cite{sobral2013} overlaps partially with the COSMOS field. The measured and predicted line fluxes are plotted in Fig. \ref{fig:ha_flux_compare}, in which no extinction correction is applied. Positional cross matches are performed for galaxies in the same redshift bins as Fig. \ref{fig:ha_lf_compare}, however there are many sources for which the estimated redshift from \cite{cosmos2020} differs by more than $|\Delta z| = 0.2$ from the nominal redshift bin (indicated by open circles). These discrepancies are most pronounced for fainter line fluxes, and could be caused by outliers in the COSMOS2020 catalog. The synthetic line fluxes are relatively unbiased for $F_{H\alpha}>10^{-16}$ \ergcms\ and at low redshift, with a slight positive bias for higher redshift bins. 

% \richard{is $\Delta z=0.2$ appropriate? what is typical photo-z uncertainty for sources with z=2.23? should have better cut on "outliers" here.}
We compare our model fluxes with two other sets of emission line measurements in the COSMOS field, namely zCOSMOS-Bright \citep{elcosmos_saito, elcosmos_lilly} and 3D-HST \citep{3DHST1,3DHST2}. These surveys have obtained spectroscopic redshifts in the COSMOS field that are magnitude-limited to $i_{AB}<22.5$. From zCOSMOS and 3D-HST, \ha\ is measured for redshifts $z\leq 0.46$ and $0.67\leq z \leq 1.6$, respectively, while for \oii\ they cover $0.47\leq z\leq 1.57$ and $1.95\leq z \leq 3.56$. \oiii\ is measured in the redshift ranges $0.11\leq z \leq 0.92$ and $1.19\leq z \leq 2.39$, respectively. The line fluxes from zCOSMOS are aperture corrected following the procedure from \cite{aperture_corr}, which uses the measured sizes of the galaxies to estimate the fraction of total flux falling within the one arcsecond slit. \cite{elcosmos_saito} estimates the line flux completenesses of zCOSMOS and 3D-HST to be $\log(F_{line}/$erg cm$^{-2}$ s$^{-1})=-15.8$ and $\log(F_{line}/$erg cm$^{-2}$ s$^{-1})=-16.5$.

\begin{figure*}
    \centering
    \includegraphics[width=0.95\linewidth]{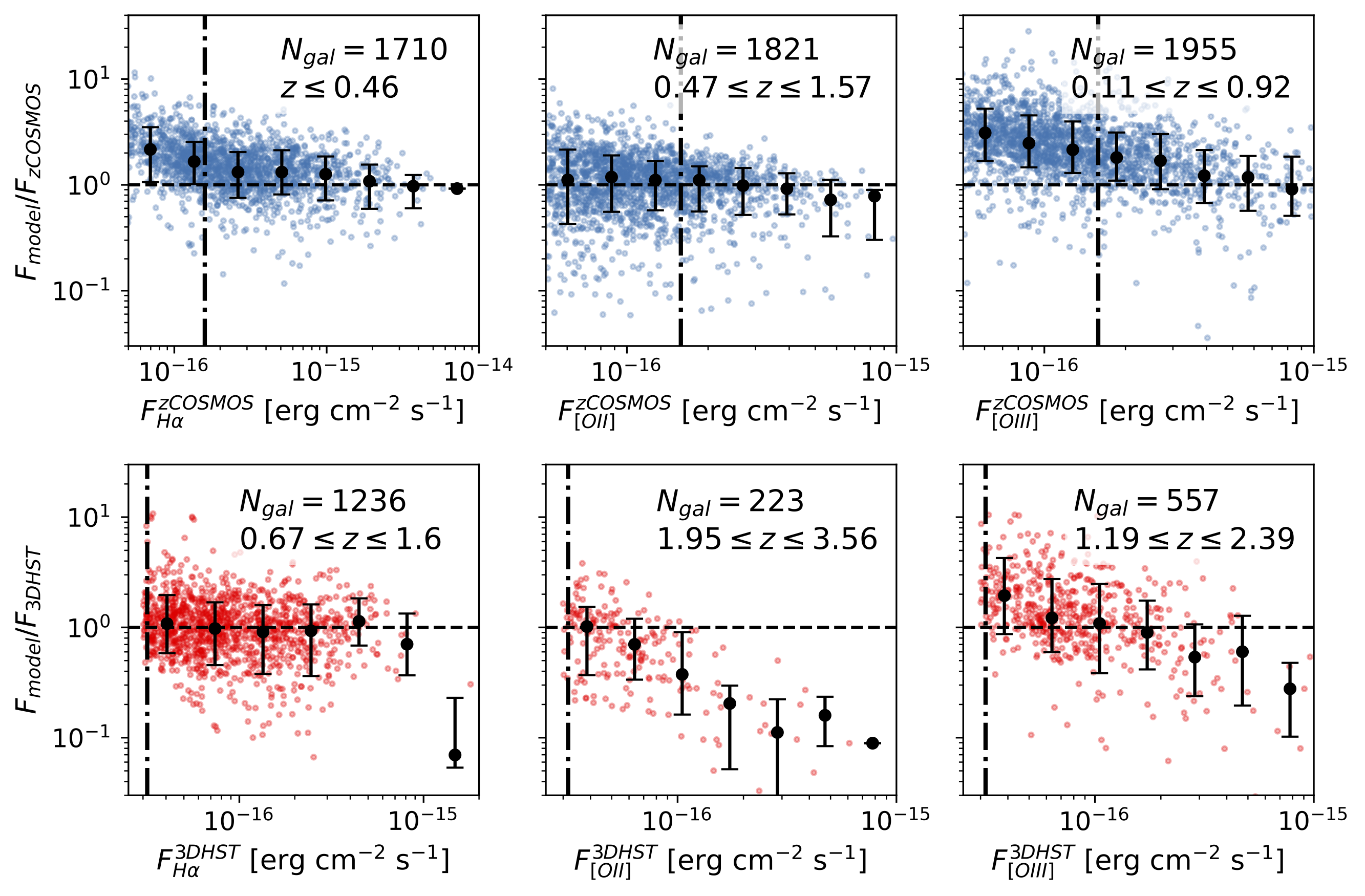}
    \caption{One-to-one line comparisons of \ha\ (left), \oii\ (middle) and \oiii\ (right) between model predicted fluxes and direct flux measurements from zCOSMOS (top row) and 3D-HST (bottom row). The black errorbars show the mean and 68 percentile range of the colored points in bins of equally spaced log-fluxes. The grey dash-dot lines indicate the completeness limits of the respective surveys as estimated in \cite{elcosmos_saito} (see Fig. 2 from that work).}
    \label{fig:zcosmos_3dhst_compare_lines}
\end{figure*}

Figure \ref{fig:zcosmos_3dhst_compare_lines} shows the ratio of predicted and observed fluxes for \ha, \oii\ and \oiii, with the mean and scatter (binned by log-flux) plotted in black. There is strong agreement with zCOSMOS-measured fluxes for all three lines, though for \oiii\ our model fluxes overestimates fluxes with $F_{[OIII]}<2\times 10^{-16}$ \ergcms. This bias is consistent with trends seen in \cite{elcosmos_saito} and our direct comparisons with HiZELS (see Fig. \ref{fig:ha_flux_compare}. For the 3D-HST sample, which contains more high redshift sources, our model \ha\ fluxes in close agreement with measured fluxes on average. However, for \oii\ and \oiii\ our model fluxes tend to underestimate measured fluxes above $F\sim 1\times 10^{-16}$ \ergcms. 

We consider several potential explanations for why our model \oii\ and \oiii\ fluxes are low relative to the observed sample:
\begin{itemize}
    \item The predicted quiescent fraction may be too high for this selection of sources when in reality they are star forming. The fraction of cross-matched zCOSMOS sources above the completeness limit (dash-dotted lines) labeled as passive by our model (i.e., $F_{line}=0$) is 3.0\%, 2.0\% and 1.3\% for \ha, \oii\ and \oiii\, respectively, and for the 3D-HST sample the corresponding fractions are 4.6\%, 9.4\% and 10.5\%.
    \item Incorrect redshift assignments could bias the fluxes of the sources low. Only a handful of COSMOS2020 sources have discrepant redshifts when compared against spec-zs from either zCOSMOS or 3D-HST. Removing sources with $\delta z/(1+z)>0.1$ from the comparison does not ameliorate the discrepancies seen in \oii\ for the $1.95\leq z \leq 3.56$ sample. \cite{3DHST2} quotes a redshift scatter of $\sim 10\%$ for $z\sim 2$ galaxies when compared against duplicate measurements and follow-up observations from the MOSDEF survey \citep{mosdef} for $JH_{IR} < 24$, which is the case for the galaxies in question with $F_{line}^{3DHST}>10^{-16}$ \ergcms.
    \item Likewise, the discrepancy could be present if the 3D-HST fluxes for these objects are biased high. There is evidence for a positive bias in \oii\ and \oiii\ on average for 3D-HST fluxes for $F_{line}^{3DHST}\lesssim 10^{-16}$ \ergcms\, however the paucity of direct comparisons prohibits us from assessing whether this fully explains the observed differences.
    \item An incomplete model of nebular attenuation may impact our \oii\ and \oiii\ predictions, in particular at higher redshifts. 
    \item If there is a non-negligible AGN fraction in the high-redshift sample it may explain the higher line fluxes compared to our model fluxes. Because the WFC3 G141 grism has moderate resolution ($R=130$) it is not possible to resolve the \ha+\nii\ complex, meaning we cannot distinguish between star forming galaxies and AGN through the BPT diagram (e.g., Fig. \ref{fig:compare_bpt}). AGN have been identified and removed in the COSMOS2020 catalog using morphological and SED criteria, however it is likely there is additional AGN contamination in particular at the higher redshifts considered.
\end{itemize}

\begin{figure*}
    \centering
    \includegraphics[width=0.95\linewidth]{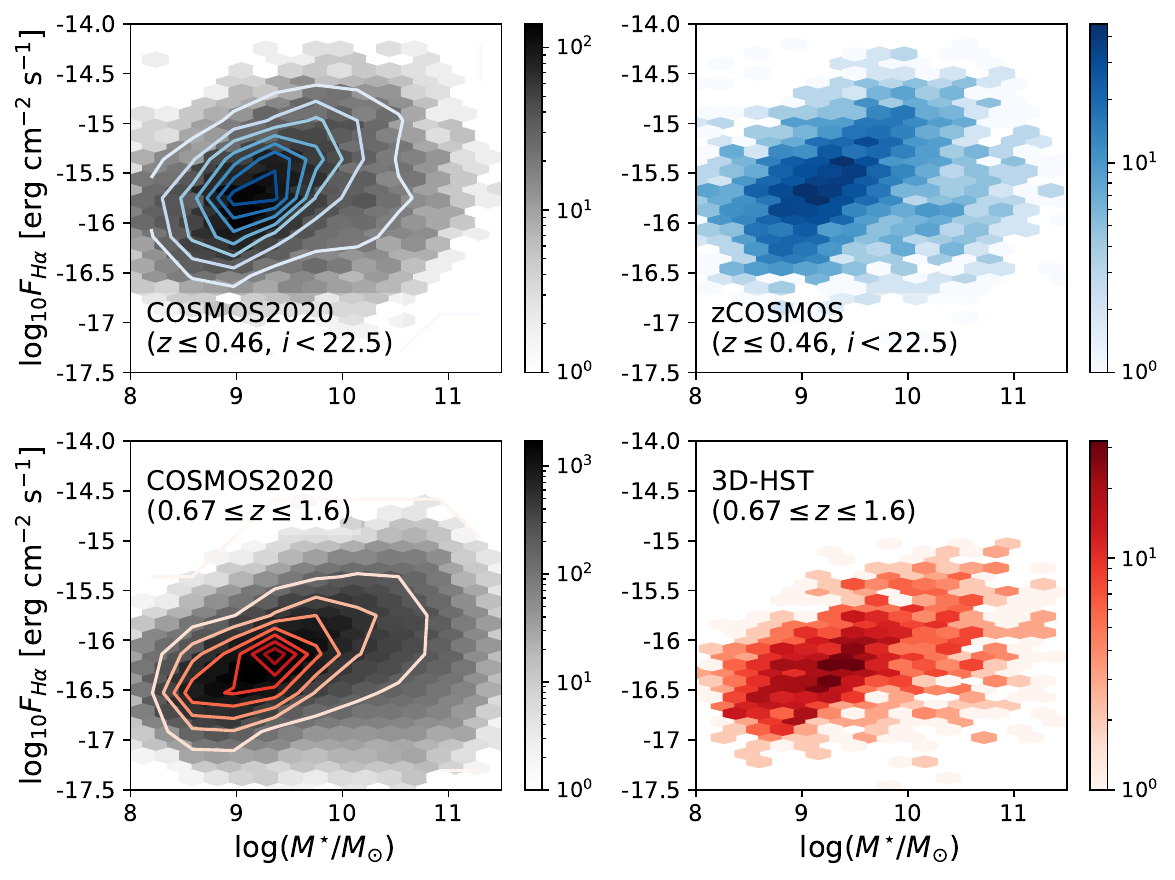}
    \caption{Log-density plot of \ha\ line fluxes and stellar masses, for synthetic fluxes from our COSMOS2020 catalog (black) and for observed samples from zCOSMOS (blue) and 3D-HST (red). The two columns show the COSMOS2020 catalog with redshift cuts matching the \ha\ ranges of zCOSMOS and 3D-HST. Density contours for the spectroscopic catalogs are overlaid on the synthetic flux distributions for comparison.}
    \label{fig:fha_spread}
\end{figure*}

At the population level, our line model captures the scatter in line strengths seen in the two measured line catalogs. Figure \ref{fig:fha_spread} compares the distribution of line fluxes from our model with those from zCOSMOS and 3D-HST as a function of stellar mass. For the low redshift sample we place an additional cut on $i<22.5$ to match the zCOSMOS selection and find that our model fluxes adequately cover the range of measured fluxes. When compared against 3D-HST, our model fluxes match both the core and tails of the measured flux distributions. While we use mean trends to predict line strengths and ratios, our synthetic line catalog are conditioned on estimates of redshift, stellar mass and dust attenuation, all of which contribute to the observed flux scatter. Reproducing this scatter in our redshift predictions is important for obtaining realistic forecasts on the number of SPHEREx sources with line detections (see \S \ref{sec:elg_prevalence}).

% \ilbert{Strange the LF of Comparat at $0.2<z<0.4$ with only bright points. TBC. Or maybe because of the combination of different surveys.}

\subsection{Line ratios/trends}

\begin{figure}
    \centering
    \includegraphics[width=\linewidth]{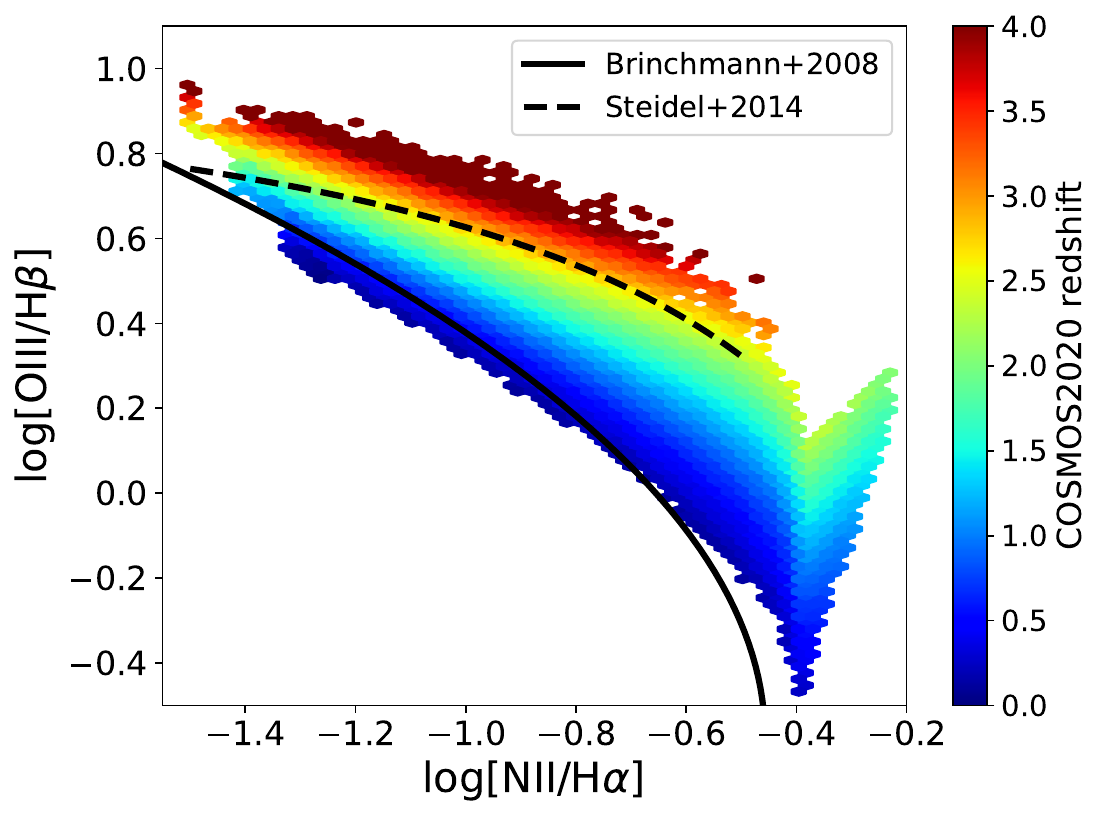}
    \includegraphics[width=\linewidth]{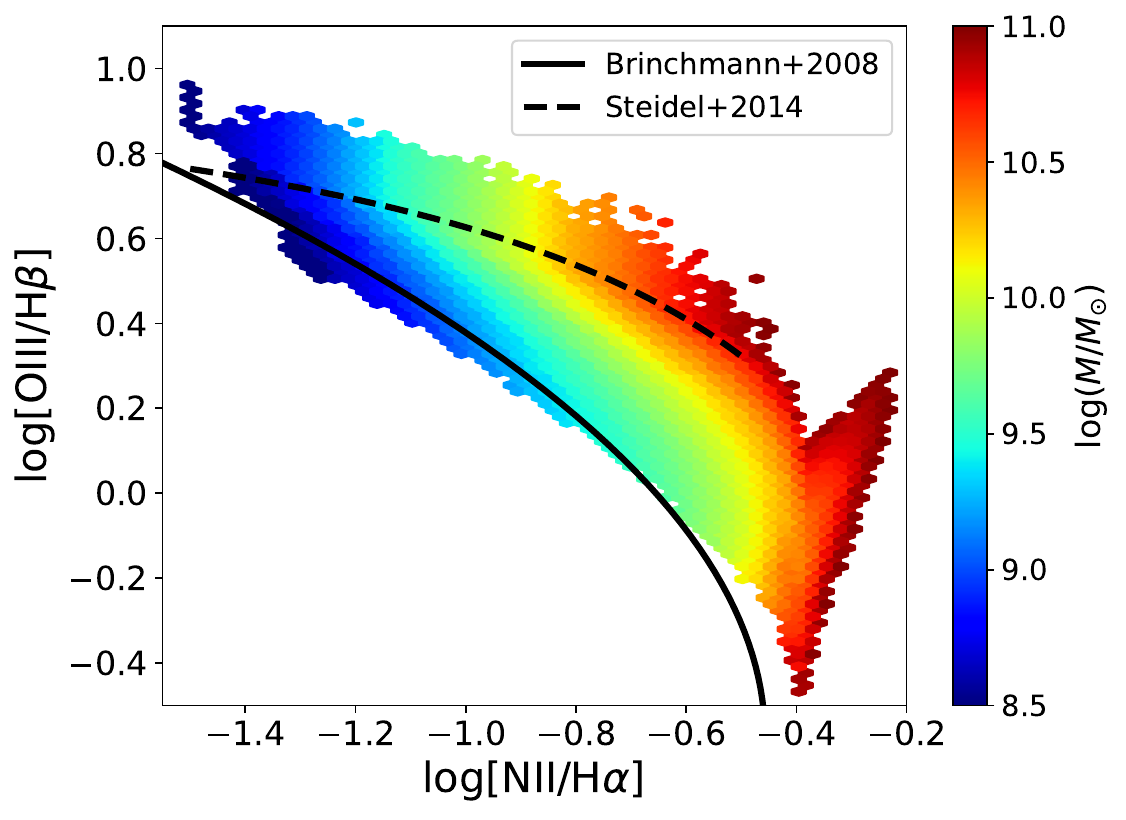}
    \caption{O3N2 BPT diagram with line ratios from the synthetic line catalog, plotted against measured trends of local ($z\sim 0$, \cite{Brinchmann08}) and higher redshift ($\langle z\rangle = 2.3$, \citealt{Steidel_2014}) galaxies. Plots show line ratios of the same sources, colored by redshift (top) and stellar mass (bottom).}
    \label{fig:compare_bpt}
\end{figure}

Evolution of the line ratios as a function of redshift can be seen by placing galaxy line ratios on the O3N2 BPT diagram. This is shown in Fig. \ref{fig:compare_bpt}, with synthetic line ratios color-coded by redshift. While the population of low-redshift galaxies resides in a locus centered on the relation from \cite{Brinchmann08} (red, dashed), moving to higher redshift has the effect of shifting this locus upward on the BPT diagram. This trend is consistent with measurements from MOSFIRE \citep{Steidel_2014} of a sample of 251 galaxies between $z=1.5$ and $z=3.5$ ($\langle z \rangle \sim 2.3$).

% \richard{Along similar lines, Ngal detected for different lines vs. redshift.}
% \ilbert{It seems that you have a tendency to overpredict the flux. You should probably quantify this trend. It could be useful if you want to give a lower limit in Fig.10.}
% \richard{For \ha\ I think I am overestimating the equivalent width, but we see overprediction for \oii\ as well, which is now independent of EW(H$\alpha$). Could be overestimation of star formation rate for certain galaxies}

% \begin{figure*}
%     \centering
%     \includegraphics[width=0.9\linewidth]{ha_oii_oiii_direct_lineflux_comparison_zCOSMOS_3DHST_101122.pdf}
%     \caption{One-to-one line comparisons between flux measurements from 3D-HST (top row) and zCOSMOS (bottom row) and model predicted fluxes.  \richard{flux completeness information, should it be quoted for surveys as a function of redshift?}}
%     \label{fig:zcosmos_3dhst_compare_lines}
% \end{figure*}
\subsection{Quiescent fraction}
The fraction of passive galaxies is important to quantify as a check that our model does not overproduce star-forming galaxies with strong emission lines. Using our synthetic line catalog, we find that 7\% of objects are either best fit by passive galaxy templates (see \S \ref{sec:tempfit}) or have \ha\ equivalent width less than 5 \AA. At low redshifts ($z < 1$), the quiescent fractions of our synthetic catalog for mass bins \logm\ $\in\ (10, 10.5]$, \logm\ $\in\ (10.5, 11.0]$ and \logm\ $\in\ (11, 11.5]$ are 20\%, 34\% and 44\%, respectively. We also compute quiescent fractions the more standard separation using the $UVJ$ diagram  \citep{williams09, daddi04}. We use the rest frame ($NUV-r$) vs. ($r-J$) selection from \cite{ilbert13} ($NUVrJ$ in shorthand):
\begin{equation}
    (NUV-r) > 3(r-J)+1 \textrm{ and } (NUV-r) > 3.1.
\end{equation}
The absolute magnitudes are calculated in the \texttt{Farmer} catalog from \cite{cosmos2020} using the best-fit photo-z solutions. Using this classification, we find the same selections of galaxies have quiescent fractions of 29\%, 41\% and 53\%, respectively. While our quiescent galaxy classification using EW(\ha) is more conservative than using $NUVrJ$, the trends of both classifications match the expectation of an increasing fraction of quiescent galaxies with higher stellar mass. Our $NUVrJ$ classification yields quiescent fractions consistent with other studies using COSMOS2020 \citep[c.f. Fig. 9 of][]{weaver22_mf}. These quiescent fractions are lower than those determined from measurements using UltraVISTA and 3D-HST which range from 25\% up to 70\% across the same range of \logm\ \citep[e.g., Fig. 2 of][]{martis16}.

% This discrepancy is likely due to our method of galaxy classification, which differs from the typical separation of active/passive galaxies from the UVJ diagram \citep{williams09, daddi04} \richard{From Jamie: So is this the reason? Otherwise it seems like a big discrepancy.} \richard{From O.I: Agree, could be investigated a bit more. Use Weaver+23 mass function which has UVJ type classification, this would explain if our passive/active classification criteria from lines are conservative}.

% \richard{Some discrepancy likely due to different classification, line EW vs. UVJ. Should I go ahead and calculate this with some filters or just compare the mean fractions? That study covers 1.7 deg$^2$}\Oli{Explain this hypothesis only maybe.}
% \Oli{Would be good for the plot to be closer to their relevant section}
% \richard{How should we reference this catalog?}
% We use the \texttt{LePhare}-derived specific star formation rate and find $\sim$6.3\% of objects have $\log_{10}sSFR < -11$. 

\section{Predicting SPHEREx Photometry}
\label{sec:sphx}

We now use the empirical model detailed above to generate synthetic SPHEREx spectrophotometry with realistic colors. In this section we describe the unique spectral scan strategy employed by SPHEREx and detail the noise properties of the full-sky and deep surveys. 

\subsection{SPHEREx}
SPHEREx uses six HAWAII-2RG (H2RG) detector arrays arranged in two mosaics, separated by a dichroic beam splitter that allows the focal plane to be simultaneously imaged \citep[details on the instrument configuration can be found in][]{korngutspherex}. A set of linear variable filters (LVFs) sit above the focal planes, which function as bandpass filters whose central wavelength varies linearly with detector position. The spectrum for a source can thus be obtained by modulating its position across the field of view in a series of exposures.

Over its nominal two-year mission, SPHEREx will complete four full-sky surveys, where each survey comprises measurements in 102 spectral channels on each sky position. Observations are shifted by half a spectral channel between the first/third and second/fourth surveys to Nyquist sample the response function. Each spectral channel is defined in steps of $\Delta\lambda = \lambda/R$ across each of the six detectors (detectors in this context are also referred to as ``bands"). The resolution of each LVF is fixed between $R=35-130$, where
\begin{equation}
    R = \frac{\lambda_c T(\lambda_c)}{\int T(\lambda)d\lambda},
\end{equation}
$\lambda_c$ is the central wavelength and $T(\lambda)$ is the filter transmission which varies continuously as a function of detector position. This will be done through a scan strategy that involves a combination of large and small slews as the spacecraft follows a low-earth sun-synchronous orbit, observing near great circles which precess over six months. As a result of this observing strategy, SPHEREx will scan the northern and southern ecliptic caps (NEP and SEP, respectively) with much higher cadence, leading to an total area of 200 deg$^2$ with $\sim200$ measurements per spectral channel after two years. The two survey depths are distinguished as ``shallow" (or ``full-sky") and ``deep" throughout this work.

\begin{figure*}
    \centering
    \includegraphics[width=0.48\linewidth]{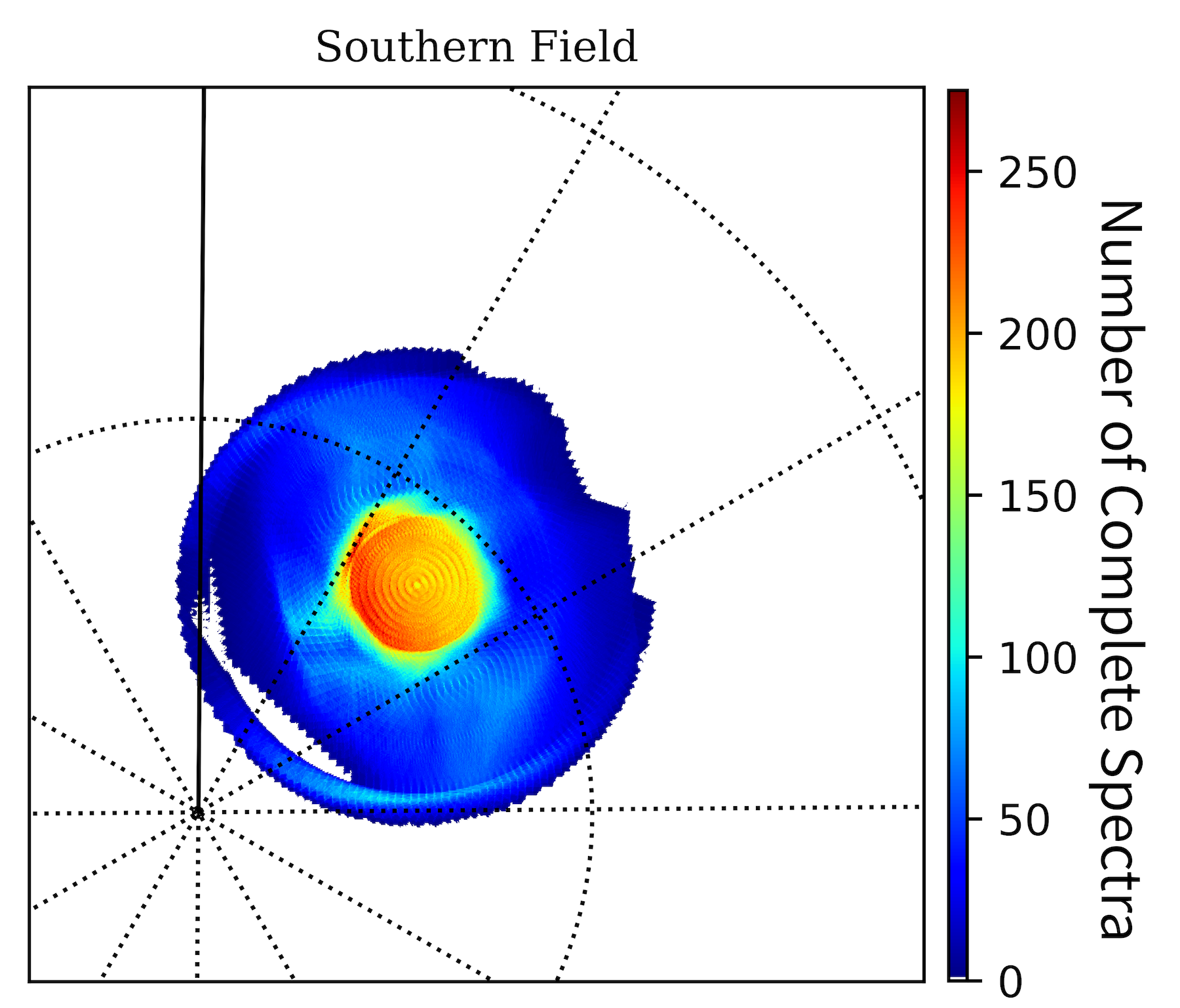}
    \includegraphics[width=0.49\linewidth]{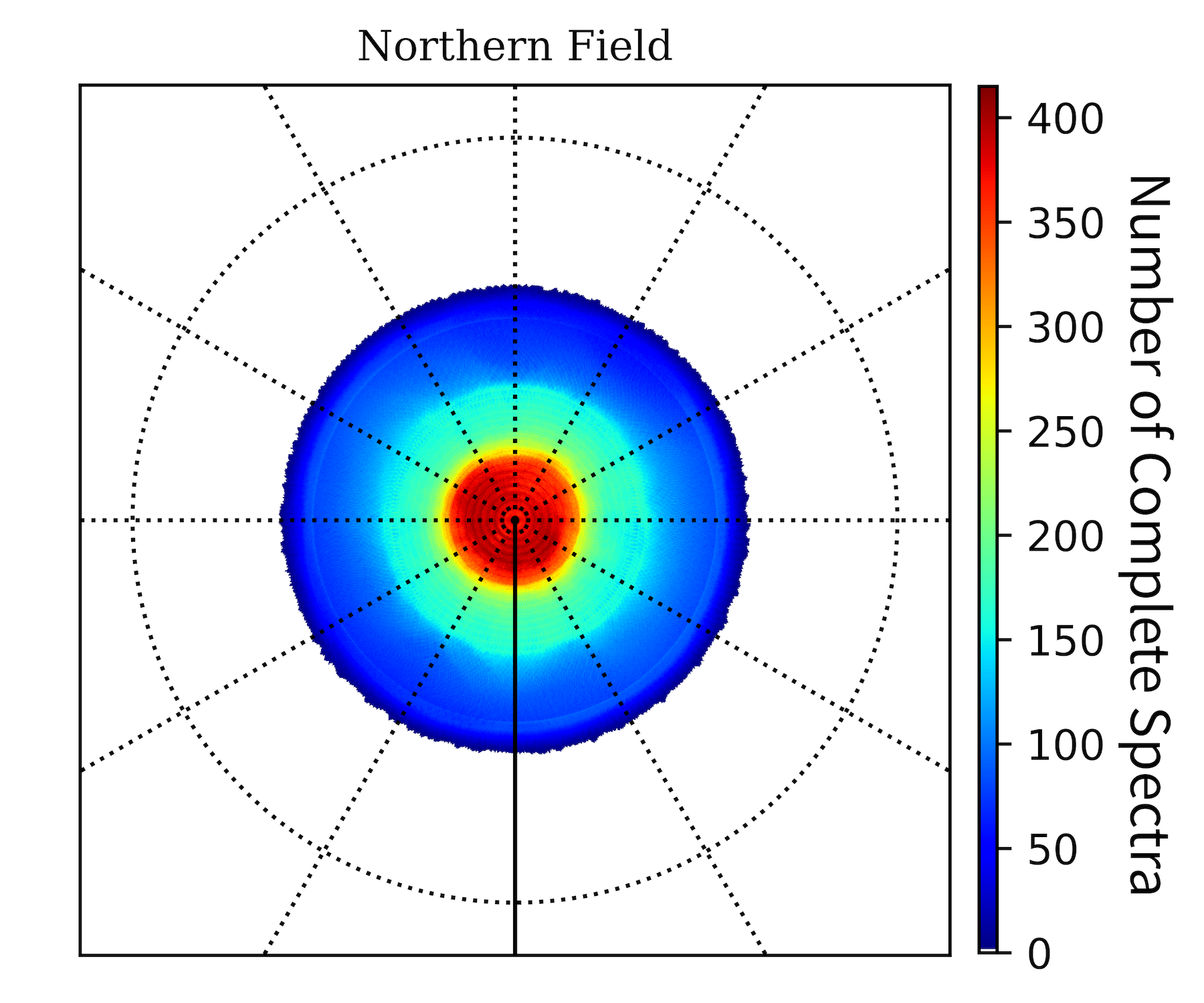}
    \caption{Coverage maps for the SPHEREx deep fields, located near the south (left) and north (right) ecliptic poles. The southern field is slightly offset from the SEP in order to avoid contamination from the Small Magellanic Cloud (SMC). Each of the two coverage maps shown covers $\sim 100$ deg$^2$, with the central red core of each map spanning a diameter of 3.5 degrees.}
    \label{fig:deep_coverage_maps}
\end{figure*}

Figure \ref{fig:deep_coverage_maps} shows coverage maps of both NEP and SEP for the simulated deep field survey. The presence of the Small Magellanic Cloud near the SEP motivates an avoidance strategy for SPHEREx which leads to slightly shallower coverage compared with NEP, along with some asymmetric structure. In these regions SPHEREx obtains considerably more measurements than of the full sky. However in the deep field regions the number of complete spectra (defined as having one measurement per channel) varies considerably as a function of ecliptic latitude, from roughly fifty complete spectra per line of sight in the outskirts of the field up to over four hundred spectra ($\sim 40000$ dithered measurements per source) in the deepest parts of the NEP field. This will yield galaxy spectro-images that are oversampled both in the spatial and spectral domains.

\begin{figure}
    \centering
    \includegraphics[width=\linewidth]{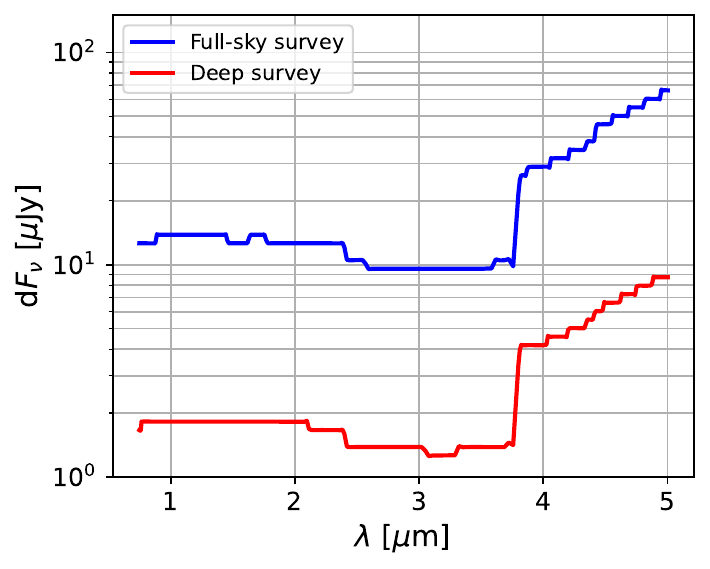}
    \caption{1$\sigma$ MEV point source sensitivity, per spectral channel. The curves represent the average sensitivity for the SPHEREx full-sky (blue) and deep (red) field depths after the nominal two-year mission.}
    \label{fig:ptsrc_uncertainty}
\end{figure}

We compute observed fluxes by integrating the set of noiseless SEDs over our nominal set of 102 SPHEREx channel bandpasses. Once this is done we then add observational noise consistent with SPHEREx's expected sensitivity. While the Zodiacal light (scattered sunlight and thermal emission by interplanetary dust grains; ZL) varies in intensity as a function of celestial position and time, for simplicity we use the conservative maximum expected value (MEV) estimates for point source sensitivity in each channel. These estimates assume a sky-averaged ZL surface brightness level informed by measurements from DIRBE \citep{kelsall98}. Many of the galaxies at full-sky sensitivity will have fluxes that are photon noise-limited (or limited by confusion noise), however we also add Poisson noise which primarily impacts the brightest galaxies.
We plot the SPHEREx point source flux sensitivities used throughout this work in Fig. \ref{fig:ptsrc_uncertainty}. At full-sky depth, the average MEV 5$\sigma$ per channel point source depths varies from 19.3 at 0.75 $\mu$m to 19.7 at 3.8 $\mu$m (bands 1-4), with reduced sensitivity for $\lambda > 3.8$ $\mu$m (bands 5 and 6). The deep fields will push roughly two magnitudes deeper in point source sensitivity, with a dependence on celestial position (see Fig. \ref{fig:deep_coverage_maps}).

% At full-sky depth, the average 5$\sigma$ per channel point source depths ranges between 19.3-19.7 across 0.75-3.8$\mu$m (bands 1-4), with reduced sensitivity for $\lambda > 3.8\mu$m (bands 5 and 6). 

The synthetic COSMOS catalog used in this work goes several magnitudes deeper than the SPHEREx full-sky sensitivity. This means the catalog can be used to make predictions for the majority of sources for which SPHEREx can measure redshifts. The catalog can also be used to simulate realistic distributions of fainter sources that contribute in the form of confusion noise, however we do not simulate source confusion in this work.

\begin{figure*}
    \centering
    \includegraphics[width=0.95\linewidth]{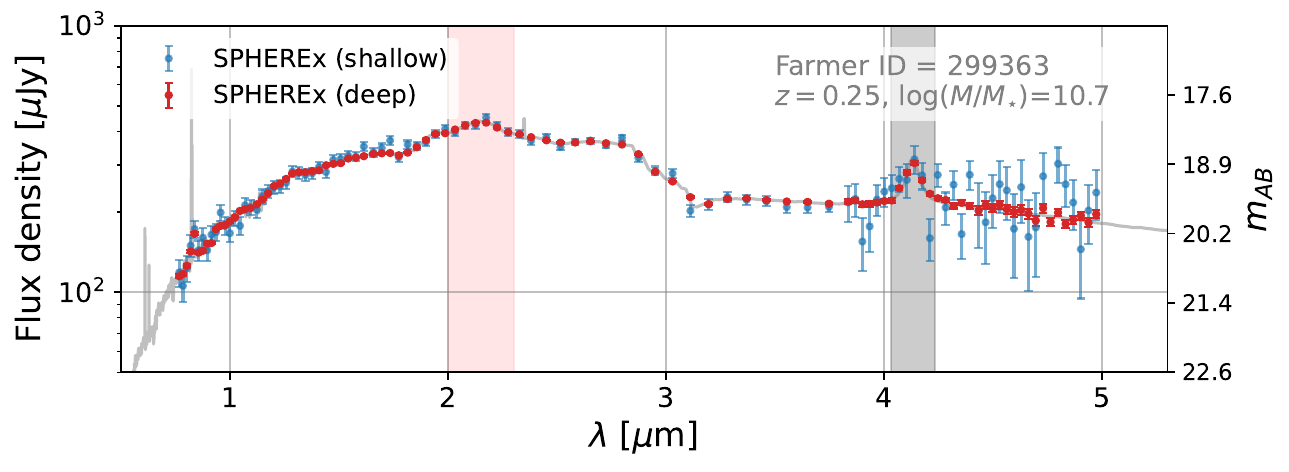}
    \includegraphics[width=0.95\linewidth]{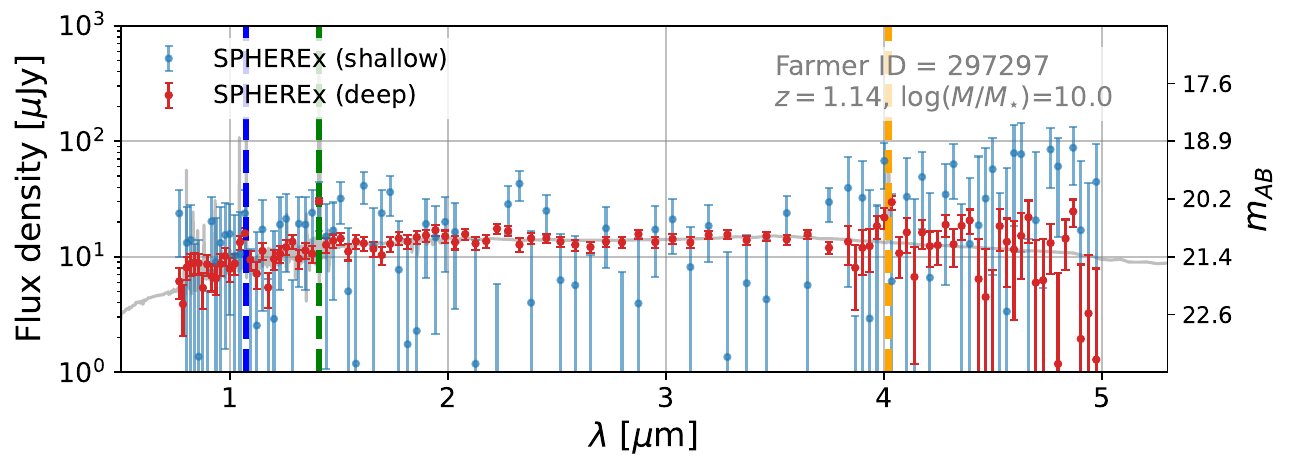}
    \caption{Mock photometry of two galaxies: a massive, quiescent galaxy at $z=0.25$ (top) and an active emission line galaxy at $z=1.14$ (bottom). Simulated flux measurement uncertainties are based on the current SPHEREx all-sky MEV point source sensitivity for 102 spectral channels. The grey lines show the underlying SEDs generated through our emprirical model, while black points show existing COSMOS photometry of the two sources. Both galaxies have photometry simulated at full-sky (blue) and deep (red) survey depths. The spectrum in the top panel shows signatures for both the rest-frame 1.6 $\mu$m bump \citep[red shaded region;][]{simpson99} and PAH emission in the black shaded region. In the bottom panel, the galaxy is detected at low significance at full sky depth while deep field photometry of the same source is sensitive to \ha+\nii\ ($\lambda_{obs}\sim1.4$ $\mu$m, green dashed line), \oiii+\hb\ ($\lambda_{obs}\sim 1.4$ $\mu$m, blue) in the rest frame optical and Paschen$-\alpha$ in the near infrared ($\lambda_{obs}\sim 4.0$ $\mu$m, orange).}
    \label{fig:active_passive_sed_compare}
\end{figure*}

Figure \ref{fig:active_passive_sed_compare} highlights two sample galaxies from the COSMOS2020 sample with simulated photometry; in the upper panel is a massive, quiescent galaxy at low redshift ($z=0.25$), while the lower panel shows an ELG observed at $z=1.14$. Like in the example shown, many quiescent galaxies have well-resolved, rest-frame 1.6\um\ bumps driven by a minimum in H$^{-}$ opacity along with PAH features at longer wavelengths. These continuum features can help derive precise redshifts for luminous red galaxies \citep{sawicki, simpson99}. In contrast, star-forming galaxies are typically less massive but contain several emission lines/line complexes which are detectable by SPHEREx, namely \ha+\nii, \oiii+\hb\ and Paschen-$\alpha$.

% \Oli{Nice plot. Not clear if SPHEREx really sees any line here. Maybe we need a zoom in? It really seems to be one or two points each times which are hardly visible. Do we want to have two panels and one galaxy per panel?}

% \Oli{that is the list of sources the spectra of which SPHEREx will measure using forced-photometry}
\subsection{Comparison with existing/near-future surveys}

Ancillary catalog data from surveys in the optical and the infrared are important for defining SPHEREx's ``reference catalog", the list of sources SPHEREx will measure the spectra for using forced photometry. The instrumental point spread function (PSF) for SPHEREx varies as a function of wavelength, with a core PSF full width at half maximum (FWHM) ranging from 2\arcsec\ at short wavelengths to a diffraction-limited 7\arcsec\ at longer wavelengths. This does not include additional spread from pointing jitter, however this is expected to be controlled at the $\sigma_{RMS}\lesssim 1\arcsec$ level. The SPHEREx PSF will in general be undersampled due to the relatively large 6.2\arcsec\ $\times$ 6.2\arcsec\ pixel size, which makes deblending adjacent sources difficult. This motivates the use of forced photometry at the locations of reference catalog sources, and is effective when the positional errors from the reference catalog are small compared to the SPHEREx pixel scale (generally the case for the ancillary catalogs mentioned in this section). \cite{symons_psf} demonstrate PSF estimation and photometry in this limit on mock SPHEREx exposures. Redshifts derived from these sources will form the basis for downstream cosmology measurements.

At optical wavelengths, several broad-band photometric surveys have been undertaken over large portions of the sky. The PanSTARRS 3$\pi$ survey has a nominal 5$\sigma$ depth of $i=23.1$ \citep{panstarrs}. The DESI Legacy Imaging Surveys combines optical data from three separate surveys (the Beijing-Arizona Sky Survey (BASS), the Dark Energy Camera Legacy Survey (DeCALS) and the Mayall $z$-band Legacy Survey (MzLS)), covering $\sim 14,000$ \sqdeg\ and reaching median $grz$ 5$\sigma$ depths of $m_{AB}=24.0$, 23.4 and 22.8, respectively \citep{desi_imaging}. Looking ahead, the \emph{Rubin} observatory LSST will obtain optical photometry in six bands ($ugrizy$) across 18000 \sqdeg\ in the southern sky \citep{lsst}. Catalogs from the first data release are expected to reach a 5$\sigma$ co-added depth of $i\sim 25$ after one year of observations, while after ten years the depth is predicted to be $i\sim 26.8$. These optical catalogs will resolve sources with considerably finer angular resolution than SPHEREx.

In the infrared, catalogs from full sky WISE imaging detect large numbers of extragalactic sources in its two broad bands centered at 3.4 and 4.5 $\mu$m (W1 and W2, respectively). The infrared catalog from 5-year WISE co-added images reaches depths of $m_{AB}\sim20.7$ and 19.9 in W1 and W2 band respectively \citep{schlafly19}, which is slightly deeper than the typical SPHEREx single-channel full-sky sensitivity. By the time of SPHEREx's launch WISE will have catalogs derived from eight years of imaging. When we cross-match WISE sources against known galaxy positions in the COSMOS field, we find a positional accuracy ranging from 0.2\arcsec\ on the bright end up to $0.5-1\arcsec$ for $W1\sim20-21$.

\subsection{Predictions from synthetic photometry}
\begin{figure*}
    \centering
    \includegraphics[width=0.48\linewidth]{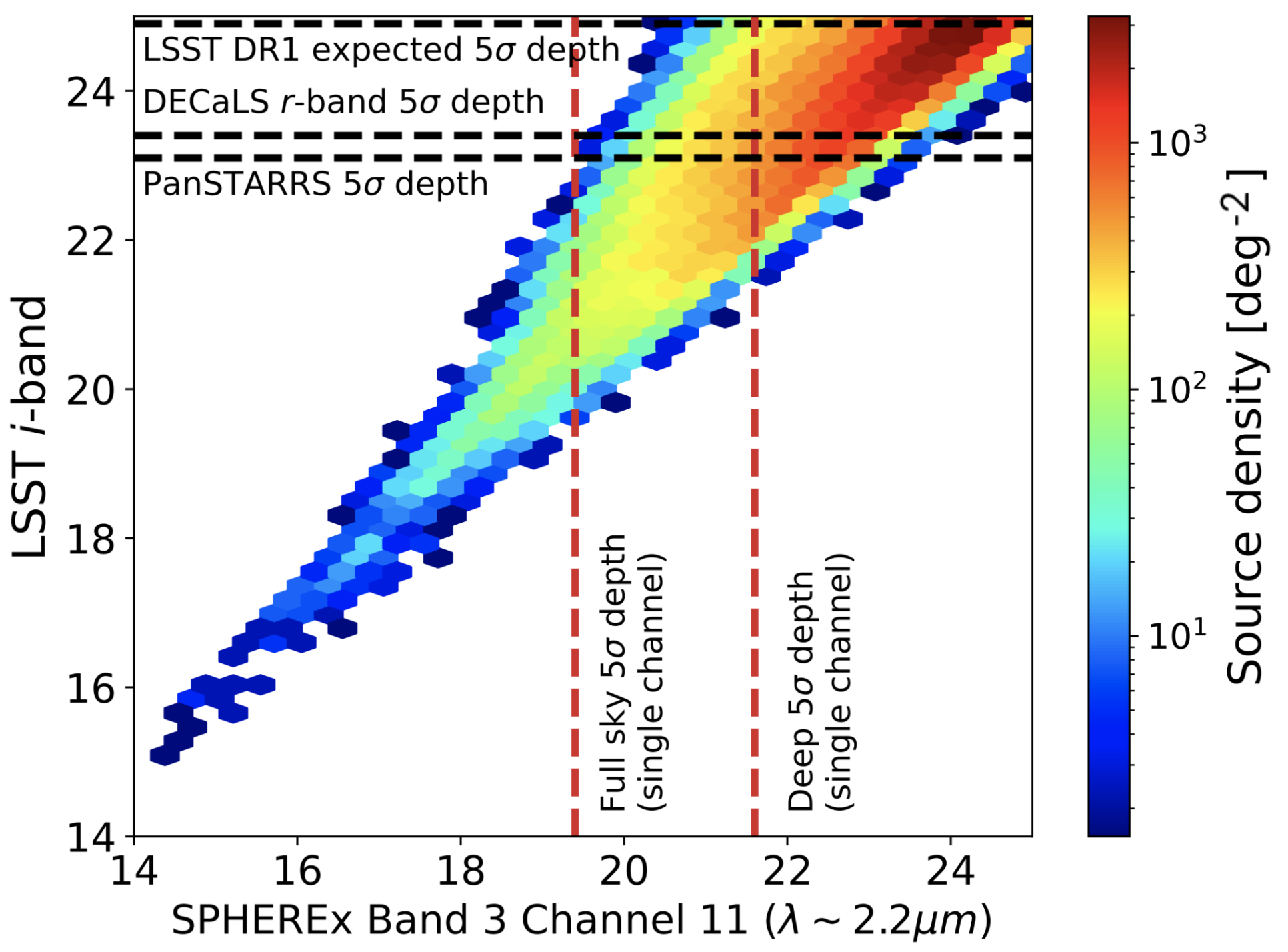}
    \includegraphics[width=0.48\linewidth]{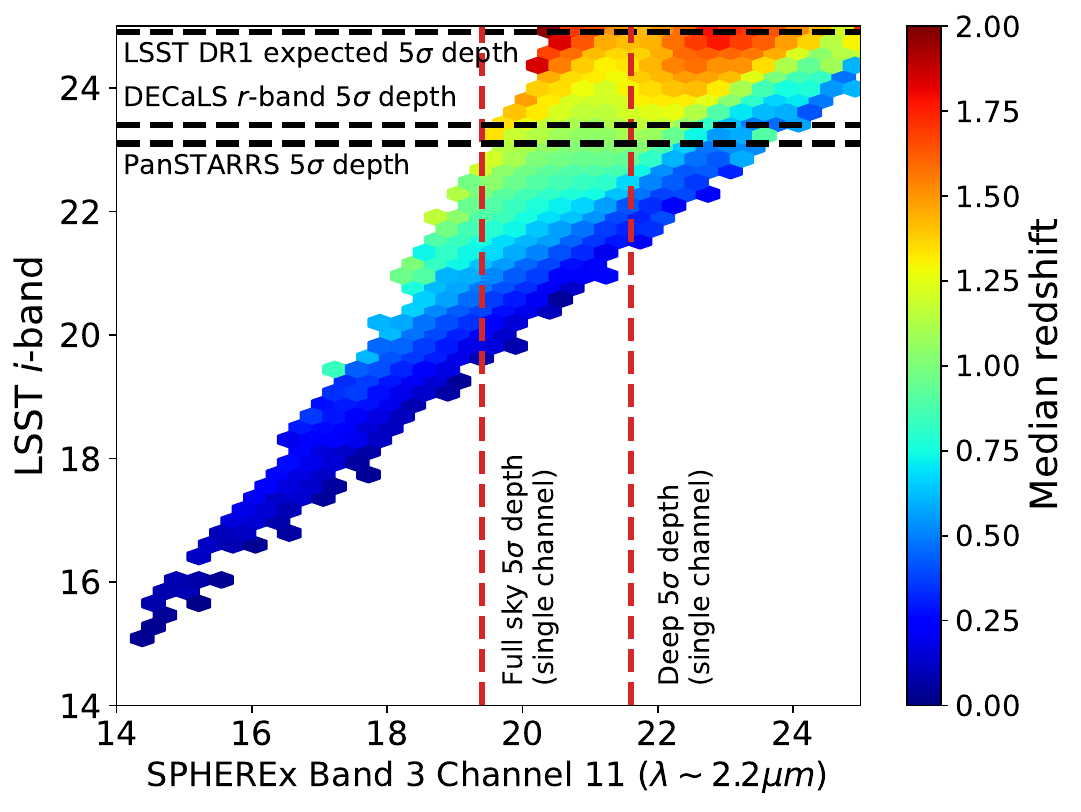}
    \includegraphics[width=0.48\linewidth]{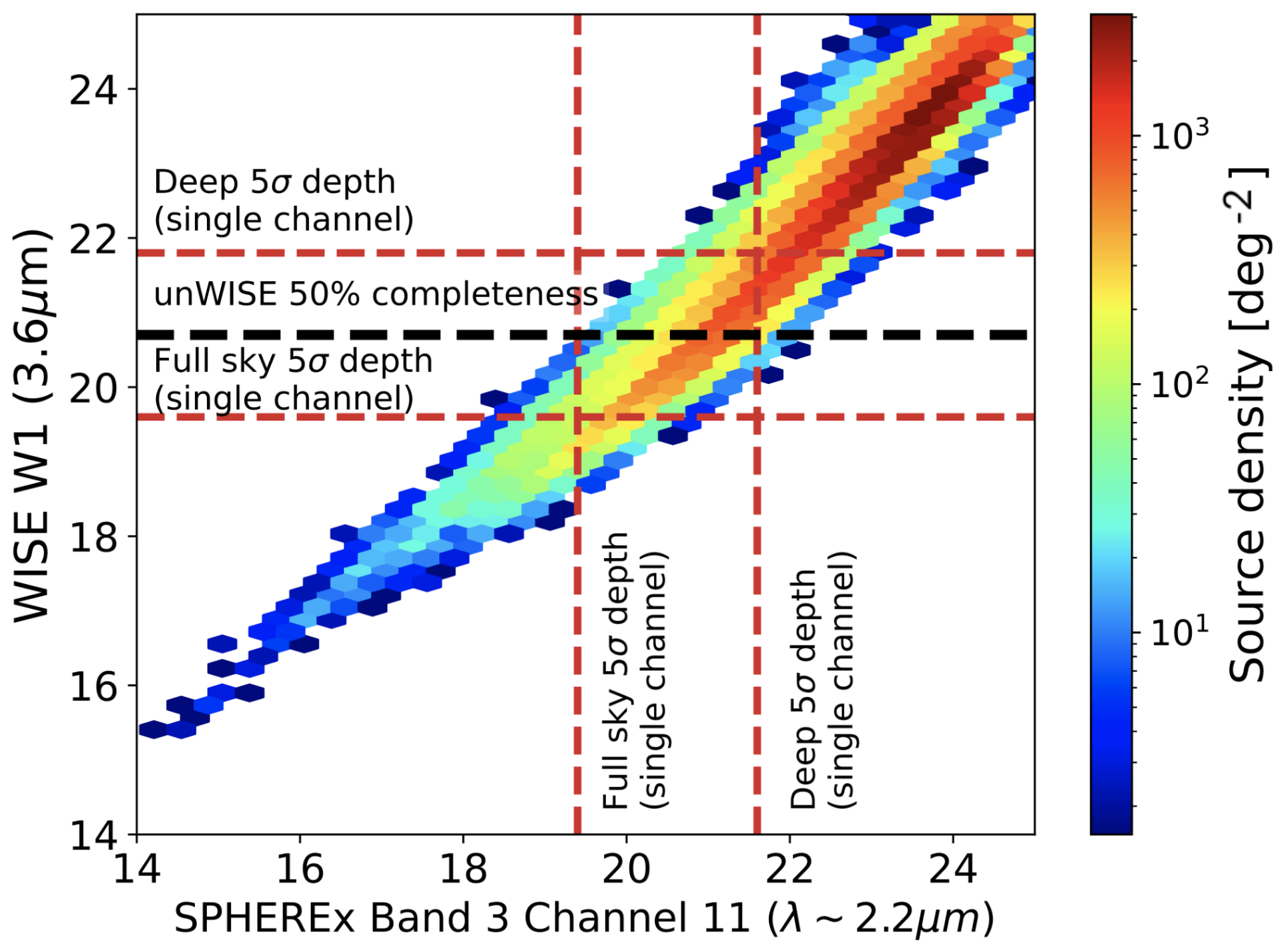}
    \includegraphics[width=0.48\linewidth]{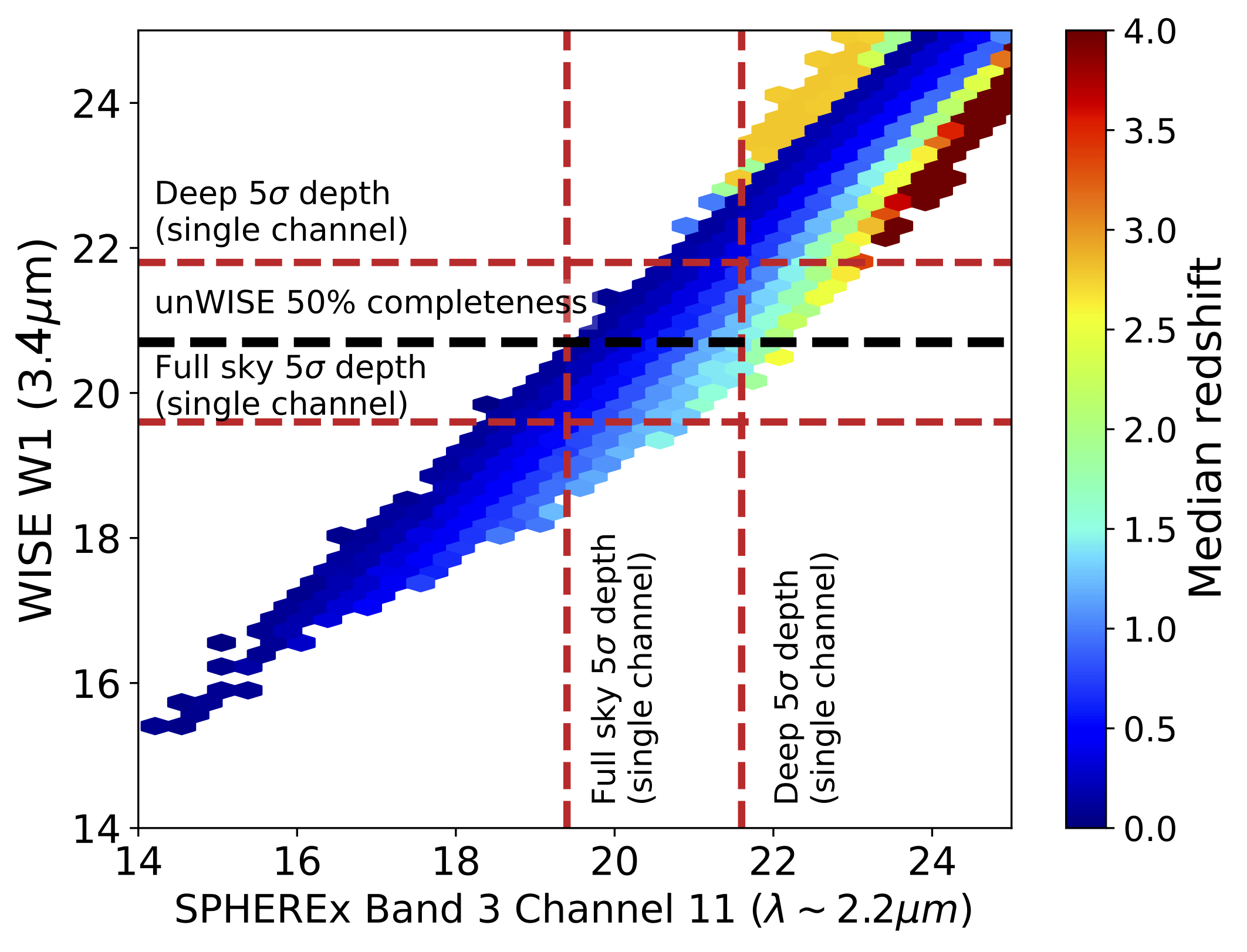}
    \caption{Magnitude-magnitude diagrams of simulated COSMOS2020 SEDs, for SPHEREx 2.2 $\mu$m magnitudes (x-axis) and LSST $i$-band (top) and WISE W1 (bottom). Also shown are expected 5$\sigma$ point source sensitivities for the three photometric surveys. The SPHEREx MEV 5$\sigma$ single channel depths are indicated by the red dashed lines for 2.2 \um\ (vertical) and 3.4 \um\ (horizontal). The SPHEREx deep field depths do not include confusion noise.}
    \label{fig:extphotdepths}
\end{figure*}

The set of high resolution model SEDs are convolved with LSST $i$-band and WISE W1 filters to obtain predicted magnitudes. These can then be compared with the nominal external survey depths to quantify the coverage afforded by the reference catalog for measurable SPHEREx sources. Figure \ref{fig:extphotdepths} shows the distribution of SPHEREx sources and synthetic \emph{Rubin}/WISE counterparts. For $i<18$ we include a subset of galaxies from our GAMA catalog matching the effective area of the COSMOS catalog. 

% \richard{Replace or remove} By plotting the magnitude distribution of individual SPHEREx channels, we can see the cross section of the full emission line galaxy population. As an example, the majority of galaxies with $(m_{[2.2]}, W1) \sim (22.5, 24)$ correspond to \ha\ line emitters with a redshift of $\sim 2.5$. \Oli{We don't see any of these guys so maybe not the best example?}

\begin{figure}
    \label{fig:red_cut}
    \centering
    \includegraphics[width=\linewidth]{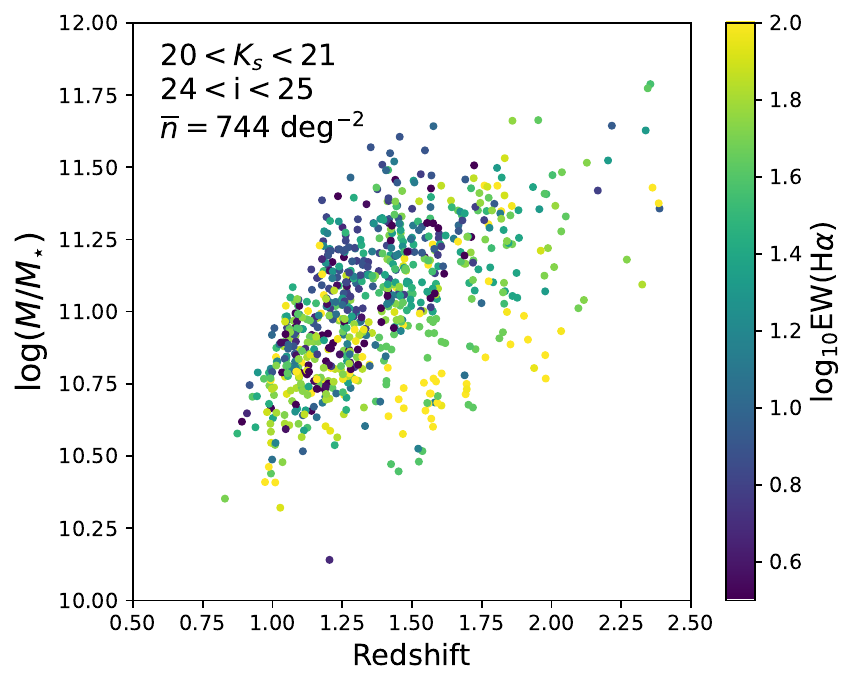}
    \caption{Distribution of galaxy redshifts and stellar masses after applying a cut on $i$- and $K$-band magnitudes. The points are colored by their \ha\ (log-) equivalent widths.}
    \label{fig:select_iK_band}
\end{figure}

SPHEREx will be effective at probing the population of infrared bright, optically faint galaxies. We inspect the population of sources in our synthetic COSMOS catalog with $24 < i < 25$ and SPHEREx $20 < m_{[2.2\mu m]} < 21$, which are measurable at SPHEREx full-sky depths. These are shown in Fig. \ref{fig:select_iK_band}. As implied by our selection, these sources are very red ($i-[2.2$ $\mu$m$]> 3$) and are typically massive (\logm\ $>10.5$), quiescent (EW(H$\alpha)<50$ \AA) galaxies with redshifts $z\sim 1-2$. Higher-redshift populations like this are important for SPHEREx cosmology measurements, which rely on precise measurement of long wavelength modes to constrain local primordial non-Gaussianity \citep[pNG;][]{dalal}.

% \richard{In the deep field, the situation is different. }
% \ilbert{Could be interesting to specify which photometric survey is expected for the deep fields.}\richard{Euclid deep field 20 deg$^2$, HSC}

% \ilbert{At this point, it could be nice to see the redshift distribution expected for the SPHEREx deep and full sky. And you could add the same redshift distribution after having imposed a cut in $i$-band.}

% \Oli{Is $N_{gal}$ the total number in COSMOS? Maybe also provide a density as in Table 2?}

\begin{figure*}
    \centering
    \includegraphics[width=0.9\linewidth]{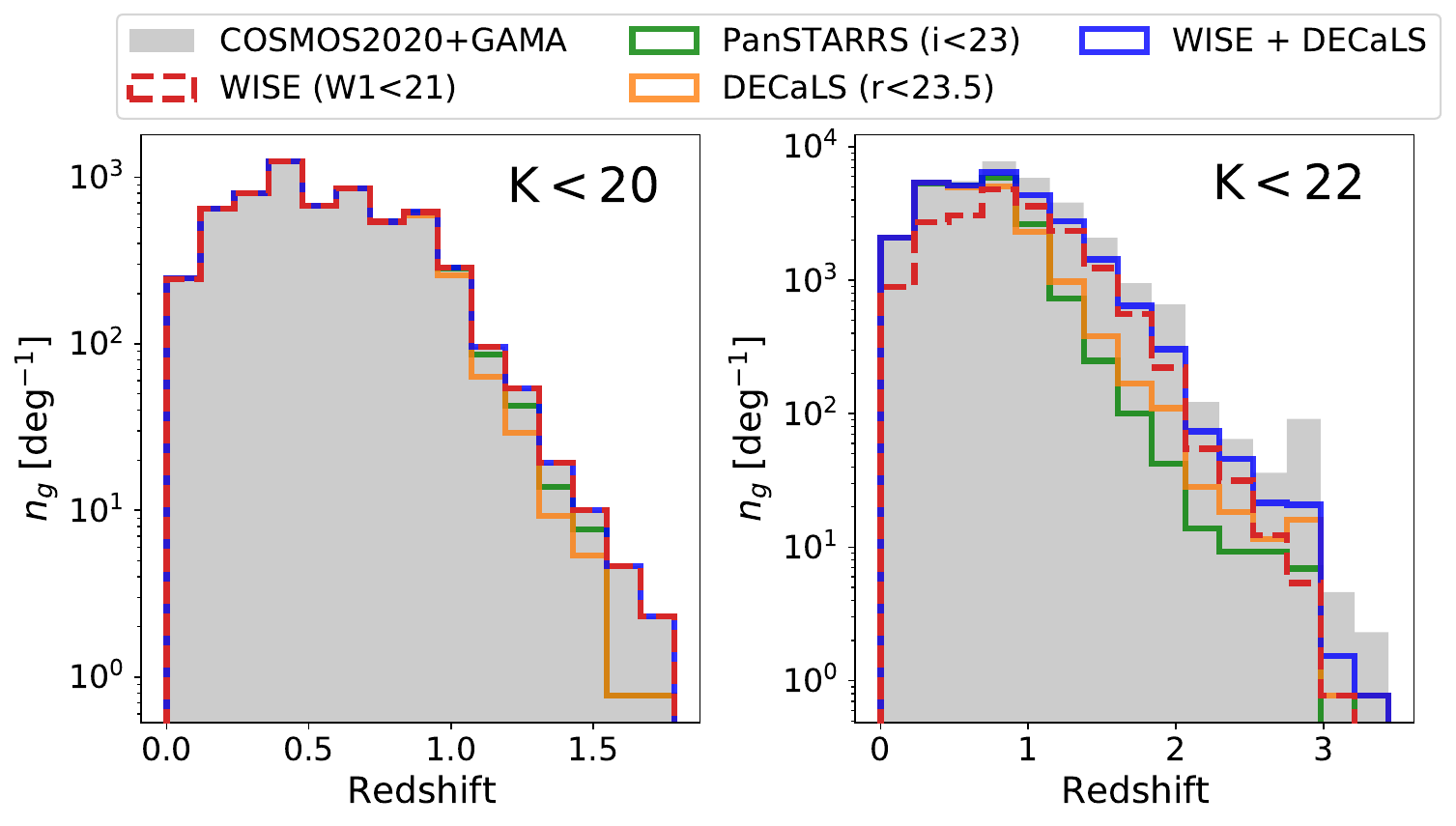}
    \caption{Comparison of the redshift distribution of our synthetic COSMOS2020+GAMA catalog (grey), along with that obtained using PanSTARRS (green), DECaLS (orange) and WISE (red) reference catalogs.}
    \label{fig:refcat_cover}
\end{figure*}

From estimated survey depths of the ancillary catalogs, the redshift distribution of sources selected by each catalog can be computed. This is shown in Fig. \ref{fig:refcat_cover} for two cuts $K<20$ and $K<22$. For $K<20$, the optical and infrared external catalogs are complete out to $z\sim 1$. For redshifts $z>1$, completeness using PanSTARRS and DECaLS falls considerably, while the WISE-selected catalog remains largely complete. For the $K<22$ cut, incompleteness of the external catalogs is more severe. The union of optical/infrared catalogs (e.g., WISE+DECaLS) complements each individual catalog (low redshifts covered by optical, higher redshifts covered by infrared), however no combination of these catalogs is complete down to SPHEREx's deep field sensitivity across the 200 deg$^2$ covering the ecliptic poles. 

\begin{figure}
    \centering
    \includegraphics[width=\linewidth]{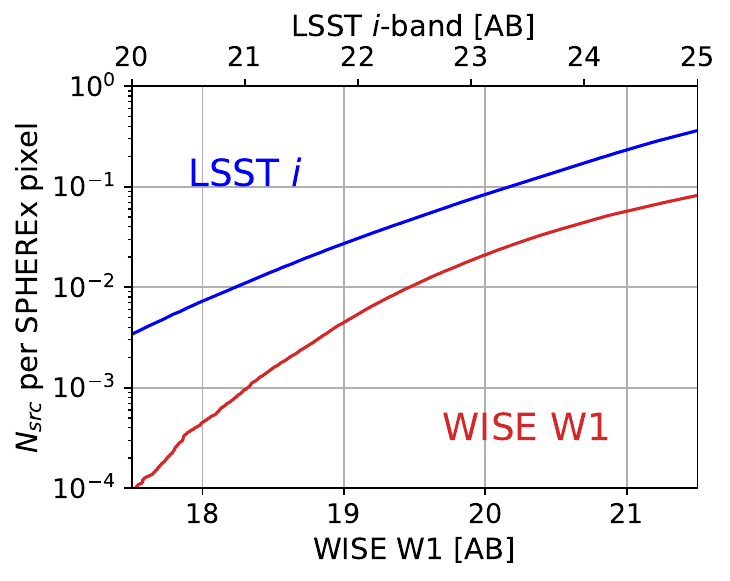}
    \caption{Source density as a function of reference catalog depth, selecting on LSST $i$-band (blue, top axis) and WISE W1 at 3.4 $\mu$m (red, bottom axis). These number densities are calculated assuming the COSMOS2020 footprint of 1.27 deg$^2$ and assuming 6.2\arcsec\ $\times$ 6.2\arcsec\  pixels.}
    \label{fig:lsst_wise_perpix}
\end{figure}

Another consideration is that the source density of the varies significantly between the optical and infrared catalogs that comprise the reference sample. This is seen clearly in Fig. \ref{fig:lsst_wise_perpix}, in which synthetic LSST $i$-band and WISE 3.4 $\mu$m densities are plotted for a range of magnitudes near each catalog's expected limiting depth. As one approaches $W1=21$ in the WISE catalog, the source density approaches 0.06 pixel$^{-1}$, or roughly one source per 16 SPHEREx pixels. In contrast, the LSST source density for all sources with $i<25$ is $\sim 0.4$ pixel$^{-1}$, i.e., one source per 2.5 SPHEREx pixels. This has implications for strategies that utilize deeper reference catalogs to define SPHEREx targets in forced photometry, and will require some parsimony in the effective number of sources that are fit simultaneously. 
% \Oli{Nice discussion}

\subsection{Prevalence of detectable emission lines in SPHEREx observations}
\label{sec:elg_prevalence}
Quantifying the fraction of galaxies where emission lines are present is important in forecasting redshift constraints. It has been shown using COSMOS 30-band photometry that accounting for emission lines in photo-z measurements treatment can increase redshift accuracy by over a factor of two \citep{ilbert09}, however it is an open question what impact emission lines will have for the SPHEREx sample. The number of detected lines and their equivalent widths will determine the redshift precision of the ELG sample. In addition, accounting for emission lines, for example \ha, can improve SFR and dust opacity estimates for many galaxies \citep{smit16}.

To compute SPHEREx line flux sensitivity we start by assuming the total flux from an emission line falls into an individual channel, i.e., $\Delta \lambda \gg \sigma_{line}$, where $\Delta \lambda$ is the channel width. For a given flux $F_{line}$ in erg cm$^{-2}$ s$^{-1}$, the flux density averaged across channel $i$ is given by
\begin{equation}
\Delta S_{\lambda_i}^{line} [\mu Jy] = \frac{10^{29} \lambda_i^2}{c}\times \frac{F_{line}}{\Delta\lambda_i},
\end{equation}
where $\lambda_i$ is the central wavelength of the bandpass in \AA\ and $c=3.0\times10^{18}$\AA\ s$^{-1}$. This expression allows us to compute the significance of detecting a signal from the line in the presence of noise. We compute this at both full-sky and deep survey depths over the full set of 102 nominal channels, assuming $R=[41, 41, 41, 35, 110, 130]$ across the six SPHEREx bands, where each band corresponds to a separate detector \citep{condon_sphx}. This calculation does not take into account the spectral dithering with which SPHEREx will sample emission lines, nor the details of line-continuum separation, which may introduce additional errors and covariances.
% \Oli{It also assumes the integrated continuum over the band is negligible or has been subtracted.}

\begin{figure}
    \centering
    \includegraphics[width=\linewidth]{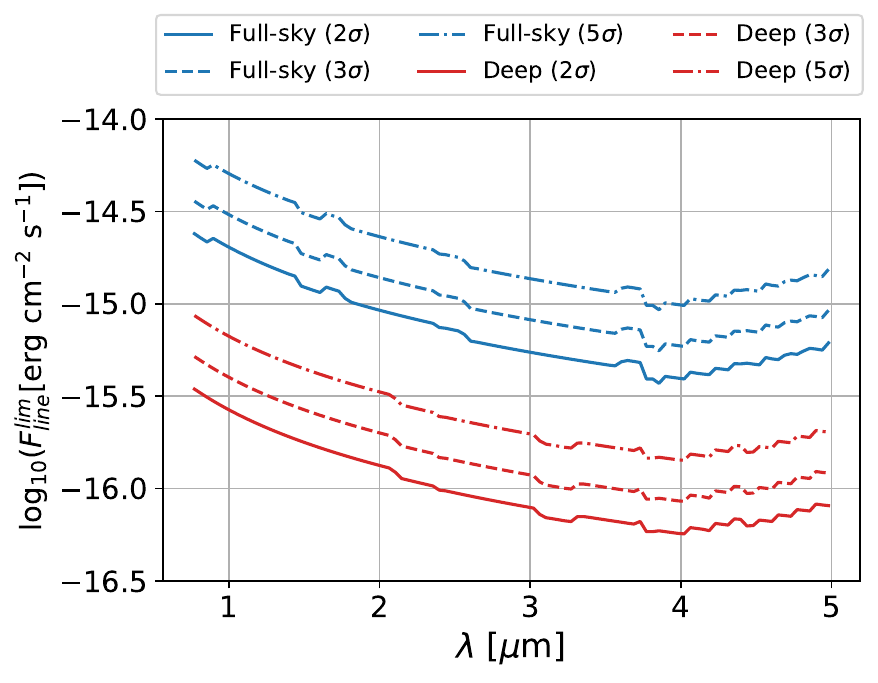}
    \caption{Line flux sensitivity as a function of observed wavelength, for SPHEREx full-sky (blue) and deep (red) surveys. The sensitivities are calculated for 2$\sigma$ (solid), 3$\sigma$ (dashed) or 5$\sigma$ (dash-dot) detections, ignoring errors from line-continuum separation and other confusion noise.}
    \label{fig:line_sens}
\end{figure}

The resulting flux sensitivities are shown in Fig. \ref{fig:line_sens}. $F_{line}^{lim}$ depends on a combination of the wavelength dependent point source sensitivity and the spectral resolution across the six LVFs. These sensitivity estimates exclude effects of confusion noise. Despite the larger instrumental noise expected at longer wavelengths, the spectral resolution is $>3$ times higher than at short wavelengths, leading to a minimum in line sensitivity around 4 $\mu$m that coincides with the minimum in Zodiacal light intensity. At full-sky depth, the 3$\sigma$ line flux sensitivity ranges from $3\times 10^{-15}$ \ergcms\ at $\lambda\sim 1$ $\mu$m down to a minimum of $6\times 10^{-16}$ \ergcms\ at $\lambda \sim 4$ $\mu$m. At the same reference wavelengths the 3$\sigma$ deep survey sensitivities range from 4$\times 10^{-16}$ \ergcms\ down to $\sim 9\times 10^{-17}$ \ergcms. This approaches the line sensitivity expected for the \emph{Euclid} and \emph{Roman} grisms and should complement these surveys through coverage beyond 2 $\mu$m.

\begin{figure}
    \centering
    \includegraphics[width=\linewidth]{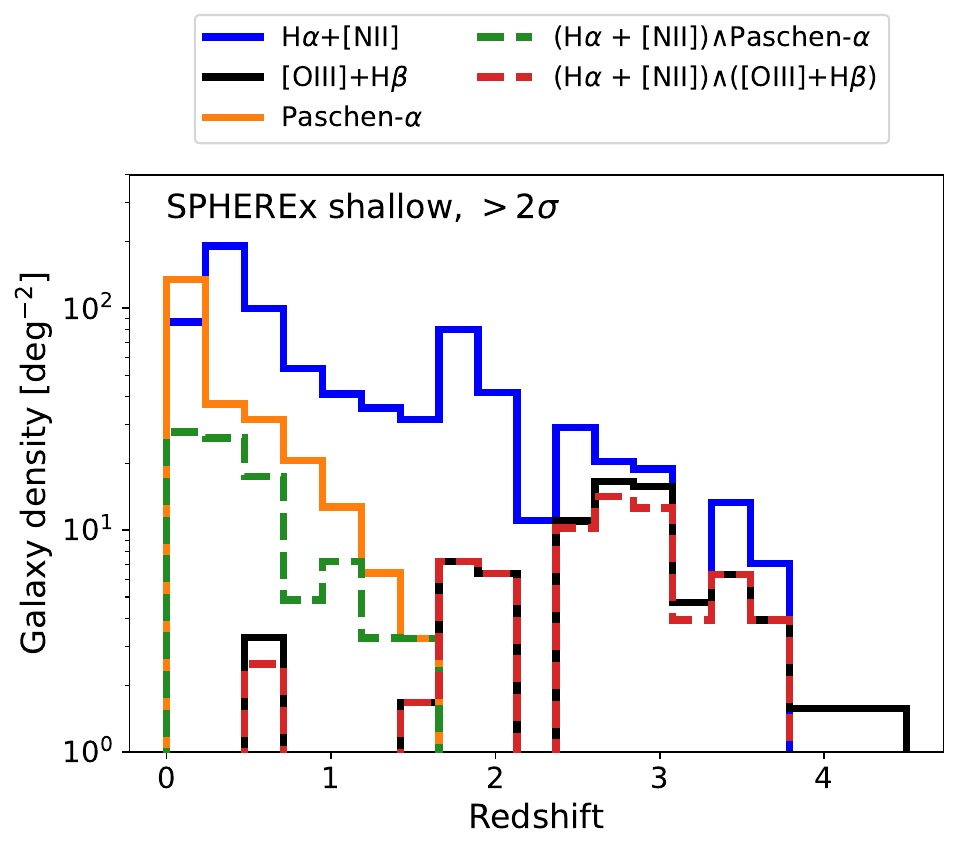}
    \includegraphics[width=\linewidth]{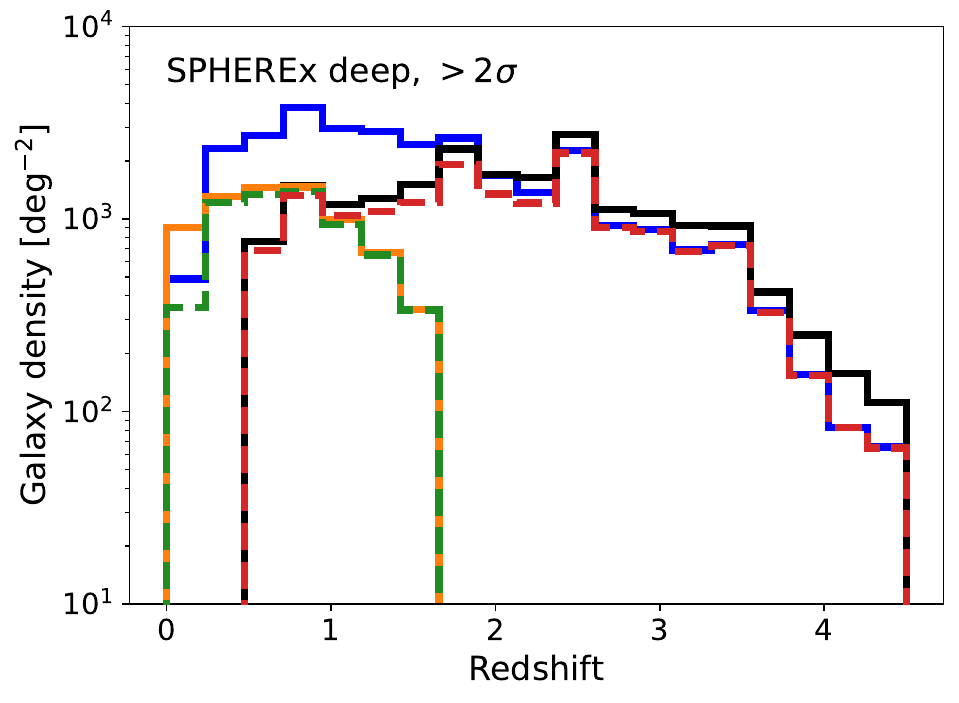}
    \caption{Redshift distribution of galaxies with detectable lines (above 2$\sigma$ significance) at full-sky (top) and deep (bottom) survey depth. This is shown for individual lines/line complexes (blue, green, orange) as well as for simultaneous line detections satisfying $>2\sigma$ in both lines.}
    \label{fig:linedetect_zdist}
\end{figure}

Using these sensitivity estimates we can predict, given a catalog of emission line fluxes and redshifts, how many sources are detectable by SPHEREx. This is done for \ha+\nii, \oiii+\hb, \oii\ and Paschen-$\alpha$. The number of $>2\sigma$ line detections at both full-sky and deep survey depth are shown as a function of redshift in Fig. \ref{fig:linedetect_zdist}. Combining line detections from both COSMOS2020 and GAMA catalogs (weighted by their effective areas), we predict full-sky number densities of  $\overline{n} \sim 770$ (260) \persqdeg\ for \ha\ detectable at 2$\sigma$ (3$\sigma$). The next most prevalent line is Paschen-$\alpha$, with $\overline{n} \sim$ 250 (140) \persqdeg. Despite the small intrinsic line flux ratio $P\alpha/H\alpha = 0.12$ (for case B recombination), Paschen-$\alpha$ is more immune to dust extinction than \ha\ and is observed at longer wavelengths where SPHEREx has better line flux sensitivity. SPHEREx is less sensitive to rest-frame optical lines blueward of \ha, due to a combination of intrinsic line ratios, more severe dust attenuation and poorer sensitivity in the blue end of the SPHEREx bandpass ($\lambda < 2$ $\mu$m). At full-sky depth, the \oiii+\hb\ complex is detectable at $>2\sigma$ for 110 COSMOS2020 catalog sources (85 deg$^{-2}$), however this drops to only 15 sources detected at $>3\sigma$. The situation is worse for \oii, for which no lines are predicted to be detectable at full-sky depth. \oii\ line emission is expected to be primarily detected by SPHEREx in the deep fields, for individual sources or in aggregate through line intensity mapping \citep[e.g.,][]{cheng_lim}.

Detecting multiple lines simultaneously with SPHEREx will enable precise and robust redshift measurements across a broad range of distances. At full-sky depth, our synthetic catalog predicts 12\% and 15\% of \ha\ lines will have a Paschen-$\alpha$ counterpart in which both lines are detected at $>2\sigma$ and 3$\sigma$ respectively. The fraction of \ha\ detections with \oiii+\hb\ are 11\% and 5\% with the same criteria, though most \oiii+\hb\ detections should have a \ha\ detection (red dashed histogram in top panel of Fig. \ref{fig:linedetect_zdist}). Simultaneous detection of \ha, Paschen-$\alpha$ and \oiii\ is rare due to limited redshift overlap. 

Our deep field predictions paint a much richer picture for the putative emission line sample near the ecliptic poles, with multiple-line detections extending from $z=0.17$ to $z=4$ and beyond. Because SPHEREx will probe the bright end of each line LF (which varies strongly with luminosity), the predicted number densities are highly sensitive to observing depth. Indeed, while the mean sensitivity of the deep survey is $\sim 10\times$ that of the full-sky, the predicted number densities are larger by factors ranging from $\mathcal{O}(10)$ up to $\mathcal{O}(1000)$. We summarize the implied ELG number densities at full-sky and deep survey depths in Table \ref{tab:line_det}. The number density estimates for the deep fields may be conservative in the sense that our initial cut on $i<25$ has a larger impact on the deep field forecasts. Due to this and our incomplete knowledge of higher-redshift line populations, we caution over-interpretation of the deep field predictions.

% \richard{This gives tantalizing prospects for finding large numbers of ELGs across the full sky, however by the same token it means predicted ELG number densities will strongly vary as a function of spatially varying depth across the sky (and other potential selection effects).}

% \begin{table*}
% \centering
% \begin{tabular}{c|c|c|c|c|c}
% Line(s) & 2$\sigma$ (full sky) & 3$\sigma$ (full sky) & 2$\sigma$ (deep) & 3$\sigma$ (deep) & 5$\sigma$ (deep)\\
% & [deg$^{-2}$] & [deg$^{-2}$] & [deg$^{-2}$] & [deg$^{-2}$] & [deg$^{-2}$]\\
% \hline
% \hline
% \ha\ + [NII] & 550 & 200 & 26600 & 15500 & 6500\\
% P$\alpha$ & 200 & 140 & 3900 & 2000 & 730 \\
% \ha\ + P$\alpha$ & 55 & 35 & 3400 & 1600 & 520 \\
% \oiii\ + \hb\ & 67 & $<10$ & 19700 & 10300 & 3200 \\
% \ha\ + \oiii\ & 55 & $<10$ & 15500 & 7400 & 2200 \\
% \oii\ & $<1$ & $<1$ & 150 & 30 & $<10$
% \end{tabular}
% \caption{Predicted number densities for galaxies with lines detectable through SPHEREx spectrophotometry. \richard{recompute}}
% \label{tab:line_det}
% \end{table*}

\begin{table*}
\centering
\begin{tabular}{c|c|c|c|c|c|c}
Line(s) & 2$\sigma$ (full sky) & 3$\sigma$ (full sky) & 4$\sigma$ (full sky) & 5$\sigma$ (full sky) & 3$\sigma$ (deep) & 5$\sigma$ (deep)\\
& [deg$^{-2}$] & [deg$^{-2}$] & [deg$^{-2}$] & [deg$^{-2}$] & [deg$^{-2}$] & [deg$^{-2}$] \\
\hline
\hline
\ha\ + [NII] & 770 & 260 & 115 & 65 & 17800 & 8000 \\
P$\alpha$ & 250 & 140 & 100 & 75 & 3800 & 1600 \\
(\ha\ + [NII]) $\wedge$ P$\alpha$ & 90 & 40 & 20 & 10 & 3100 & 1100 \\
\oiii\ + \hb\ & 85 & 12 & $<10$ & $<10$ & 10500 & 3600 \\
(\ha\ + [NII]) $\wedge$ (\oiii\ + \hb) & 74 & 12 & $<10$ & $<10$ & 7900 & 2600 \\
\oii\ & $<1$ & $<1$ & $<1$ & $<1$ & 30 & $<10$
\end{tabular}
\caption{Predicted number densities for galaxies with lines detectable through SPHEREx spectrophotometry. Number densities for combinations of lines correspond to sources where both lines/line complexes are detected above the noted significance.}
\label{tab:line_det}
\end{table*}

\section{Redshift recovery}
\label{sec:lp}

We test redshift recovery using the photometric redshift estimation code implemented in \cite{Stickley2016}, which is similar in spirit to the widely used template fitting code \lp\  \citep{lephare1, lephare2}. The code performs a $\chi^2$ minimization across a pre-specified grid of models, which we construct from the same underlying set of 160 templates used in \S \ref{sec:tempfit} to generate our synthetic observations. For each template, we deploy a grid of models with $E(B-V)=0-1$ in steps of $\delta E(B-V) = 0.05$ for three dust extinction laws (Prevot, Calzetti and Allen) with redshifts spanning $z=0-3$ with $\delta z = 0.002$. We assume flat priors over these parameters and the set of templates. In \S \ref{sec:temp_sensitivity} we test reducing the set of templates used in redshift estimation as a measure of robustness for our results.
% \Oli{Remind the reader than the 160 templates are the same than the input COSMOS2020 ones but we will revisit this in section XXX to test robustness (if I am correct!)}
% For each source, after marginalizing over the model parameters at each redshift step, we compute a redshift point estimate $\hat{z}$ and uncertainty $\sigma_z$ from the maximum likelihood solution and second moment of the redshift PDF $p(z)$, respectively.

While sufficient for broad band photometric redshift measurements, the emission line model implementations of these codes have shortcomings with stronger implications for intermediate spectral resolution SPHEREx measurements. As a result we choose in this work to assess the redshift information from continua and lines separately, which can be combined in a hybrid line-continuum redshift estimation approach that will be the subject of a future publication.

To emulate the selection of SPHEREx target galaxies, we evaluate redshift recovery for galaxies pre-selected using optical and infrared ancillary catalogs. In particular, we select any galaxies detected by DECaLS ($g<24.0$, $r<23.4$ or $z<22.8$) or WISE ($W1<20.5$ or $W2<20.5$), which constitutes fifty four thousand galaxies of the full 160K simulation catalog. We include synthetic DECaLS $grz$ and WISE $W1/W2$ photometry with representative noise in the fits, adding a 1\% noise floor to capture additional photometric errors in the external catalogs.
% \Oli{Maybe remind the reader what defined these thresholds}

% \richard{We run the photometric redshift estimation code \lp\ on mock SPHEREx full-sky photometry, which implements a $\chi^2$ minimizaztion across a pre-specified grid of models \citep{lephare1,lephare2}.
% The same set of 160 templates described in \S \ref{sec:tempfit} are used to fit the mock observations, so our results do not include systematic effects related to template mismatch. \lp\ incorporates a separate line prescription based on scaling relations from \cite{kennicutt98} between $L_{UV}$ and \oii\ \citep[as used in][]{ilbert09}. This \oii\ relation is the same as we use for model-based COSMOS templates (i.e., when we do not have a direct line equivalent width measurement from a \brown\ template), however our mocks have intrinsic line ratios that scale as a function of redshift and stellar mass as opposed to being fixed. The \lp\ runs use a single dust law \citep{calzetti2000} while in our simulations we apply five different laws to source SEDs (the dust law for a given source is determined by \texttt{Fitcat}, see \S \ref{sec:sedlib}) and include a redshift-dependent differential nebular extinction correction. For each template, we deploy a grid of models with $E(B-V)=0-0.5$ in steps of $\Delta E(B-V) = 0.1$ and redshifts from $z=0-6$ with a resolution of $\delta z = 10^{-4}$. Our results assume flat priors over these parameters and we use the median of the likelihood distribution ($z_{ML}$ in \lp) for the following results.} 

\subsection{Continuum redshift results}

\subsubsection{COSMOS2020}
\label{sec:cos_cont_results}
We calculate the mean $\chi^2$ statistic of the fits to be 101.7, corresponding to a reduced chi-squared of $\chi^2_{red} \approx 0.99$ assuming four model parameters (redshift, $E(B-V)$, template scale, template index) and 107 total bands (SPHEREx + external). We confirm that the distribution of best fits follows a $\chi^2$ distribution, with few galaxies having $\chi^2_{red}>1.5$. Given the use of the same galaxy templates used to fit the photometry as used to generate the SEDs, this level of agreement indicates that our fits are well behaved.  
% \Oli{That's really nice}

\begin{figure}
    \centering
    \includegraphics[width=\linewidth]{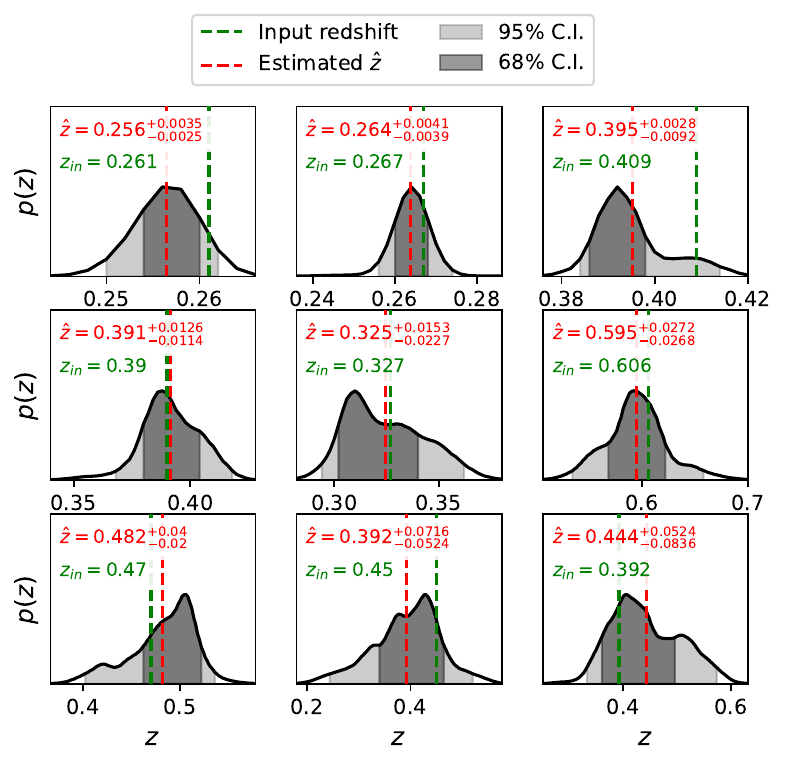}
    \caption{Normalized redshift probability density functions (PDFs) for a collection of examples from the COSMOS2020 sample. For each source, the true redshift, the reported redshift estimate and 68\%/95\% credible intervals are indicated.}
    \label{fig:zpdfs}
\end{figure}

We plot a random selection of redshift PDFs with increasing $\sigma_{z}$ in Figure \ref{fig:zpdfs}. Each redshift point estimate is computed from the first moment of the redshift $p(z)$, which roughly coincides with the maximum \emph{a posteriori} estimate for unimodal distributions. It can be seen that many of the redshift PDF estimates have non-Gaussian structure, including heavy tails and often more than one local $\chi^2$ minimum. This is to be expected given the complexity of the template set, for which several templates may be degenerate, along with other parameters in the model space. Motivated by this, we include two-sided redshift uncertainty estimates derived from the highest (posterior) density interval (HPDI), defined as the shortest interval on a posterior density for some given confidence level. In some cases where the uncertainty is comparable to the redshift step size ($\sigma_z \sim \delta z=0.002$, e.g., top left panel of Fig. \ref{fig:zpdfs}), discretization effects may impact the redshift estimates, suggesting that for the highest accuracy sample some refinement of $p(z)$ will be necessary. 

% \richard{YKC and AF suggest colorbars for this plot and Fig. 24. "I know it is going to be messy as you have different normalizations in each panel. Two options: one is to use a simple grayscale (or any single color sequential cmap), the second option is to do scatter plots, which is better in the panels when you have very small scatters and all galaxies are on the one-to-one line."}

\begin{figure*}
    \centering
    \includegraphics[width=\linewidth]{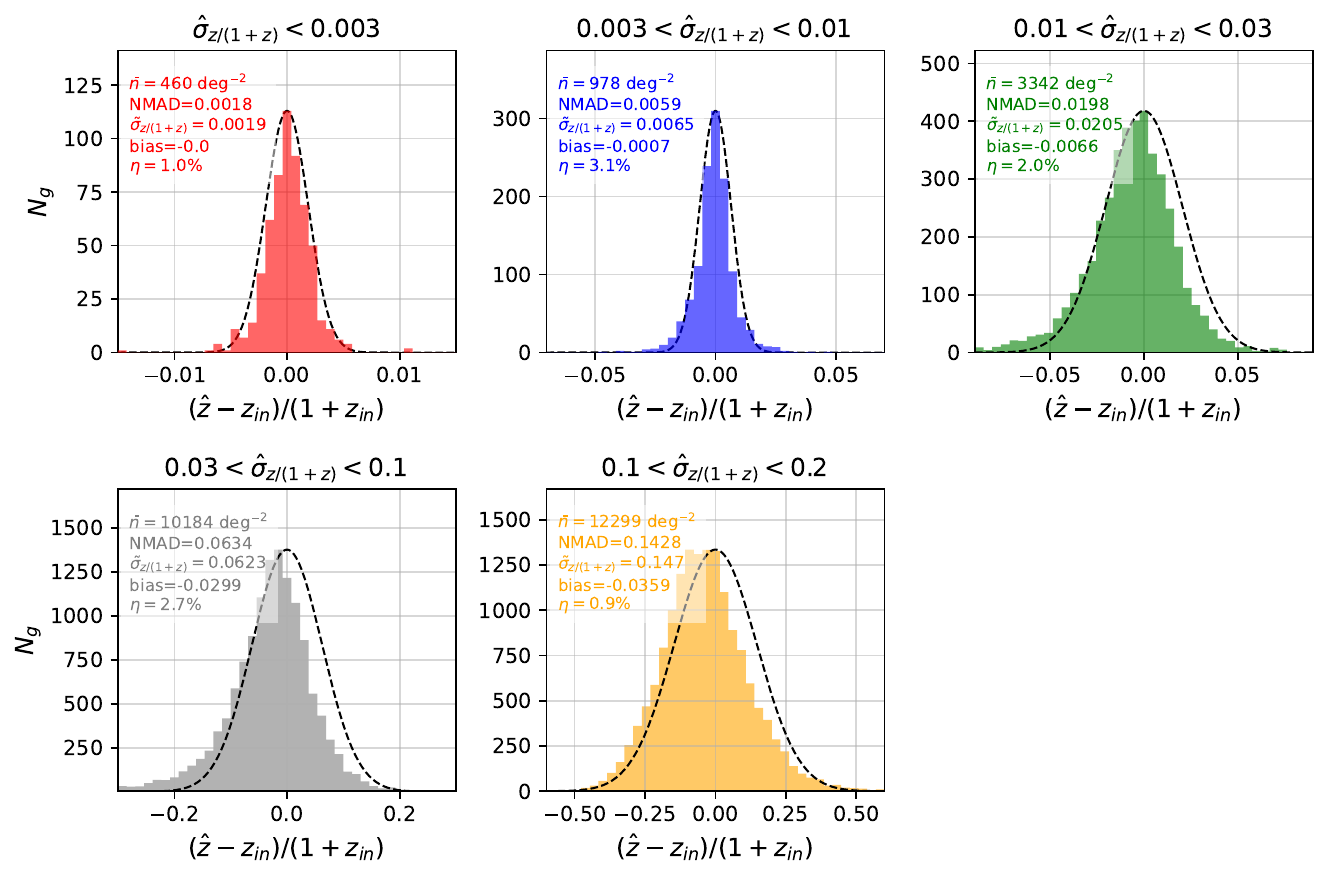}
    \caption{Redshift error distributions for COSMOS sources, binned by the reported redshift uncertainty $\hat{\sigma}_{z/(1+z)}$. The dashed Gaussian curves have widths that correspond to the median reported uncertainties within each redshift uncertainty bin, $\tilde{\sigma}_{z/(1+z)}$ and are normalized to the peaks of the histograms. For each sample we report $\tilde{\sigma}_{z/(1+z)}$, the mean redshift bias, the normalized median absolute deviation (NMAD) and the 3$\sigma$ outlier fraction $\eta$. These results do not account for effects of source confusion.}
    \label{fig:lp_zml_sort_sigmaz_zerr}
\end{figure*}

Using these estimates we plot redshift error distributions for the COSMOS sample in Fig. \ref{fig:lp_zml_sort_sigmaz_zerr} relative to the true redshifts of the samples. These are binned by reported fractional redshift uncertainties, $\hat{\sigma}_{z/(1+\hat{z})}$, which can be compared with the true errors to assess the fidelity of the redshift estimates. For this comparison we compute redshift uncertainties corresponding to one-half the width of the 68\% credible interval (i.e., $\hat{\sigma}_z^{68}$). To quantify redshift errors in each bin, we calculate the normalized median absolute deviation (NMAD), a measure of dispersion that is robust to outliers. We also quantify the outlier fraction $\eta$, which is defined as the fraction of $3\sigma$ outliers given $\hat{\sigma}_z$ and the true error. These results demonstrate that SPHEREx will measure a wide range of high- and low-accuracy galaxy redshifts. In this test configuration the reported uncertainties closely match the true errors (which can also be evaluated in terms of the z-score distribution, see Appendix \ref{sec:zscore_appendix}). The 3$\sigma$ outlier fraction remains at the few percent level for all redshift uncertainty bins. We note that there is a mild negative bias that becomes larger for the lower-accuracy samples, which upon further inspection is largely driven by the quiescent galaxy samples. This motivates further investigation into the parameter degeneracies within our template fitting, which may lead to multi-modal redshift solutions for some fits.

\begin{figure*}
    \centering
    \includegraphics[width=0.95\linewidth]{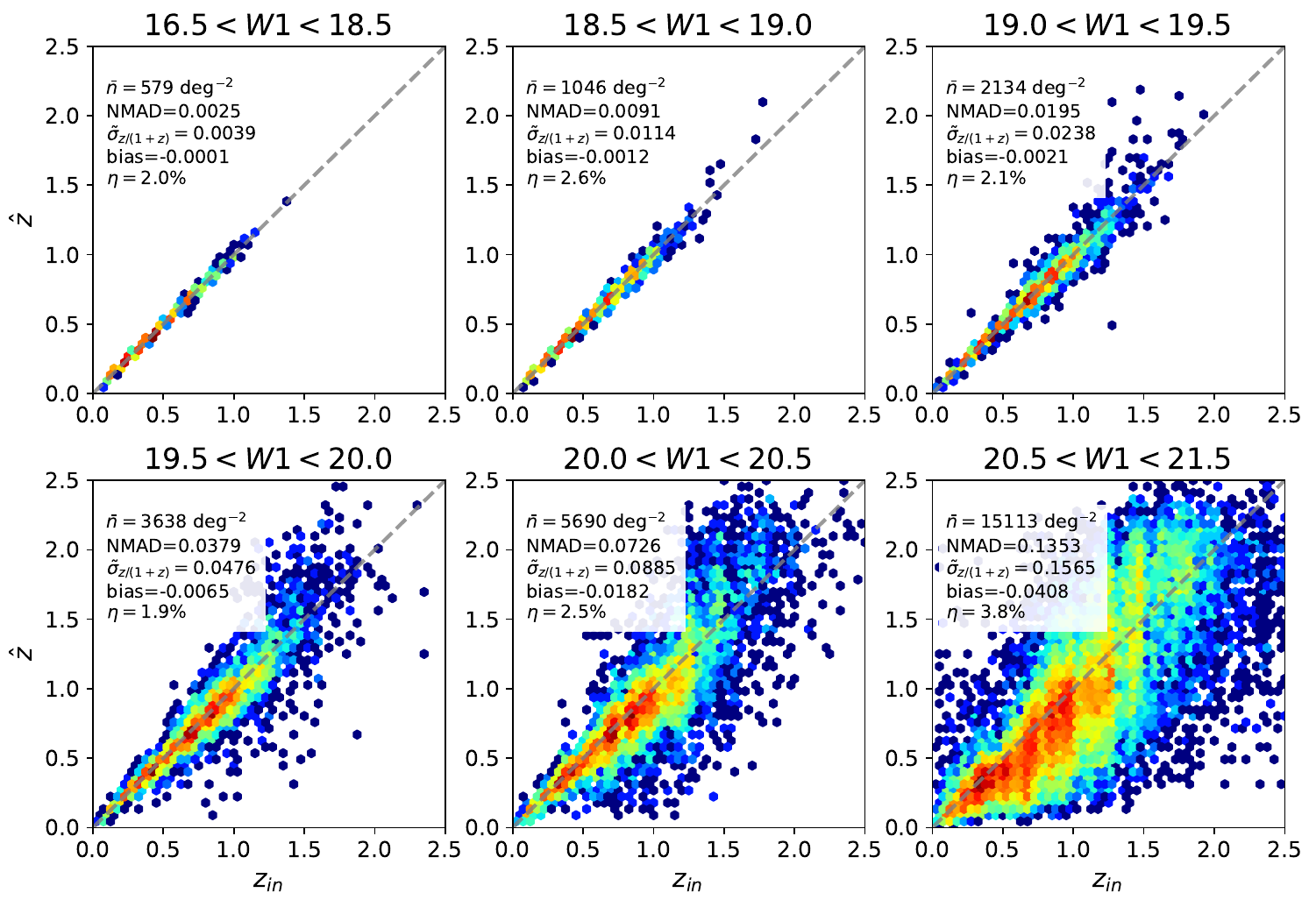}
    \caption{Recovered redshifts for synthetic SPHEREx photometry derived from the COSMOS2020 catalog from sources selected in the $W1$ band. For each magnitude bin the galaxy density $\overline{n}$, normalized median absolute deviation (NMAD), median redshift uncertainty $\bar{\sigma}_{z/(1+z)}$, mean bias and 3$\sigma$ outlier fraction $\eta$ are included. The colormap corresponds to the (log-) number density of galaxies in each bin.}
    \label{fig:lp_zml_zspec_C20}
\end{figure*}

To further understand the redshift results we also plot the input and recovered redshifts for the COSMOS2020 sample as a function of $W1$ magnitude in Fig. \ref{fig:lp_zml_zspec_C20}. Within each $W1$ bin there is a wider distribution of redshift uncertainties. Nonetheless there is a clear trend between the redshift accuracy and $W1$, along with for the mean bias and outlier fraction. Redshift measurements in the bin $20.5 < W1 < 21.5$, for which long-wavelength SPHEREx data are largely uninformative, show a clear bias toward lower redshift values which warrants further investigation.

By combining the recovered number densities from COSMOS2020 and GAMA (see Appendix \ref{sec:gama} for redshift recovery results of the GAMA sample) we can forecast the number of galaxy redshifts accessible to the SPHEREx full-sky cosmology sample. Assuming an effective area of 3$\times 10^4$ deg$^2$ and removing 3$\sigma$ outliers, our results imply a sample of 19 million galaxies with $\delta z<0.003(1+z)$, which primarily occupy redshifts $z\lesssim 1$. There are many more intermediate and low-accuracy redshifts; we forecast a sample of 445 million galaxies with $\delta z<0.1(1+z)$, and this grows to 810 million galaxies with $\delta z<0.2(1+z)$. While the loss of information on $f_{NL}$ from galaxies becomes more significant beyond $\sigma(z)=0.1(1+z)$ \citep{putter}, the number density of galaxies grows significantly across this range, meaning these low-accuracy galaxies do still provide useful information for pNG constraints. Other studies may also benefit from such low-accuracy samples, for example cross correlations with CMB lensing \citep{krowleski_lensing, act_lensing}. 

\subsubsection{Sensitivity to choice of template library}
\label{sec:temp_sensitivity}
\begin{figure*}
    \centering
    \includegraphics[width=0.95\linewidth]{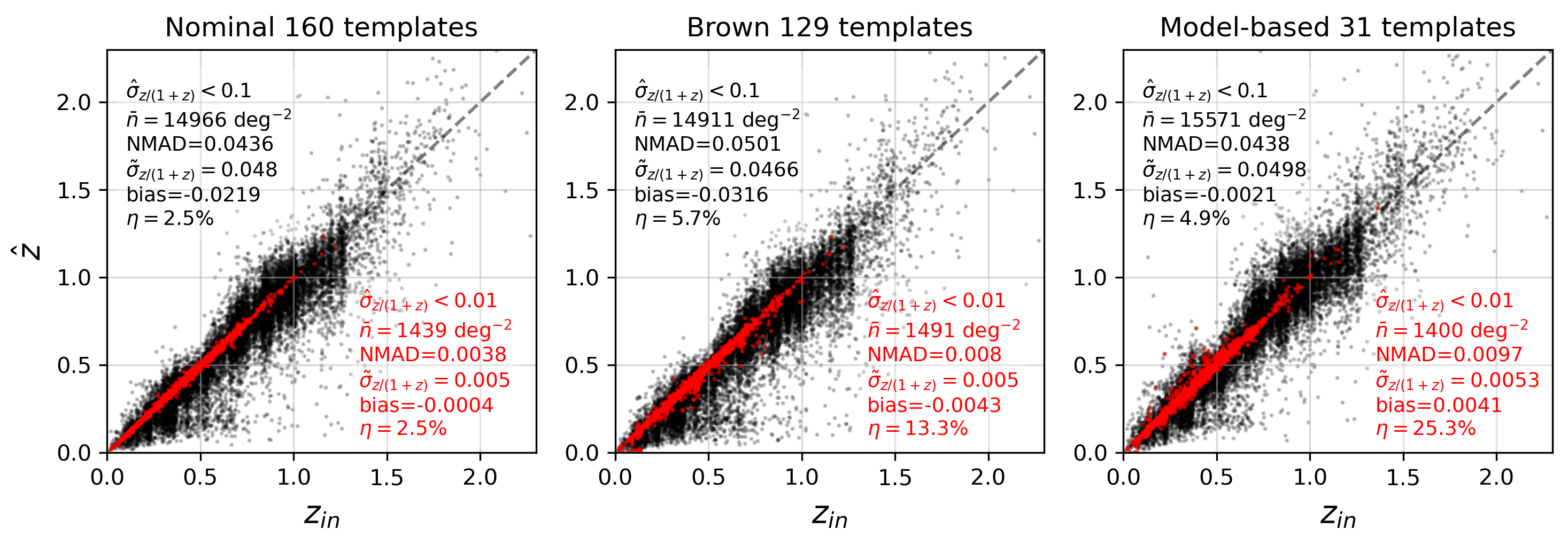}
    \caption{Input vs. recovered redshifts on the COSMOS2020 sample, using three different template library sets: the full 160 template set used to generate the photometry (left), the 129 empirical \texttt{B14} templates (middle) and the 31 model-based templates used in \cite{cosmos2020} (right). We distinguish high- and low-accuracy (red, black) samples whose redshift recovery statistics are noted as well.}
    \label{fig:compare_zin_zout_temp_subsets}
\end{figure*}

Thus far we have assumed the same set of templates used to fit the redshifts as were used to generate galaxy SEDs. However in practice a reduced set of templates may suffice, both for recovering reliable redshifts and for computational performance. To test the effects of different template sets on the recovered redshifts, we perform similar fits using the empirical Brown templates (129 in total) and the model-based COSMOS templates (31 in total) separately. These results are shown in Figure \ref{fig:compare_zin_zout_temp_subsets}. Looking at sources with $\hat{\sigma}_{z/(1+z)}<0.1$ (black points), both the Brown-only and COSMOS-only template sets perform reasonably well compared to the full template set, albeit with slightly higher outlier fractions and NMAD. Interestingly, the mean bias from using the 31 model-based templates is much smaller than the other two cases. In contrast, the high-accuracy results ($\hat{\sigma}_{z/(1+z)}<0.01$, red points) are more sensitive to template coverage. Compared to the full template set results, those using \texttt{B14} templates alone have a 5$\times$ higher outlier fraction, a $10\times$ larger bias, and a NMAD that is much larger than the reported uncertainty. The high-accuracy results degrade further when using just the 31 model-based templates, with a 3$\sigma$ outlier fraction that rises to 25.3\% and a NMAD that is nearly twice as large as the reported uncertainties. These results confirm our intuition that high SNR fits are much more sensitive to template coverage than the lower-accuracy samples. 

\subsection{Redshifts from spectrally dithered emission line measurements}

Our line detection and redshift results thus far have utilized photometry in 102 homogenized spectral bandpasses, corresponding to the 102 SPHEREx ``channels". However in practice, each observed SPHEREx source will have flux measurements sampled at a unique set of sub-channel positions which roughly Nyquist sample the spectral response function (see \S \ref{sec:sphx}). This additional complexity comes with an opportunity. In this section we demonstrate that by modeling emission lines with the native flux measurements it is possible to go beyond the naive redshift accuracy implied by the per-channel resolution.

To illustrate the potential of SPHEREx's low-resolution spectroscopy we simulate spectrally dithered line flux measurements consistent with the nominal full-sky survey strategy. We focus in this work on the \ha+\nii\ complex and Paschen-$\alpha$, using line fluxes from our GAMA catalog. For each source we simulate four measurements per channel, and assume the filters are separated at twice the channel resolution. We consider an idealized setup in which the continuum is perfectly subtracted and the continuum measurements constrains the positions of emission lines with a redshift accuracy $\lesssim 10\%$, i.e., the position of each lines are known to within a few SPHEREx channels. Within this range we assume a uniform prior over redshift.

We employ a $\chi^2$ minimization to fit the flux measurements from one or several lines, evaluated over a grid of redshifts using Gaussian line profiles. To model the \ha+\nii\ complex we use a fixed prior on the line ratio \nii/\ha$=0.35$, which is informed by the distribution of detectable lines at full-sky depth. At each redshift, we marginalize over the amplitude of the line(s) (denoted $\lbrace A_i\rbrace$) to obtain the conditional maximum \emph{a posteriori} (MAP) estimate:
\begin{align}
    \lbrace\hat{A_i}\rbrace^{MAP} &= \max \left[\ln p(\vec{F}|\lbrace{A_i}\rbrace)+\ln p(\lbrace{A_i}\rbrace|z)\right] \\ &\approx \max\left[\ln p(\vec{F}|\lbrace{A_i}\rbrace)\right].
\end{align}
We do not impose a prior on the line amplitudes (beyond a fixed \ha/\nii\ line ratio), i.e., we do not enforce positive solutions for the line amplitudes. Once the model is evaluated over the pre-determined redshift range we compute the global MAP estimate and 68\% credible interval of the 1D redshift PDF $p(z|\vec{F})$. 

% From these estimates we can assess the redshift precision as a function of line flux. 

\begin{figure*}
\centering
\includegraphics[width=\linewidth]{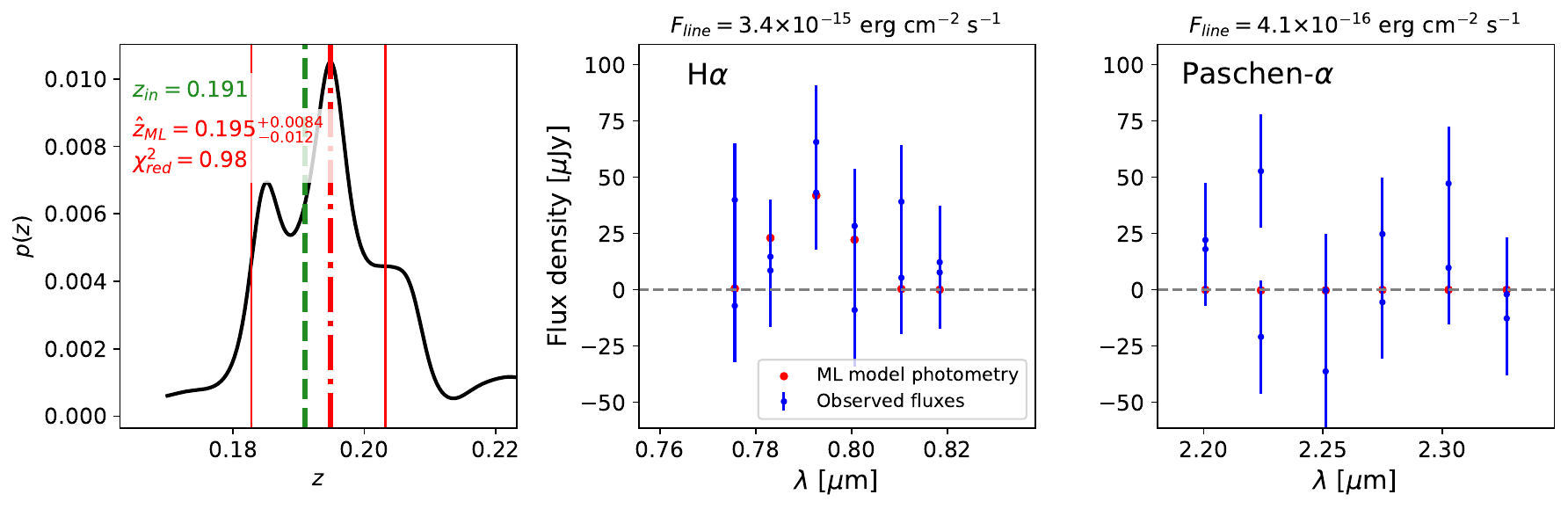}
\includegraphics[width=\linewidth]{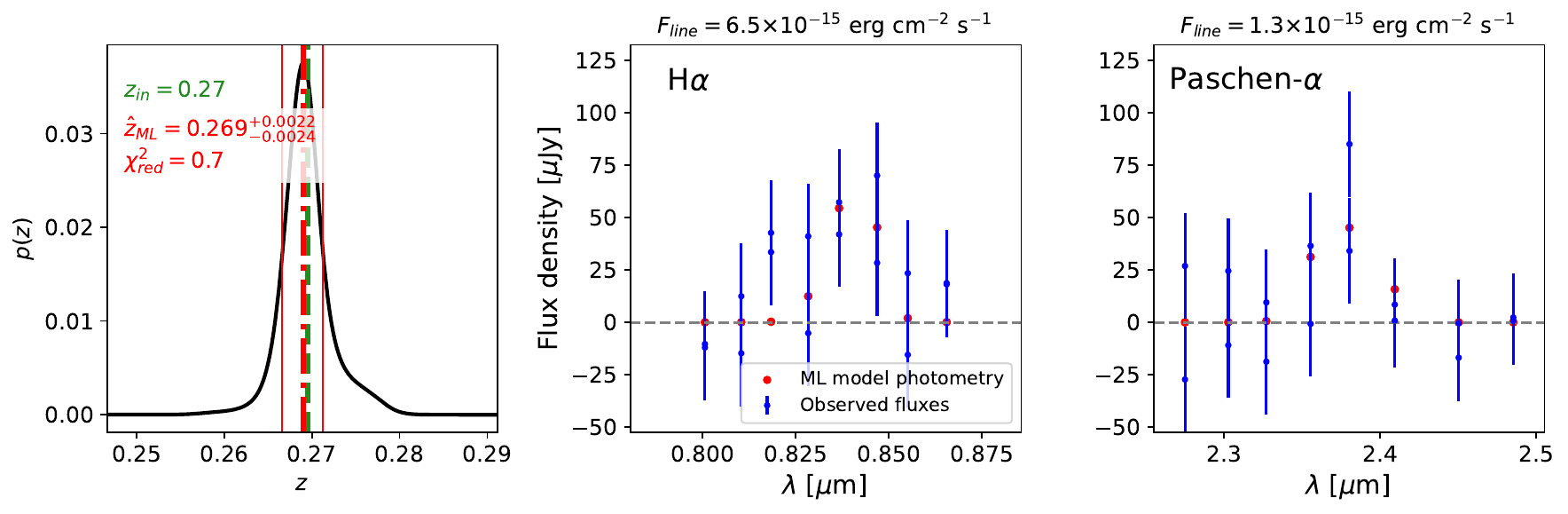}
\includegraphics[width=\linewidth]{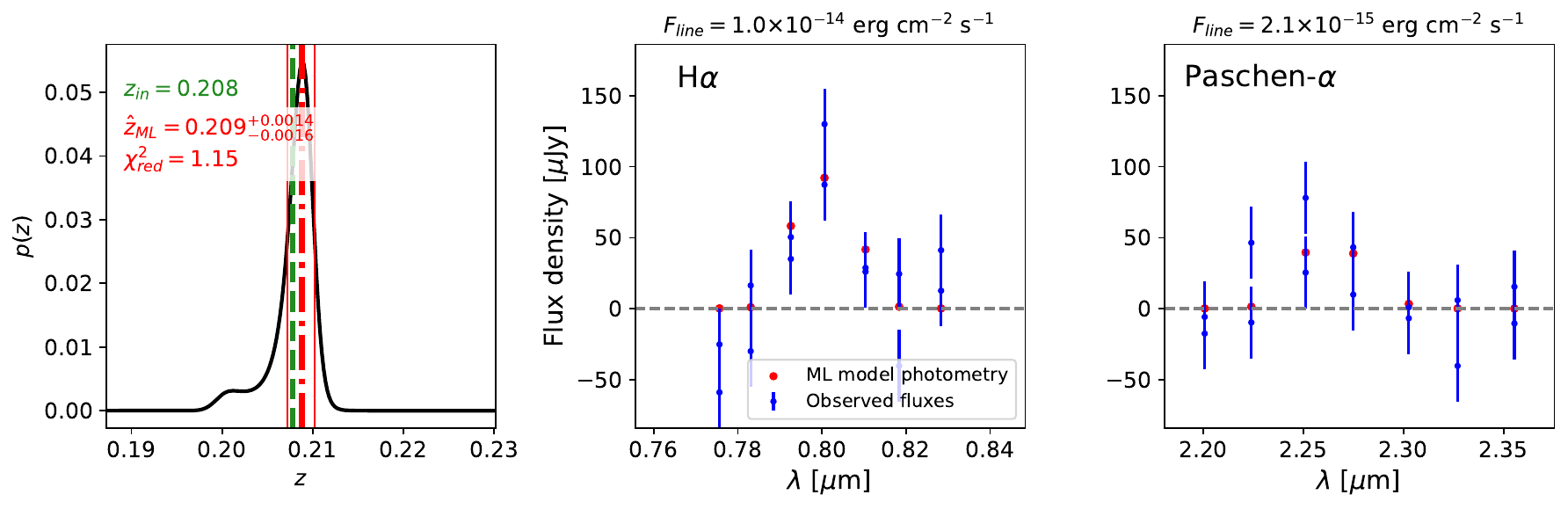}
\caption{Line fits and redshift PDFs for simulated SPHEREx measurements of three isolated, low-redshift galaxies. The left column shows the derived line redshift PDFs -- the green and red dashed lines indicate the input and recovered redshifts for each case, while the solid red lines bound the 68\% credible interval of each PDF. The middle and right columns show the synthetic fluxes (blue) and best-fit model photometry (red) for measurements near \ha\ and Paschen-$\alpha$, respectively.}
\label{fig:indiv_line_likelihood}
\end{figure*}

% We select GAMA sources with either detectable \ha\ or Paschen-$\alpha$ flux $F_{line}>5\times 10^{-16}$ \ergcms. This cut corresponds to the minimum of the 2$\sigma$ line flux sensitivity at $\lambda \sim 4\mu$m (assuming 102 binned channels, see Fig. \ref{fig:line_sens}). 
Figure \ref{fig:indiv_line_likelihood} shows the line fitting results for three $z\sim 0.2$ galaxies with varying levels of detectable \ha\ and Paschen-$\alpha$ emission. The simulated flux measurements use the nominal 102 channel filters with central wavelengths spaced at twice the channel resolution, which approximates Nyquist sampling of the spectral response function. In reality, the observations will have more dispersion in filter locations that depend on the sub-channel (pixel) positions of the sources and the overall survey strategy. The redshift estimates become more precise as the total line SNR increases, with uncertainties ranging from $\sigma_{z}=0.01$ for our faint example down to $\sigma_{z}=0.0015$ for the brightest example.

\begin{figure}
    \centering
    \includegraphics[width=\linewidth]{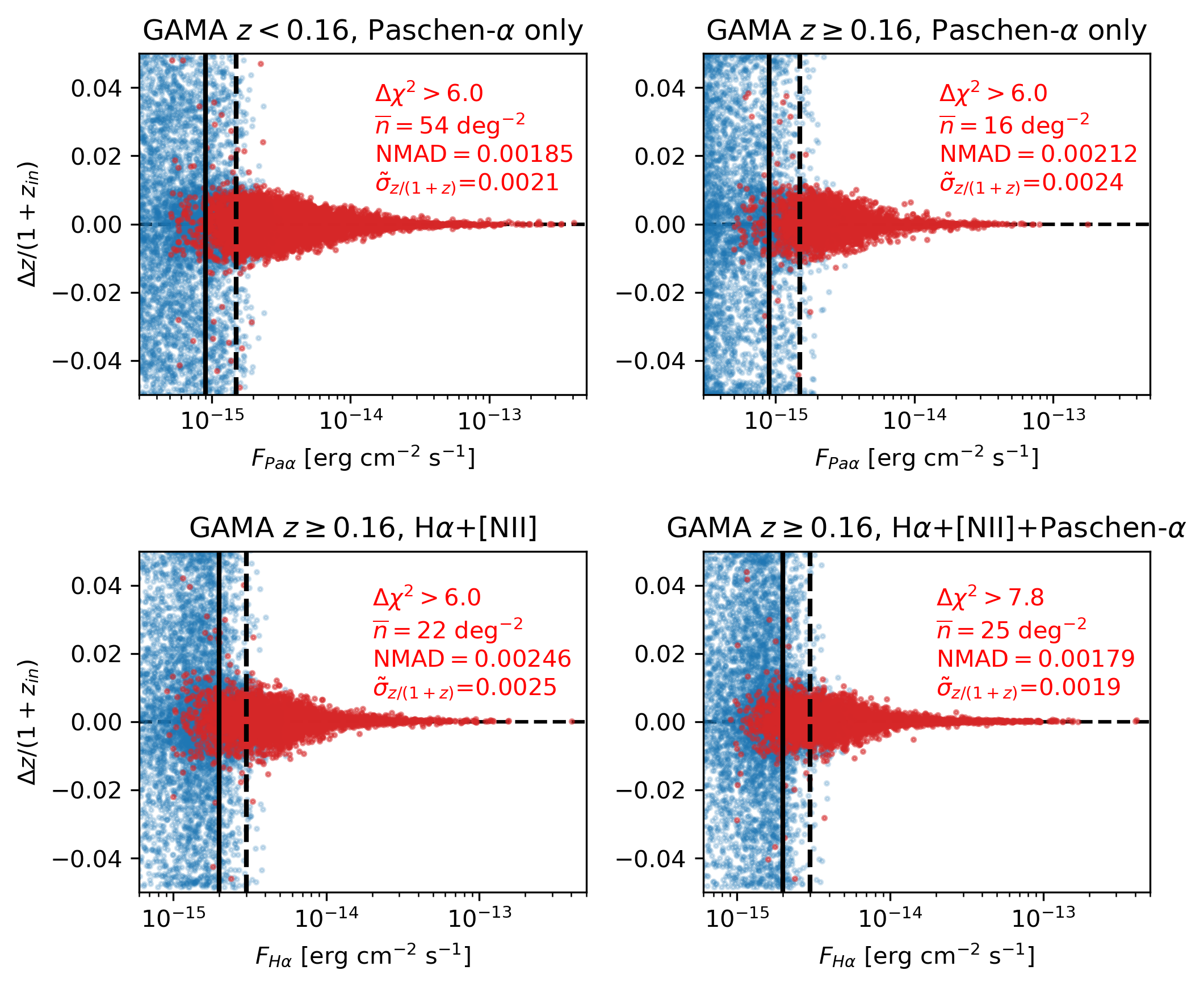}
    \caption{Fractional redshift errors as a function of line flux, for different one- and two-line fits. The red and blue points indicate galaxies with and without ``high-quality fits", respectively, where high-quality fits have a sufficient improvement in the delta log-likelihood compared to a model with no lines ($\Delta \chi^2 > 6$ for single line (complex) fits, $\Delta \chi^2 > 7.8$ for two lines). The mean redshift uncertainty and NMAD are calculated for each high-quality fit sample. The solid and dashed black lines indicate the 2$\sigma$ and 3$\sigma$ line flux sensitivities from Fig. \ref{fig:line_sens}, respectively, at observed wavelengths of 0.8 $\mu$m for \ha\ and 2.0 $\mu$m for Paschen-$\alpha$. We assume a fixed ratio [NII]/H$\alpha$ when fitting the native photometry, though the true line ratios in the simulated lines vary.}
    \label{fig:zdist_one_twoline}
\end{figure}

Figure \ref{fig:zdist_one_twoline} shows redshift errors plotted as a function of line flux for one- and two-line fits. Our redshift estimates are unbiased with redshift errors that decrease from $\delta z/(1+z)\sim 0.01$ down to sub-percent precision for brighter lines. In each case there is a flux limit below which the redshift errors effectively revert to the original redshift range considered, i.e., the lines are uninformative. These thresholds roughly correspond to the 2$\sigma$-3$\sigma$ line flux sensitivities at 0.8 and 2.0$\mu$m. Despite SPHEREx's coarse spectral resolution, in our tests we found that fitting a single Gaussian to the \ha+\nii\ complex resulted in a mild bias $\delta z/(1+z) = -0.001$. This motivates the use of line models that account for the full line complex, as done for \ha+\nii\ in this work, and will be relevant for other cases such as the \oiii+\hb\ complex.

To prevent spurious effects of overfitting we calculate the improvement in $\chi^2$ from the best line model relative to the null model case (i.e., no lines) as a test statistic to place an example cut on the sources with line fits. Assuming a likelihood ratio $\Lambda$ and invoking Wilks' theorem with $n_p=2$ and 3 model parameters for the one- and two-line cases, respectively, the test statistic $ -2\ln \Lambda$ should be $\chi^2$-distributed with $n_p$ degrees of freedom. We identify the subset of lines with best fit models above the 95th percentile of their expected $\chi^2$ distributions (plotted in red), while those below are plotted in blue. These criteria are flexible in the sense that we can specify the desired likelihood ratio threshold, however in general the cut is effective at separating line fits with high accuracy from those unconstrained by the photometry. 

Table \ref{tab:redshift_nmad_vs_lineflux} summarizes the redshift errors as a function of \ha\ and Paschen-$\alpha$ line fluxes after making cuts on $\Delta\chi^2$. As seen by eye in Fig. \ref{fig:zdist_one_twoline}, the errors decrease monotonically with increasing line flux. The single-line redshift errors for \ha\ and Paschen-$\alpha$ are of similar size at fixed line SNR. When \ha\ and Paschen-$\alpha$ are fit together, the dispersion of redshift errors is $\sim 30\%$ smaller than that from fitting \ha\ alone to the same set of sources. Broadly speaking, these results demonstrate that when one or several lines are detectable and correctly identified, it is possible to recover highly accurate redshifts.  

% Within the GAMA sample 61\% of star-forming galaxies have redshifts $z<0.16$, meaning the primary emission line in the SPHEREx bandpass is Paschen-$\alpha$. 

% \richard{ The accuracy increases for the Paschen-$\alpha$ sample with line SNR: of the 80\% of sources with $z<0.16$ and $F_{P\alpha}>10^{-15}$ \ergcms\, the NMAD values for subsamples $F_{P\alpha}>\lbrace 10^{-15}, 2\times10^{-15}, 5\times 10^{-15}, 10^{-14}\rbrace$ \ergcms\ are $\lbrace 0.0017, 0.0013, 0.0007, 0.0005\rbrace$, respectively. For $z\geq 0.16$, 6993/14043 galaxies have either $F_{H\alpha}>5\times 10^{-15}$ \ergcms\ or $F_{P\alpha}>1\times 10^{-15}$ \ergcms\ (compared to 6917 with $F_{H\alpha}>5\times 10^{-15}$ \ergcms\ alone).} 

% \Oli{Nice. Do we have examples of these at higher cosmologicaly interesting redshifts?}

\begin{table*}[]
    \centering
    \begin{tabular}{c|c|c|c|c|c|c|c|c}
        $F_{H\alpha}$ [$\times 10^{-15}$ erg cm$^{-2}$ s$^{-1}$] & &
        $[2, 3)$ & $[3,5)$ & $[5,7)$ & $[7, 10)$ & $[10,50)$ & $[50,100)$ & $[100,1000)$ \\
        \hline 
        \hline
    NMAD ($H\alpha$ only) & & 0.0048 & 0.0041 & 0.0028 & 0.0022 & 0.0011 & 0.0003 & 0.0001 \\
    NMAD ($H\alpha$+P$\alpha$) & & 0.0049 & 0.0038 & 0.0023 & 0.0014 & 0.0005 & 0.0001 & $<0.0001$  \\
    $F_{P\alpha}$ [$\times 10^{-15}$ erg cm$^{-2}$ s$^{-1}$] & 
        $[0.8, 2)$ & $[2,3)$ & $[3,5)$ & $[5, 7)$ & $[7,10)$ & $[10,50)$ &$[50,100)$ & $[100,1000)$ \\
        \hline 
        \hline
    NMAD (P$\alpha$ only) & 0.0038 & 0.0027 & 0.0018 & 0.0012 & 0.0008 & 0.0004 & 0.0001 & 0.0001 \\ NMAD ($H\alpha$+P$\alpha$) & 0.0020 & 0.0014 & 0.0010 & 0.0007 & 0.0006 & 0.0003 & 0.0001 & $<0.0001$
    \end{tabular}
    \caption{Normalized median absolute deviation (NMAD) of redshift errors $(\hat{z}-z_{in})/(1+\hat{z})$ for the GAMA sample, separated as a function of $H\alpha$ and P$\alpha$ line fluxes. These galaxies come from our quality cuts based on the line fits ($\Delta \chi^2 > 6$ for single lines, $\Delta \chi^2 > 7.8$ for two lines).}
    \label{tab:redshift_nmad_vs_lineflux}
\end{table*}

Isolating the emission line measurements from the full SED fits enables redshift estimates that are more robust to the details of the SED model, however this technique relies on some prior estimate for the redshift and thus the line center(s). The quality of the continuum redshift and errors in the continuum model near the lines will primarily impact the purity of the recovered line redshifts. Using continuum-driven line ratio priors can potentially improve redshift estimates for many ELGs, however they may not be appropriate in all cases. This procedure is being considered by the flight software pipeline team to account for as-observed spectrophotometry, however a detailed investigation of the technique and its limits are left to future work.

% \richard{The flux covariance of the raw measurements will depend on the details of the scan strategy and the overlap of the transmission profiles within a channel.}

% \Oli{Do we have a well established procedure to implement this? More of an internal questions? Is it well defined after this nice exercise? If so, maybe worth laying out here as a reference point. }

\subsection{Redshift validation}
While our use of synthetic data allows us to directly quantify redshift errors, spectroscopic data from existing surveys will be used in practice to validate SPHEREx redshifts. This approach has been used to perform redshift validation for existing surveys including the Dark Energy Survey \citep[DES;][]{DES_zcal} and near-future surveys conducted with \emph{Euclid} \citep{euclid_zcal}. Such a procedure relies on a set of independent, high-resolution spectroscopic measurements that span the spectro-photometric color space occupied by SPHEREx galaxies, and may motivate targeted spectroscopic surveys such as the Complete Calibration of the Color-Redshift Relation survey \citep[C3R2;][]{C3R2} to fill in observational gaps. 

\section{Conclusion}
\label{sec:conclusion}

In this work we have presented a set of galaxy SEDs that combine multi-wavelength fits to existing photometry with an emission line prescription rooted in empirical scaling relations. After validating that the line model is consistent with a number of existing measurements, we generate mock photometry from the synthetic SEDs observed across the SPHEREx bandpass using existing sensitivity estimates. We then demonstrate that precise, accurate redshifts can be obtained using continuum and line information from simulated full-sky depth photometry. These redshift simulations will form the basis for downstream clustering measurements through the power spectrum and bispectrum. 

% Characterizing redshift errors will be a major effort for SPHEREx as the selection functions for clustering analyses are developed.

% \Oli{One thing we should check is that these results are consistent with the number densities available in our public website. Could you please use the same bins in $z$ and $\sigma(z)/(1+z)$ and check that? You can for now clip at 3 $\sigma$ and ignore the outlier issue. You just need to compute a comoving volume for each redshift slide for the effective COSMOS area. It would be good to have these numbers and make sure we are consistent or even better. We could then write so here.}

Due to the SPHEREx survey strategy, the observing depth in the NEP and SEP (100 deg$^2$ each) will be considerably deeper than that of the full sky survey, with measurements along each line of sight that are both spatially and spectrally dithered, oversampling the SPHEREx response function. Given the dearth of existing NIR spectroscopic measurements, galaxy spectra from the deep fields will be useful for refining the galaxy templates used for the full sky survey (the results of this work assume by construction that the templates are representative of the SPHEREx photometry). Forecasts on emission lines will also become more refined with observations from upcoming spectroscopic surveys such as DESI and PFS and at higher redshift by JWST. However, obtaining reliable spectra will rely on a proper treatment of source confusion, which will be much more pronounced. This is not addressed in this work as we do not directly perform forced photometry on mock observations. A dedicated study of deep field photometry and implications for calibrating the full sky survey are left to future work.

We show (under idealized conditions) that by directly fitting the spectrally dithered flux measurements and assuming a mild redshift prior derived from the galaxy continuum, it is possible to recover reliable redshifts with accuracy approaching that of higher-resolution spectroscopic surveys. More work is needed to implement this technique for more realistic cases that incorporate errors in line-continuum separation, errors due to line interlopers and confusion noise from sources along the same line of sight. Despite these additional complexities, it should be possible to obtain accurate redshifts when lines are detected at moderate significance. These results may also be improved in the limit of more dithered measurements (as in the deep fields).

% \Oli{No prospect for higher redshift over the all-sky you think?}

The synthetic catalogs from this work do not include other objects such as stars and active galactic nuclei (AGN). As we only simulate galaxies in this work, our redshift predictions assume that source classes are properly separated. A similar empirical approach to this work may be used with a set of star and/or AGN templates \citep[e.g.,][]{brownagn} to generate synthetic SEDs. While not explored in this work, SPHEREx has an advantage for star-galaxy-AGN separation because of its broad spectral coverage and intermediate spectral resolution, however external information may be needed for certain cases. 

Recent studies have shown that the assumption of universality in the halo mass function is a poor description for relating the pNG bias $b_{\phi}$ to linear galaxy biases $b_1$ \citep{barreira20}. However, by relaxing the universality assumption and choosing informed galaxy sub-samples, it may be possible to improve constraints on $f_{NL}$ beyond those currently assumed \citep{barreira23, sullivan23}. The diversity of SPHEREx galaxy types provides opportunities to isolate samples with different linear biases $b_1$ and non-Gaussian biases $b_{\phi}$ for multi-tracer analyses using the power spectrum and potentially higher-order statistics such as the bispectrum.

We release our full synthetic COSMOS and GAMA catalogs to the public (these along with high-resolution SEDs are available upon request from the corresponding author). While these simulations will form the basis for tests of the SPHEREx redshift pipeline, the synthetic line catalogs and mock spectra should also be useful for surveys beyond SPHEREx. Our modeling framework is implemented in the tool \texttt{CLIPonSS}. The tool is flexible and can be tailored to other use cases which require consistent modeling of line features and continuum properties.

While more work is needed to match the realism of the full SPHEREx survey and to soon process observed samples, we have demonstrated a redshift procedure which is effective and meets the SPHEREx science requirements with margin, laying the groundwork for measurements of galaxy clustering on both small and large scales.

% \Oli{Maybe a short section saying that the redshift procedure works and meets requirements even if more work is needed.}

\section*{acknowledgements}
Part of this work was done at the Jet Propulsion Laboratory, California Institute of Technology, under a contract with the National Aeronautics and Space Administration. We also acknowledge support from the SPHEREx project under a contract from the NASA/GODDARD Space Flight Center to the California Institute of Technology. We thank Sylvain de la Torre for providing cross-matched versions of the zCOSMOS/3D-HST catalogs with our synthetic catalog for direct source comparison. We also thank Sean Bryan and the SPHEREx survey team for providing the deep field coverage maps shown in this work. More information on the COSMOS survey is available at https://cosmos.astro.caltech.edu. GAMA is a joint European-Australasian project based around a spectroscopic campaign using the Anglo-Australian Telescope. The GAMA input catalogue is based on data taken from the Sloan Digital Sky Survey and the UKIRT Infrared Deep Sky Survey. Complementary imaging of the GAMA regions is being obtained by a number of in-dependent survey programmes including GALEX MIS, VST KiDS, VISTA VIKING, WISE, HerschelATLAS, GMRT and ASKAP providing UV to radio coverage. GAMA is funded by the STFC (UK), the ARC (Australia), the AAO, and the participating institutions. The GAMA website is http://www.gama-survey.org/.
\newpage
\appendix 
\section{SED fits of GAMA sources and redshift recovery}
\begin{figure*}
\label{fig:gamafitcat}
  \centering
    \includegraphics[width=0.95\linewidth]{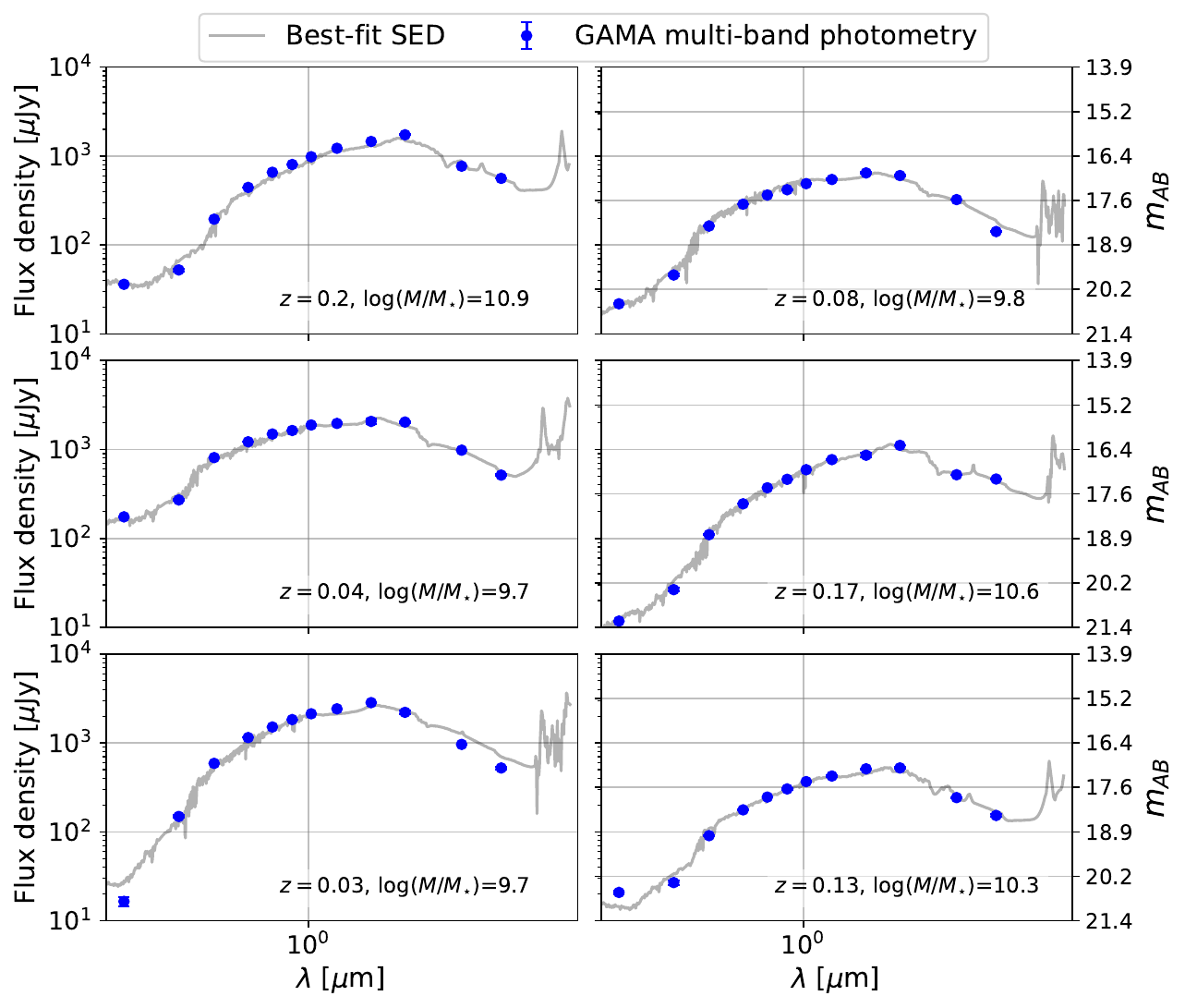}
    \caption{Six example fits to 12-band photometry of GAMA sources with $i < 18$ (blue points), with best fit SEDs shown in grey. These sources will be measured by SPHEREx at high SNR across many channels.}
\end{figure*}

\begin{figure*}
    \centering
    \includegraphics[width=0.8\linewidth]{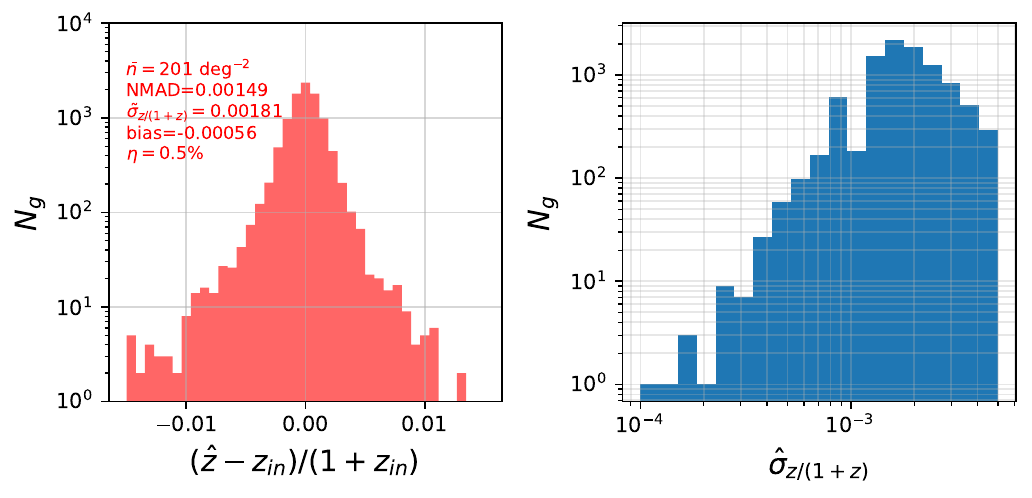}
    \caption{GAMA bright source catalog redshift results. The redshift error distribution is shown on the left, with the black dashed curve following a Gaussian whose width matches $\bar{\sigma}_{z/(1+z)}$. The right panel shows the distribution of reported uncertainties for the same sample.}
\label{fig:lp_zml_sort_sigmaz_zerr_gama}
\end{figure*}

In this section we include example SED fits to multiband photometry of GAMA galaxies described in \S \ref{sec:gama}. Each galaxy in the catalog we use has spectroscopic redshifts and cross-matched photometry from GALEX (FUV/NUV), the VST Kilo-Degree Survey (KiDS, $ugri$), the near-infrared VISTA VIKING survey (ZYJH$K_s$) and WISE all-sky infrared data (W1 and W2) \citep{GAMAphot}. A sample of our fits is shown in Fig. \ref{fig:gamafitcat}. 
The synthetic GAMA catalog represents a bright, low-redshift population for which SPHEREx will measure many precise redshifts. We perform template fitting on synthetic SPHEREx 102-band continuua for ten thousand galaxies of the 44K GAMA sample, also including WISE W1/W2 and DECaLS $grz$ photometry. To account for photometric errors of bright sources we include a 1\% uncertainty floor on the external photometry. In addition, given the GAMA sample is comprised of low-redshift galaxies, we deploy a finer grid from $z=0$ to $z=1$ with $\delta z=0.001$. We show the results of this test in Fig. \ref{fig:lp_zml_sort_sigmaz_zerr_gama}. The majority of sources from this sample (85\%) have redshift uncertainties $\sigma_{z/(1+z)}<0.003$, contributing an additional $\bar{n}=170$ deg$^{-2}$ to our high accuracy sample. This sample has a similar NMAD to our COSMOS sample, with a slightly lower outlier fraction to our COSMOS sample but a considerably smaller NMAD, which is expected since this comprises the bright end distribution of the SPHEREx galaxy sample. There is a small $-0.3\sigma$ bias in the sample which warrants further investigation.
\section{Validation of redshift estimates using z-scores}
\label{sec:zscore_appendix}
\begin{figure*}
    \centering
    \includegraphics[width=0.95\linewidth]{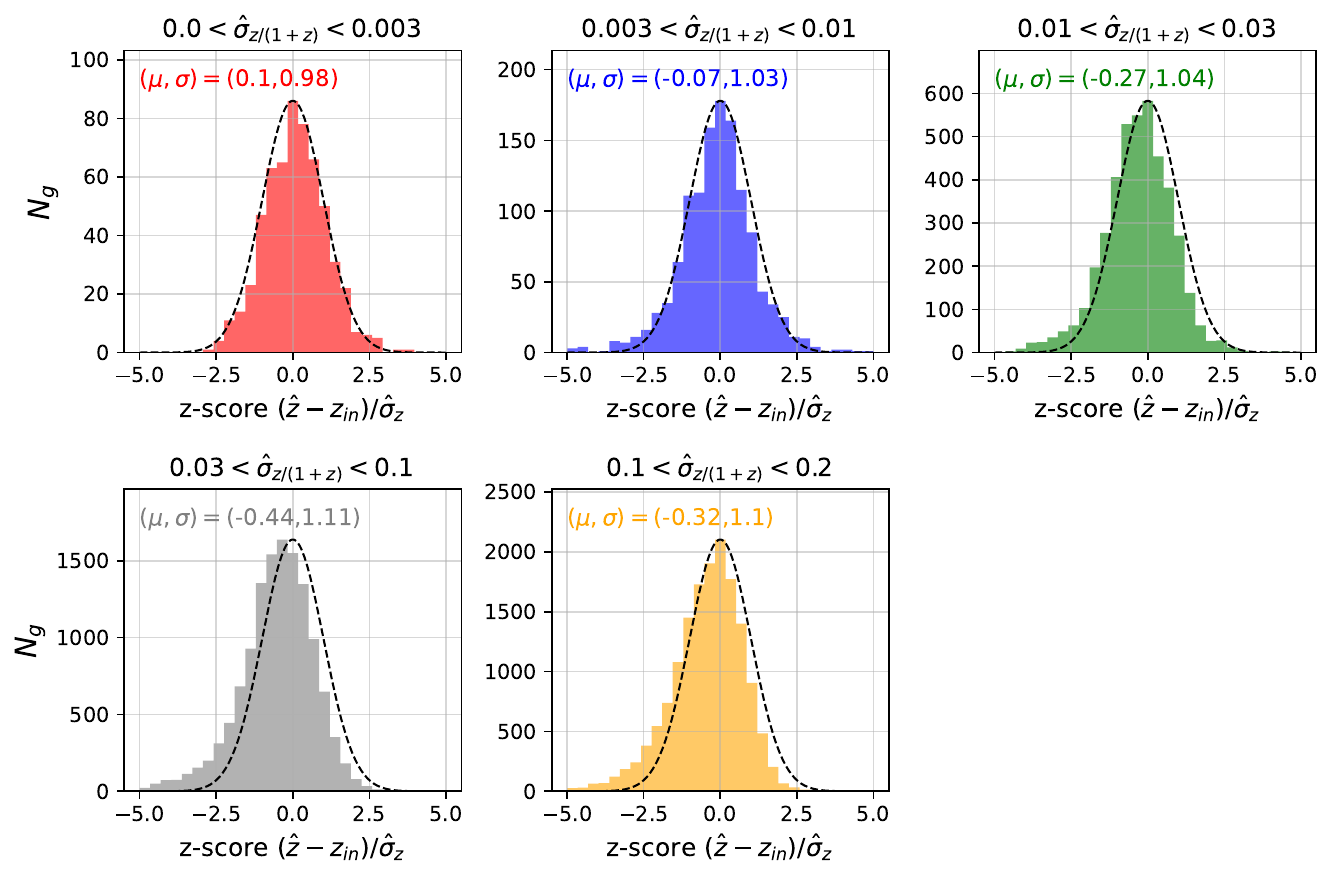}
    \caption{Distribution of z-scores for COSMOS2020 continuum redshift estimates, binned in redshift uncertainty $\hat{\sigma}_{z/(1+z)}$. The black dashed curves indicate a Gaussian with unit variance, i.e., the desired z-score distribution. The $(\mu, \sigma)$ labels indicate the mean and standard deviation of the z-score histograms.}
    \label{fig:zscore_dist_C20}
\end{figure*}

To test the reliability of photometric redshift uncertainties we calculate the z-score distribution for our redshift results. The z-score for an individual estimate is given by $Z=(\hat{z}-z_{in})/\hat{\sigma}_z$ (with capital $Z$ indicating a z-score, not a redshift) and should be unit Gaussian-distributed if the uncertainties statistically match the true errors. In Figure \ref{fig:zscore_dist_C20} we show the z-score distributions for the nominal COSMOS2020 results in \S \ref{sec:cos_cont_results}, grouped in the same redshift uncertainty bins. This distribution is qualitatively similar in shape to the redshift error distributions, with a negative tail of outliers for the medium- and low-accuracy samples. For the first three uncertainty bins, the distribution of z-scores indicates consistency between reported uncertainties and true errors within 5\%. For the two lowest accuracy samples, the widths of the z-score distributions suggest that the reported uncertainties are $\sim10\%$ overconfident, though this may also be driven by the larger mean biases and outlier tails.
Stricter tests, such as those utilizing the probability integral transform (PIT), can be used to test the reliability of the full $p(z)$ distribution, for example near the tails of the distribution. These metrics will be important for assessing the reliable $p(z)$ information that gets passed downstream to clustering measurements.
\bibliography{biblio}{}
\bibliographystyle{aasjournal}
\end{document}